\documentclass[reprint, floatfix, superscriptaddress, amsmath, amssymb, longbibliography]{revtex4-2}

\pdfoutput=1
\usepackage[utf8]{inputenc}
\usepackage[T1]{fontenc}

\usepackage{graphicx}
\usepackage{dcolumn}
\usepackage{bm}
\usepackage{verbatim}
\usepackage{booktabs}
\usepackage{footnote}
\usepackage{xcolor}
\usepackage{float}
\usepackage{amsmath}
\usepackage[most]{tcolorbox}
\usepackage{soul}
\usepackage{mathtools}
\usepackage{siunitx}
\usepackage{braket}
\usepackage{circuitikz}
\usepackage{tikz}
\usetikzlibrary{arrows.meta, positioning, calc, decorations.pathmorphing, shapes.multipart}
\usepackage{blkarray}
\usepackage[colorlinks=true,linkcolor=blue,allcolors=blue]{hyperref}

\usepackage{mleftright}

\newcommand{\figpanel}[2]{Fig.~\hyperref[#1]{\ref*{#1}(#2)}}
\newcommand{\figpanels}[3]{Fig.~\hyperref[#1]{\ref*{#1}(#2)--(#3)}}
\newcommand{\figpanelNoPrefix}[2]{\hyperref[#1]{\ref*{#1}(#2)}}

\begin{document}

\author{Mohammed Alghadeer}
\email{mohammed.alghadeer@physics.ox.ac.uk} 
\affiliation{Department of Physics, Clarendon Laboratory, University of Oxford, OX1 3PU, UK}

\author{Simon Pettersson Fors}
\affiliation{Department of Microtechnology and Nanoscience, Chalmers University of Technology, 412 96 Gothenburg, Sweden} 

\author{Shuxiang Cao}
\affiliation{Department of Physics, Clarendon Laboratory, University of Oxford, OX1 3PU, UK}

\author{Simone D. Fasciati}
\affiliation{Department of Physics, Clarendon Laboratory, University of Oxford, OX1 3PU, UK}

\author{Haru Ishizaka}
\affiliation{Department of Physics, Clarendon Laboratory, University of Oxford, OX1 3PU, UK}

\author{Anton Frisk Kockum}
\affiliation{Department of Microtechnology and Nanoscience, Chalmers University of Technology, 412 96 Gothenburg, Sweden}

\author{Peter Leek}
\affiliation{Department of Physics, Clarendon Laboratory, University of Oxford, OX1 3PU, UK}

\author{Mustafa Bakr}
\email{mustafa.bakr@physics.ox.ac.uk} 
\affiliation{Department of Physics, Clarendon Laboratory, University of Oxford, OX1 3PU, UK}

\title{Crosstalk Dispersion and Spatial Scaling in Superconducting Qubit Arrays}

\begin{abstract}  

Crosstalk between qubits fundamentally limits the scalability of quantum processors, necessitating physics-based models that can handle the complexity of large qubit arrays. Here, we develop a comprehensive theoretical and experimental framework that captures residual interactions between both adjacent and non-adjacent qubits in fixed-frequency transmon lattices. The model integrates the combined effects of exponential localization in banded capacitance matrices, suppression of virtual couplings through detuning products across intermediate modes, and evanescent decay of below-cutoff electromagnetic fields, yielding predictive scaling relations for coupling strength as a function of spatial separation and spectral detuning. Experimental characterization of a $4 \times 4$ superconducting-qubit lattice with inductive shunt pillars reveals exponential spatial decay and frequency-dependent suppression consistent with theoretical predictions, achieving quantitative agreement for all nearest-neighbor couplings across the \qtyrange[range-phrase = --, range-units = single]{4}{6}{\giga\hertz} operating range. Our results show that standard dispersive Hamiltonian approximations systematically overestimate long-range coupling when spatial and spectral dependencies are neglected; these errors propagate into circuit simulation and design strategies. Our framework provides design guidance for crosstalk mitigation in larger-scale quantum processors under realistic fabrication constraints, addressing a bottleneck in scalability.

\end{abstract}

\maketitle

\section{Introduction}

Large-scale integration of superconducting qubits has enabled remarkable progress in quantum computing, with demonstrations of quantum supremacy and error correction reaching hundreds of physical qubits~\cite{Arute2019, megrant2025scaling, google2023suppressing, acharya2025quantum}. However, this scaling has revealed fundamental system-level challenges that were negligible in smaller devices, particularly parasitic couplings and control-signal crosstalk, which become increasingly relevant as devices scale~\cite{mohseni2024build}. 
These interactions generate correlated errors that violate the fundamental assumption of independent error channels in quantum error correction, making them a major obstacle to fault-tolerant operation. In fixed-frequency devices, circuit-mediated residual ZZ interactions~\cite{krinner2020benchmarking} are a particularly important source of such correlations. Unlike independent stochastic gate errors that can be addressed through improved fabrication and calibration, these correlated effects persist even as individual gate fidelities approach theoretical limits~\cite{shor1995scheme, lacroix2024scaling}. This elevates crosstalk mitigation from a performance optimization concern to a fundamental requirement for maintaining logical qubit coherence in error-corrected quantum processors~\cite{eickbusch2024demonstrating}.

Addressing these correlated errors at the hardware level requires careful circuit engineering, navigating seemingly conflicting requirements: qubits should be robustly coupled for fast, high-fidelity gates while at the same time suppressing spurious interactions~\cite{le2023scalable}. Increasing qubit counts while maintaining gate fidelity and minimizing hardware complexity under these demands remains an unresolved engineering challenge~\cite{mohseni2024build}. For example, superconducting transmon qubits~\cite{transmon} are typically coupled via lithographically defined capacitors~\cite{alghadeer2025characterization}, and these couplings must be carefully tuned to implement high-fidelity two-qubit gates~\cite{yan2018tunable}. However, any residual parasitic coupling between qubits~\cite{ketterer2023characterizing, tripathi2022suppression, murali2020software} gives rise to persistent, state-dependent ZZ interactions~\cite{krinner2020benchmarking, zhao2022quantum, fors2024comprehensive}. While strong ZZ coupling can be harnessed for implementing CPHASE or CZ gates~\cite{chow2011simple, collodo2020implementation, sung2021realization, stehlik2021tunable, long2021universal, chu2021coupler}, any non-negligible ZZ shift degrades the fidelity of simultaneous single- and two-qubit gates and, by extension, limits overall device performance~\cite{chen2023voltage, ganzhorn2020benchmarking, mckay2019three, bharti2022noisy}. Suppressing such residual interactions is therefore essential for maintaining high two-qubit gate fidelities in large-scale processors. 

Beyond direct qubit–qubit coupling, other architecture-dependent pathways---such as residual interactions through shared control~\cite{Bakr2025JJMetasurfaces, Bakr2025LongRangeJJM} and readout lines~\cite{Bakr2025ReentrantReadout}, DC and AC flux crosstalk in tunable devices~\cite{Fasciati2024GeometricShunt}, and spurious enclosure modes---can also mediate crosstalk~\cite{kosen2024signal, das2024reworkable, huang2021microwave, spring2020modeling, spring2022high}. While important in certain architectures, these mechanisms are not the focus of this work, which concentrates on static qubit–qubit interactions in fixed-frequency-qubit arrays. Throughout this work, we use the term crosstalk to refer to static cross-couplings as they contribute to unwanted, state-dependent interactions during circuit operation. For this class of interactions, parasitic couplings can scale unfavorably as circuit size and complexity increase, amplifying interactions negligible in small systems. ZZ-type crosstalk now constitutes one of the dominant sources of coherent error in fixed-frequency architectures, limiting simultaneous gate performance and constraining logical qubit overheads in surface codes~\cite{acharya2025quantum} and low-density-parity-check codes~\cite{Mathews2025}. Although it is known that parasitic ZZ interactions can increase with system size and architectural complexity, their detailed spatial scaling, strength, and architecture dependence are not yet quantitatively understood. Establishing reliable scaling laws is therefore of direct practical importance for error-corrected quantum computing. 

To develop such predictive models for crosstalk scaling, we require a description that captures both intended nearest-neighbor interactions and parasitic long-range couplings in a unified framework. The challenge lies in systematically accounting for coupling mechanisms that span multiple energy and length scales, from direct capacitive interactions at the device level to cavity-mediated processes involving the electromagnetic environment. Existing approaches often rely on simplified two-qubit models or phenomenological coupling constants, limiting their applicability to realistic multi-qubit architectures where multiple interaction pathways contribute simultaneously to the overall system dynamics.

In this work, we develop a comprehensive theoretical framework that integrates electromagnetic and quantum crosstalk analysis for fixed-frequency transmon arrays. Our approach combines nodal circuit analysis and electromagnetic Green’s-function theory with perturbation theory in circuit quantum electrodynamics to predict coupling strengths as a function of inter-qubit separation, frequency detuning, and device geometry. We validate this framework through detailed comparison with experimental data from a $4 \times 4$ transmon lattice, demonstrating quantitative agreement across multiple coupling channels and frequency configurations. The resulting model establishes design principles for engineering crosstalk-resilient quantum architectures, showing how strategic frequency allocation, lattice spacing, and enclosure design can suppress state-dependent static interactions and preserve simultaneous-gate performance. This combined approach provides the predictive capability needed for scaling to larger quantum processors while maintaining error-correction viability.

This article is organized as follows. In Section~\ref{probstat}, we formulate a unified Hamiltonian that combines circuit-mediated and enclosure-mediated exchange processes in fixed-frequency transmon arrays. In Section~\ref{emcircuitcrosstalk}, we develop two complementary classical frameworks---nodal circuit analysis and electromagnetic Green's-function theory---to derive the spatial and spectral scaling of residual couplings. Then, in Section~\ref{sec:models_for_zz_couplings}, we apply quantum perturbation theory to obtain analytical expressions for nearest- and next-nearest-neighbor ZZ interactions based on these residual coupling strengths. In Section~\ref{sec:results}, we present experimental measurements from a $4 \times 4$ transmon lattice and compare them directly with the theoretical scaling model. We conclude in Section~\ref{sec:discussion_conclusion} by discussing architectural implications of our results for crosstalk mitigation and scalable superconducting quantum processors. Technical derivations and experimental details are provided in Appendices~\ref{matrixinversion}--\ref{app:simulations}.

\section{Circuit- and enclosure-mediated crosstalk}
\label{probstat}
\begin{figure}
\centering
   \includegraphics[width=\linewidth]{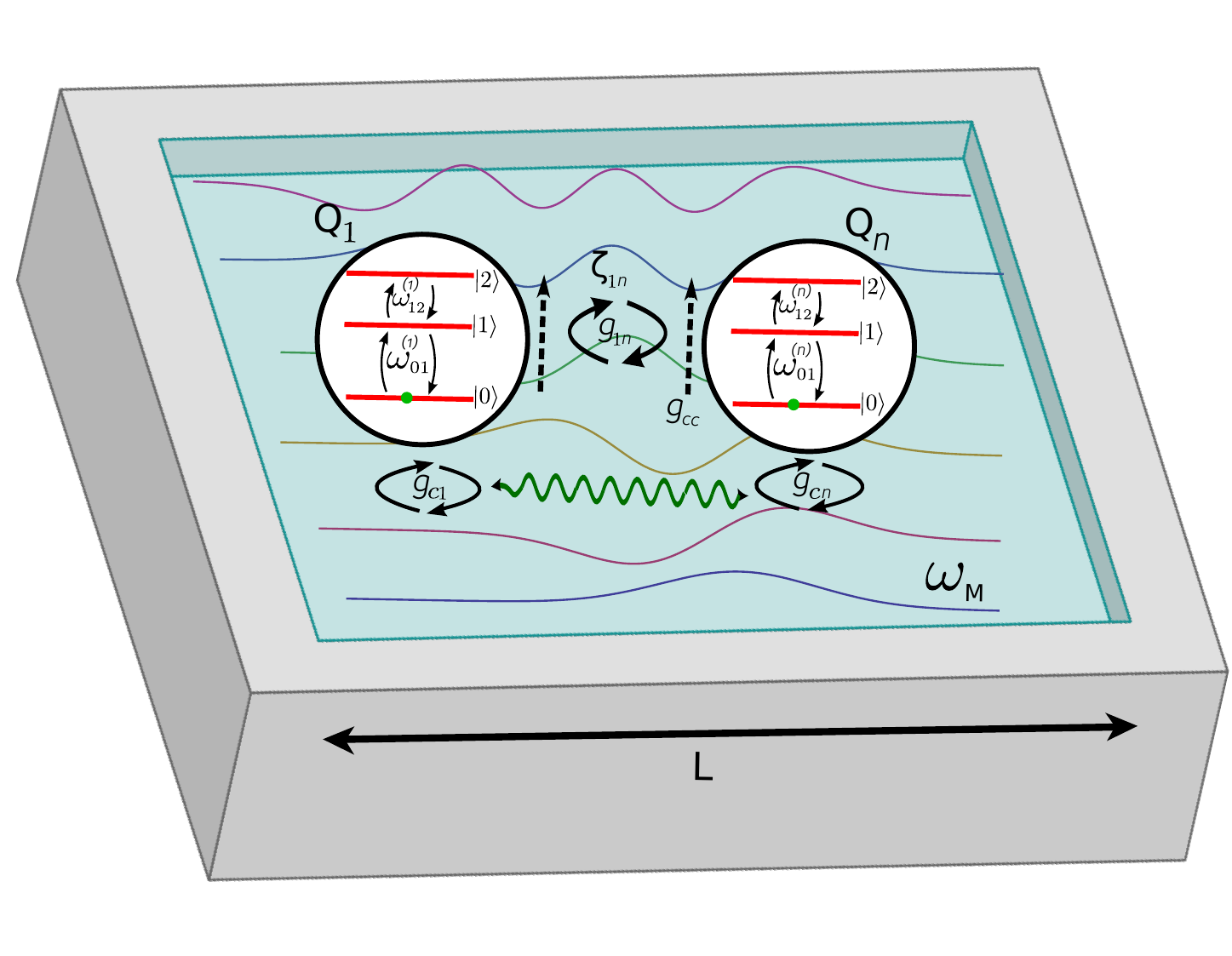}
   \caption{Cross-sectional and three-dimensional schematic of a rectangular cavity enclosure containing transmon qubits $Q_{1}$ and $Q_{n}$, where $Q_{n}$ denotes an arbitrary qubit among the total of $N$ qubits. Two primary crosstalk pathways are illustrated: (i) cavity-mediated electromagnetic coupling through evanescent enclosure modes below cutoff (green arrow), characterized by coupling strengths $g_{c1}$, $g_{cn}$, and the cavity--cavity coupling $g_{cc}$; and (ii) direct circuit-mediated exchange interactions $g_{1n}$ arising from the inverse capacitance matrix $C^{-1}$. Each transmon displays its anharmonic energy-level structure---ground state $\lvert 0\rangle$, first excited state $\lvert 1\rangle$, and second excited state $\lvert 2\rangle$---along with the transition frequencies $\omega_{01}^{(1)}$ and $\omega_{12}^{(1)}$, etc. The colored standing-wave patterns represent electromagnetic enclosure modes that mediate long-range interactions, and $g_{1n}$ denotes the effective direct coupling between distant qubits $Q_{1}$ and $Q_{n}$. The parameter $\zeta_{1n}$ denotes the corresponding static ZZ interaction strength, defined as the state-dependent frequency shift $\zeta_{1n} = \omega_{1n}^{(11)} - \omega_{1n}^{(10)} - \omega_{1n}^{(01)} + \omega_{1n}^{(00)}$, which quantifies the correlated error induced by residual couplings.
   }
   \label{fig1:circuit}
\end{figure}

We consider a two-dimensional lattice of $N$ fixed-frequency transmon qubits embedded in a microwave enclosure (schematically illustrated for two qubits in Fig.~\ref{fig1:circuit}). Crosstalk arises through two pathways: (i) \emph{direct} circuit-mediated exchange set by the inverse capacitance matrix of the layout, and (ii) \emph{indirect} enclosure-mediated exchange via the cavity electromagnetic modes.

To model the direct couplings through the circuit lattice and the indirect couplings via the intermediate cavity modes, we assume, here, that the cavity states have large excitation energies compared to the transmons. In our devices, the inductively shunted enclosure raises the lowest cavity-mode frequencies to $\gtrsim$~\qtyrange[range-phrase = --, range-units = single]{30}{35}{\giga\hertz}, placing them far above the \qtyrange[range-phrase = --, range-units = single]{4}{6}{\giga\hertz} transmon band. This large detuning ensures that all enclosure modes are far in the dispersive regime, remain unpopulated during operation, and contribute only through virtual transitions. We therefore treat the cavity modes as being in their ground states, so that their internal dynamics is negligible and their only effect on the transmons is to mediate indirect couplings. These mediated interactions between the transmons through the cavity states are sometimes referred to as virtual interactions. Under these assumptions, we write the model Hamiltonian as
\begin{equation}
    H_\mathrm{model} = H_\mathrm{circuit} + V_\mathrm{enclosure},
    \label{eq:Hamiltonian_model_1}
\end{equation}
where $H_\mathrm{circuit}$ models the circuit physics and $V_\mathrm{enclosure}$ describes the enclosure-mediated couplings between the transmons. 

\subsection{Circuit Hamiltonian and direct exchange}
We now examine the first component: the circuit Hamiltonian $H_\mathrm{circuit}$. Each transmon $i$ consists of a Josephson junction with energy $E_{J,i}$ shunted by a capacitor with charging energy $E_{C,i} \approx e^2/(2C_i)$, where $C_i$ is the total capacitance to ground. These capacitances are the diagonal elements of the capacitance matrix $\mathrm{C}$, which characterizes the classical circuit. The off-diagonal elements $C_{ij}$ ($i \neq j$) are the mutual capacitances between the transmons labeled with $i$ and $j$. We use single indices to denote the diagonal elements $C_i \equiv C_{ii}$, and also for the charging-energy matrix $E_{C,i} = E_{C,ii}$.

Following standard circuit quantization procedures~\cite{blaiscqed}, we promote the classical node charges $n_i$ and superconducting phases $\phi_i$ in the circuit to quantum operators satisfying the commutation relation $\mleft[ \mathrm{e}^{\pm \mathrm{i} \phi_i}, n_j \mright] = \mp \delta_{ij} \mathrm{e}^{\pm \mathrm{i} \phi_i}$, where $\delta_{ij}$ is the Kronecker delta. The circuit Hamiltonian then takes the form
\begin{equation}
    H_{\text{circuit}} = \sum_{i,j=1}^N 4 E_{C,ij} (n_i - n_{g,i})(n_j - n_{g,j}) + \sum_{i=1}^N E_{J,i} \cos \phi_i,
\label{eq:circuit_hamiltonian}
\end{equation}
where $n_{g,i}$ is the charge offset and $E_{C,ij} = e^2 \mathrm{C}^{-1}_{ij}/2$ is given by the inverse capacitance matrix $\mathrm{C}^{-1}$. A critical observation is that even when the capacitance matrix $\mathrm{C}$ exhibits sparse structure (e.g., only nearest-neighbor capacitive coupling), its inverse $\mathrm{C}^{-1}$ is generally dense, introducing long-range interactions throughout the circuit. In a one-dimensional chain with comparable mutual capacitances, these long-range interactions described by the entries of $\mathrm{C}^{-1}$ decay exponentially with distance~\cite{YaoBakrHunter2019Modal, MusondaBakr2023Singlet}. More generally, on two-dimensional grids this exponential off-diagonal decay holds in the Manhattan (graph) distance provided the capacitance matrix is finite-range and gapped (e.g., strict diagonal dominance / below-cutoff packaging); see Appendix~\ref{matrixinversion} for Demko-type bounds, a 1D Toeplitz inversion, and conditions under which exponential locality can fail. 
 
\subsection{Effective circuit Hamiltonian}
Given the charging and Josephson energies, the circuit Hamiltonian in Eq.~\eqref{eq:circuit_hamiltonian} gives an exact description of the circuit physics. However, the exact model does not lend itself, in general, to developing simple analytical models for measurable quantities connected to the crosstalk, e.g., the ZZ couplings in the system. The analytical models are key since we use them to infer the mutual couplings, i.e., $E_{C,ij}$, from experimentally measured ZZ couplings. Here, we discuss the approximate Hamiltonian, also referred to as an effective Hamiltonian, which we later apply in Section~\ref{sec:models_for_zz_couplings} to relate the mutual couplings and the ZZ couplings. 

We approximate each transmon in Eq.~\eqref{eq:circuit_hamiltonian} with an anharmonic oscillator. By then applying a Schrieffer--Wolff transformation~\cite{swtransform}, we find the approximate Hamiltonian ($\hbar = 1$ here and in the rest of the article)
\begin{equation}
\begin{aligned}
    H'_\mathrm{circuit} \approx &\sum_i^N \mleft( \omega'_i a_i^\dagger a_i + \frac{\alpha_i}{2} a_i^\dagger a_i^\dagger a_i a_i \mright) + \\ 
    &\sum_{i < j}^N g'_{ij} \mleft( a_i^\dagger a_j + a_j^\dagger a_i \mright) + \mathcal{O} \mleft( \frac{g^3}{\omega^2} \mright),
    \label{eq:approximate_Hamiltonian}
\end{aligned}
\end{equation}
where $a_i$ ($a_i^\dag$) are bosonic annihilation (creation) operators; we refer the reader to Appendix~\ref{app:SWT} for the full computation. The model parameters in Eq.~\eqref{eq:approximate_Hamiltonian} derive from the circuit parameters in Eq.~\eqref{eq:circuit_hamiltonian}. The transmon frequencies and mutual coupling strengths are, respectively,
\begin{align}
    \omega'_i &= \omega_i - \sum_{k \neq i}^N \frac{g_{ik}^2}{\omega_i + \omega_k} + \mathcal{O} \mleft( \frac{g^3}{\omega^2} \mright) , 
    \label{eq:omega_prime} \\ 
    g'_{ij} &= g_{ij} - \frac{1}{2} \sum_{k \neq i,j}^N \mleft( \frac{g_{ik} g_{jk}}{\omega_i + \omega_k} + \frac{g_{ik} g_{jk}}{\omega_j + \omega_k} \mright) + \mathcal{O} \mleft( \frac{g^3}{\omega^2} \mright),
    \label{eq:g_prime}
\end{align}
where 
\begin{align}
    \omega_i &\approx \sqrt{8 E_{C,i}  E_{J,i}} - E_{C,i} , \\
    g_{ij} &= \frac{E_{C,ij}}{ \sqrt[4]{4E_{C,i} E_{C,j}}}  \sqrt[4]{E_{J,i} E_{J,j}} , \label{eq:g} 
\end{align}
while the transmon anharmonicity is given by $\alpha_i \approx - E_{C,i}$. The approximate Hamiltonian is valid when each transmon is well described by its lowest few levels; higher excited states contribute only at higher perturbative orders and can be neglected to leading accuracy in the ZZ interaction. 

Using Eqs.~\eqref{eq:omega_prime}--\eqref{eq:g}, we note that the Josephson energies relate to the transmon frequencies through $\omega'_i \approx \sqrt{8 E_{C,i}  E_{J,i}} - E_{C,i}$. Note that Eq.~\eqref{eq:omega_prime} shows that mutual couplings
produce a small renormalization of order $g^2/\omega$, yielding the dressed frequency $\omega'_i = \omega_i + O(g^2/\omega)$.
Consequently, the coupling strengths retain the frequency dependence $g'_{ij} \approx E_{C,ij} / \sqrt{16E_{C,i} E_{C,j}} \sqrt{\omega'_i \omega'_j} $, which arises directly from the circuit quantization. This dependence distinguishes the circuit-mediated coupling mechanism from simplified models that treat exchange couplings as frequency-independent. As a result, nominally identical coupling capacitors yield different effective exchange strengths depending on individual transmon frequencies, suggesting that cross-couplings cannot be fully characterized by geometry alone and  necessitating frequency-aware modeling for accurate crosstalk prediction in experimental systems. We return to this point in Section~\ref{emcircuitcrosstalk}, where we analyze the frequency and scale dependencies of the circuit- and enclosure-mediated coupling strengths. 

\subsection{Enclosure-mediated exchange}
Turning from the circuit to the enclosure model, we adopt the simplest possible model for the enclosure-mediated couplings. Although the cavity modes are assumed to be in their ground states due to their large excitation energies, they still mediate virtual interactions between distant transmons. Inspired by the exchange interactions in Eq.~\eqref{eq:approximate_Hamiltonian}, we model this interaction through an excitation-preserving Hamiltonian ansatz
\begin{equation}
    V_\mathrm{cavity} = \sum_{i<j}^N G_{ij} \mleft( a_i^\dagger  a_j + a_j^\dagger  a_i \mright) ,
    \label{eq:V_cavity}
\end{equation}
where $ G_{ij} $ is the cavity-mediated exchange between transmons $i$ and $j$, which we assume to be real-valued. At this stage, we remain agnostic about the parameter dependence of $G_{ij}$, e.g., with respect to the transmon frequencies or the physical distance between the transmons. In the next section, we study the classical crosstalk mediated by the enclosure modes and derive the parameter $G_{ij}$ in our Hamiltonian ansatz.

Altogether, Eqs.~\eqref{eq:Hamiltonian_model_1},~\eqref{eq:approximate_Hamiltonian}, and~\eqref{eq:V_cavity} give the model Hamiltonian
\begin{equation}
\begin{aligned}
    H_\mathrm{model} \approx &\sum_i^N \mleft( \omega’_i a_i^\dagger a_i + \frac{\alpha_i}{2} a_i^\dagger a_i^\dagger a_i a_i \mright) + \\ &\sum_{i<j}^N J_{ij} \mleft( a_i^\dagger  a_j + a_j^\dagger  a_i \mright) + \mathcal{O} \mleft( \frac{g^3}{\omega^2} \mright),
\end{aligned}
\label{eq:Hamiltonian_model_2}
\end{equation}
where $J_{ij} = g_{ij}' + G_{ij}$ is the total coupling strength. A brief remark is appropriate here. A similar model Hamiltonian can be achieved more directly by applying the rotating-wave approximation~\cite{rw1, rw2} on the circuit Hamiltonian in Eq.~\eqref{eq:circuit_hamiltonian}. The difference is that the rotating-wave approximation does not capture the corrections to the capacitive coupling strengths (second term) in Eq.~\eqref{eq:g_prime}. These corrections are significant to include as they are otherwise wrongly ascribed to the enclosure-mediated couplings $G_{ij}$.

\section{Classical lattice crosstalk}
\label{emcircuitcrosstalk}
Building on the Hamiltonian model in Section~\ref{probstat} [Eq.~\eqref{eq:Hamiltonian_model_1}] and the total coupling-strength model in Eq.~\eqref{eq:Hamiltonian_model_2}, we develop two complementary frameworks to quantitatively predict classical, chip-level crosstalk as a function of inter-qubit separation and frequency detuning in fixed-frequency transmon arrays. For simplicity, we here model the transmons as linear harmonic oscillators. The first framework employs nodal circuit analysis to capture virtually mediated coupling through chains of intermediate electromagnetic modes. The second framework utilizes electromagnetic Green's-function theory to describe evanescent-field coupling in environments operating below cutoff. Under the far-detuned, weak-loss, reciprocal assumptions, both approaches yield identical scaling laws, providing independent validation and physical insight into the mechanisms governing crosstalk suppression. 

\subsection{Nodal circuit analysis}
\label{nodalgreen}
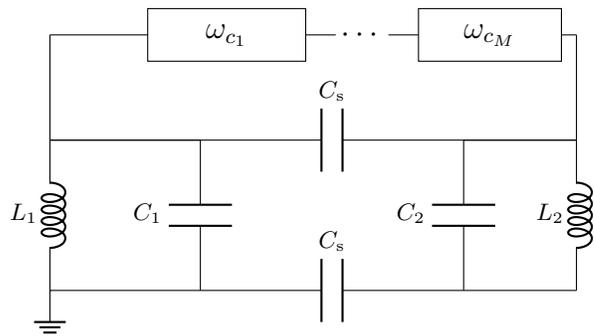
\begin{figure}[t]
\centering
\begin{circuitikz}[american voltages]
\draw (1,0) node[ground]{} -- (1,0);
\draw (1,2) to[L,l_=$L_1$] (1,0);
\draw (1,2) -- (3,2);
\draw (3,2) to[C,l_=$C_1$] (3,0);
\draw (6.5,2) -- (8,2);
\draw (6.5,2) to[C,l_=$C_2$] (6.5,0);
\draw (8,2) to[L,l_=$L_2$] (8,0);
\draw (1,0) -- (3,0);
\draw (3,0) to[C,l=$C_{\mathrm{s}}$] (6.5,0);
\draw (6.5,0) -- (8,0);
\draw (1,2) -- (3,2);
\draw (3,2) to[C,l=$C_{\mathrm{s}}$] (6.5,2);
\draw (6.5,2) -- (8,2);
\draw (1,2) -- (1,3.4);
\draw (8,2) -- (8,3.4);
\def\xLboxA{2.3}
\def\xLboxB{4.4}
\def\xRboxA{5.9}
\def\xRboxB{7.8}
\def\yEnv{3.4}
\draw (1,\yEnv) -- (\xLboxA,\yEnv);
\draw (\xLboxA,3.05) rectangle (\xLboxB,3.75);
\node at ({(\xLboxA+\xLboxB)/2},\yEnv) {\large $\omega_{c_1}$};
\draw (\xLboxB,\yEnv) -- (4.8,\yEnv);
\node at (5.15,\yEnv) {\large $\cdots$};
\draw (5.5,\yEnv) -- (\xRboxA,\yEnv);
\draw (\xRboxA,3.05) rectangle (\xRboxB,3.75);
\node at ({(\xRboxA+\xRboxB)/2},\yEnv) {\large $\omega_{c_M}$};
\draw (\xRboxB,\yEnv) -- (8,\yEnv);

\end{circuitikz}
 \caption{Equivalent circuit model representing qubits as linear LC resonators coupled through both direct capacitive coupling (capacitors \(C\)) and virtual processes mediated by intermediate electromagnetic modes through intermediate cavity modes with frequencies $\omega_{c_M}$ to $\omega_{c_1}$. The circuit elements represent the effective inductances and capacitances of the resonators-cavity system, where the intermediate modes act as harmonic oscillators that facilitate long-range coupling between spatially separated resonators.}
  \label{fig2:circuit}
\end{figure}

We start with the nodal framework and treat all modes as linear harmonic oscillators. For simplicity, we consider two harmonic oscillators, labelled $1$ and $2$, separated by distance $d_{ij}$ and coupled through virtual processes involving $M$ intermediate cavity modes. We model the system as a linear chain of $M+2$ coupled harmonic oscillators, where the endpoints $b_1$ and $b_2$ represent the transmons and $c_1, ..., c_M$ represent the mediating modes (Fig.~\ref{fig2:circuit}). Under the assumption that intermediate modes are far detuned from the transmon frequencies ($\omega_{c_n}-\omega_{1,2} \gg g$), they remain unexcited and only mediate virtual coupling. Hence, their dynamics can be adiabatically eliminated~\cite{AE, Bakr2025IntrinsicMMI}. The chain Hamiltonian takes the form

\begin{equation}
\begin{aligned}
H_{\text{ext}}
&= \omega_{b_1}\, b_1^\dagger b_1
 + \omega_{b_2}\, b_2^\dagger b_2
 + \sum_{n=1}^{M} \omega_{c_n}\, c_n^\dagger c_n \\
&\quad + g_{b_1,c_1}\!\left( b_1^\dagger c_1 + c_1^\dagger b_1 \right)
      + g_{c_M,b_2}\!\left( c_M^\dagger b_2 + b_2^\dagger c_M \right) \\
&\quad + \sum_{n=1}^{M-1} g_{c_n,c_{n+1}}\!\left( c_n^\dagger c_{n+1} + c_{n+1}^\dagger c_n \right).
\end{aligned}
\label{eq:H_ext}
\end{equation}
where $b_{1,2}$ and $c_n$ are bosonic annihilation operators for the two harmonic oscillators and the intermediate cavity modes, $\omega_{1,2}$ are the bare frequencies of the two harmonic oscillators, $\omega_{c_n}$ represents the frequency of the $n$th intermediate mode, and $g_{ij}$ denotes the coupling strength between adjacent modes in the chain. The coupling coefficients arise from electromagnetic field overlap and are determined by the chip and enclosure geometries.

To derive the effective coupling mediated by virtual transitions through intermediate modes, we use the node-flux equations of motion for coupled resonators. With flux variables \(\Phi_i\), capacitance matrix \(C\), and inductance matrix \(L\) (including mutual inductances \(M_{ij}\)), the linear dynamics are
\begin{equation}
\sum_j C_{ij}\,\ddot{\Phi}_j  +  \sum_j (L^{-1})_{ij}\,\Phi_j  =  0,
\label{eq:coupled_flux_eq}
\end{equation}
so that mutual inductive couplings enter through the off-diagonal elements of \(L^{-1}\). For weak mutual inductances and negligible mutual capacitances, this reduces the component form to
\begin{equation}
C_i\,\ddot{\Phi}_i  +  \frac{\Phi_i}{L_i}  -  \sum_{j\neq i}\frac{M_{ij}}{L_i L_j}\,\Phi_j  =  0.
\label{waveeq}
\end{equation}

We parameterize the inductive coupling by the dimensionless coefficient \(k_{ij}=M_{ij}/\sqrt{L_i L_j}\), and define mediator detunings \(\Delta_n = \lvert \bar{\omega} - \omega_{c_n}\rvert\) with \(\bar{\omega}=(\omega_1+\omega_2)/2\).
For completeness, if the coupling is capacitive rather than inductive, it enters through off–diagonal elements of the capacitance matrix: 
\(\sum_j C_{ij}\,\ddot{\Phi}_j + (L^{-1})_{ii}\Phi_i = 0\) for \(C_{ij}\neq 0\) and \((L^{-1})_{ij}=0\) (\(i\neq j\)).
Thus capacitive coupling generates cross–terms (\(C_{ij}\ddot{\Phi}_j\)), whereas mutual inductance generates cross–terms \((L^{-1})_{ij}\Phi_j\). 
The two descriptions are related by node–loop duality (\(C\!\leftrightarrow\! L\), \(\Phi\!\leftrightarrow\! Q\)) and, in the weak–coupling dispersive regime, yield the same effective exchange Hamiltonian up to this parameter mapping. 

For a chain of $M$ intermediate modes, systematic algebraic elimination (detailed in Appendix~\ref{nodalanalysis}) yields the effective coupling
\begin{equation}
J_{\mathrm{eff}}(M) = \mleft( \frac{{\bar{g}}}{\bar{\Delta}} \mright)^M J_0,
\label{eq:Jeff_scaling}
\end{equation}
where $J_0 =$ $\bar{g} = (\prod_{n=0}^{M} g_{n,n+1})^{1/(M+1)}$ is the geometric mean coupling strength, and $\bar{\Delta} = (\prod_{n=1}^{M} \Delta_n)^{1/M}$ is the geometric mean detuning of intermediate modes. Thus the mediator chain yields the expected algebraic suppression $J_{\mathrm{eff}}(M)\simeq J_0\,(g/\Delta)^M$. Next, we show that the enclosure path contributes a spatial evanescent envelope, and together these produce the unified scaling used.

\subsection{Electromagnetic Green's function}
\label{nodalgreen1}
We now develop the complementary electromagnetic framework by relating exchange to the enclosure’s two-port response, operating below cutoff under the far-detuned regime. A step-by-step derivation is provided in Appendix~\ref{greenfunction}. 

Following network methods, we characterize the enclosure by its transimpedance between the two qubit ports. For weakly anharmonic transmons, the enclosure-mediated exchange can be written in terms of the transimpedance as~\cite{solims}
\begin{equation}
\label{eq:Jij_impedance1}
J_{ij}
= -\,\frac{1}{4}\sqrt{\frac{\omega_i\,\omega_j}{L_i L_j}} 
\mathrm{Im}\mleft[\frac{Z_{ij}(\omega_i)}{\omega_i}+\frac{Z_{ij}(\omega_j)}{\omega_j}\mright],
\end{equation}
where $L_i$ and $\omega_i$ are the effective inductance and frequency of transmon $i$. Equivalently, one may work with the transfer admittance $Y_{ij}(\omega)$ via $J_{ij}\propto \sqrt{\omega_i\omega_j}\,\mathrm{Re}\,Y_{ij}(\bar\omega)\sqrt{C_i C_j}$, evaluated at the mean frequency $\bar\omega=(\omega_i+\omega_j)/2$.

To compute the transfer function, we use an enclosure Green's-function description. For capacitive (TM-like) coupling it is convenient to work with the scalar electric Green’s function $G_\phi$, which solves the Helmholtz problem with the enclosure boundary conditions. The transfer admittance can then be expressed as
\begin{equation}
\label{eq:transfer_admittance}
Y_{ij}(\omega)= i\omega\,\varepsilon_{\mathrm{eff}}\!\iint \rho_i(\mathrm r;\omega)\,
G_\phi(\mathrm r,\mathrm r';\omega)\,\rho_j(\mathrm r';\omega)\,d^3\mathrm r\,d^3\mathrm r',
\end{equation}
where $\rho_{i}$ is the displacement charge associated with the electrodes of transmon $i$ (the current formulation $J=-i\omega\varepsilon\nabla\phi$ is equivalent). 

Below the enclosure cut-off $\omega_c$, fields are evanescent. In an effective two-dimensional (2D) description, the Green’s function takes the form
\begin{equation}
\label{eq:Green_2D_correct}
G_\phi(\mathrm r,\mathrm r';\omega)
= \frac{1}{2\pi\,\varepsilon_{\mathrm{eff}}}\,
K_0 \mleft(\kappa(\omega)\,|\mathrm r-\mathrm r'|\mright),
\end{equation}
where
\begin{equation}
\label{eq:kappa_def}
\kappa(\omega)=\frac{\sqrt{\omega_c^{2}-\omega^{2}}}{v_p},
\end{equation}
with $K_0$ the modified Bessel function of the second kind and $v_p$ the relevant phase velocity in the medium.
Since $K_0(x)\sim e^{-x}/\sqrt{x}$ for large $x$, the enclosure-mediated amplitude inherits an evanescent envelope $K_0\mleft(\kappa(\omega)d_{ij}\mright)\sim e^{-\kappa(\omega)d_{ij}}/\sqrt{\kappa(\omega)d_{ij}}$ at large separations. Consequently, the enclosure-mediated amplitude inherits the Bessel-$K_0$ spatial law
\begin{equation}
\label{eq:J_spatial}
J_{ij}(\bar\omega) \propto K_0\mleft(\kappa(\bar\omega)\,d_{ij}\mright) \sim 
\frac{e^{-\kappa(\bar\omega)\,d_{ij}}}{\sqrt{\kappa(\bar\omega)\,d_{ij}}}\quad(d_{ij}\to\infty),
\end{equation}
i.e., exponential suppression with distance (modulo a $1/\!\sqrt{d}$ prefactor) set by the evanescent penetration length $\kappa(\bar\omega)^{-1}$.

In the dispersive regime $|\Delta_{ij}|\equiv|\omega_i-\omega_j|\gg J(\bar\omega)$, the time-dependent exchange averages out and produces a static interaction
\begin{equation}
\zeta_{ij} \approx \frac{2\,J(\bar\omega)^{2}}{\Delta_{ij}},
\label{zzscale}
\end{equation}
so detuning suppresses exchange algebraically ($\propto 1/\Delta_{ij}$).

\subsection{Identification of $G_{ij}$}
In the Hamiltonian ansatz $V_{\mathrm{cavity}}=\sum_{ij} G_{ij}\,a_i^\dagger a_j$, the enclosure-mediated hopping is the exchange amplitude computed from the enclosure transfer function. Consistent with Eqs.~(\ref{eq:coupled_flux_eq})--(\ref{eq:transfer_admittance}), we take
\begin{equation}
\label{eq:Gij_def_admittance}
G_{ij}\ \equiv\ J_{ij}^{(\mathrm{env})}(\bar\omega)
\ =\
\frac{C_{ij}^{\mathrm{eff}}(\bar\omega)}{2\sqrt{C_i C_j}} \sqrt{\omega_i \omega_j}.
\end{equation}
where
\begin{equation}
\label{eq:Ceff_def}
C_{ij}^{\mathrm{eff}}(\omega)\ \equiv\ \frac{\mathrm{Im}\,Y_{ij}(\omega)}{\omega},
\end{equation}
with $Y_{ij}(\omega)$ given by Eq.~\eqref{waveeq} via the scalar Green’s function $G_\phi$ [Eqs.~\eqref{eq:Jeff_scaling}--\eqref{eq:Jij_impedance1}].
Equivalently, using the transimpedance form [Eq.~\eqref{eq:coupled_flux_eq}] and assuming $\omega_i\approx\omega_j=\bar\omega$ and a slowly varying $Z_{ij}(\omega)$,
\begin{equation}
\label{eq:Gij_def_impedance}
G_{ij}\ \simeq\ -\,\frac{1}{2}\sqrt{\frac{\omega_i\omega_j}{L_iL_j}} 
\mathrm{Im}\mleft[\frac{Z_{ij}(\bar\omega)}{\bar\omega}\mright].
\end{equation}
Below cut-off, the Green's-function evaluation in Eqs.~\eqref{eq:Jeff_scaling}–\eqref{eq:transfer_admittance} yields the spatial envelope
\begin{equation}
\label{eq:Gij_scaling_env}
G_{ij}\ =\ J_0^{(\mathrm{env})}\,S_{\mathrm{dist}}\mleft(d_{ij};\bar\omega\mright),\qquad
S_{\mathrm{dist}}(d;\bar\omega)\ \propto\ K_0\mleft(\kappa(\bar\omega)\,d\mright),
\end{equation}
with $\kappa(\omega)=\sqrt{\omega_c^2-\omega^2}/v_p$ and $J_0^{(\mathrm{env})}\equiv J_{ij}^{(\mathrm{env})}(d_0;\bar\omega)$ the enclosure-mediated coupling at the reference spacing $d_0$.
In a reciprocal, weak-loss enclosure, $G_{ij}=G_{ji}$ and is real to leading order; we evaluate $S_{\mathrm{dist}}$ at $\bar\omega$ throughout.
In the full device, the total hopping is
$J_{ij}^{\mathrm{tot}}=J_{ij}^{(\mathrm{circ})}+G_{ij}$,
combining circuit-level exchange with the enclosure contribution.

\subsection{Unified scaling framework}
\label{nodalgreen3}
We now combine circuit-mediated near/fringing/far-field trends with the enclosure-mediated evanescent envelope to obtain a single factorized law for $J_{ij}$. The spatial dependence of enclosure- and circuit-mediated coupling follows from electrode geometry and field profiles. We identify three practical regimes, set by the separation relative to device dimensions:
\begin{itemize}
    \item \textit{Near field} ($d_{ij}\!\sim\! a$, with lattice pitch $a$): direct capacitive overlap dominates, $C_{ij}\propto \varepsilon A/d_{ij}$, giving $J_{ij}\propto (d_0/d_{ij})$ after normalization at a reference spacing $d_0$.
    \item \emph{Fringing regime} ($d_{ij}$ comparable to, but larger than, the electrode width $w$): coplanar fringing yields $C_{ij}\propto \varepsilon/\ln(2d_{ij}/w)$, so $J_{ij}\propto \mleft[\ln(2d_{ij}/w)\mright]^{-1}$.
    \item \textit{Dipolar far field} ($d_{ij}\!\gg$ electrode dimensions): electric dipole–dipole scaling, $C_{ij}\propto \varepsilon A_iA_j/d_{ij}^3$, hence $J_{ij}\propto (d_0/d_{ij})^3$.
\end{itemize}

\noindent In enclosures operated below cut-off, the enclosure contribution picks up the evanescent law
\begin{equation}
\label{eq:Sdist_K0}
S_{\mathrm{dist}}(d;\omega) \propto K_0\mleft(\kappa(\omega)\,d\mright) \sim \frac{e^{-\kappa(\omega)\,d}}{\sqrt{\kappa(\omega)\,d}},
\end{equation}
which multiplies the quasi-static trends above when the enclosure dominates.

A convenient way to summarize these cases is to factor the coupling into a spatial envelope evaluated at the mean qubit frequency:
\begin{equation}
\label{eq:unified_J}
J_{ij}(\omega_i,\omega_j) \simeq J_0 \,S_{\mathrm{dist}}\mleft(d_{ij};\bar\omega\mright),
\end{equation}
with
\begin{equation}
\label{eq:Sdist_piecewise}
S_{\mathrm{dist}}(d;\bar\omega)=
\begin{cases}
(d_0/d) & \text{near field},\\[2pt]
\displaystyle \frac{\ln(2d_0/w)}{\ln(2d/w)} & \text{fringing regime},\\[8pt]
(d_0/d)^3 & \text{dipolar far field},\\[4pt]
\displaystyle \frac{K_0\mleft(\kappa(\bar\omega)d\mright)}{K_0\mleft(\kappa(\bar\omega)d_0\mright)} & \text{below cut-off}.
\end{cases}
\end{equation}
Here $J_0\equiv J_{ij}(d_0,\bar\omega)$ fixes the overall scale at a chosen reference spacing $d_0$.

\medskip
Combining these spatial dependencies with spectral suppression from both theoretical approaches and assuming the qubit frequencies are far below the cut-off frequency, we obtain the unified scaling law
\begin{align}
J_{ij} &= J_0\, K_0 \mleft( \kappa d_{ij} \mright)\, f(\Delta\omega_{ij}) \nonumber\\
& \approx J_0\, \frac{e^{-\kappa d_{ij}}}{\sqrt{\kappa d_{ij}}}\, f(\Delta\omega_{ij}),
\qquad \text{for large } d_{ij},
\label{eq:J_ij}
\end{align}
where $J_0$ sets the overall coupling scale (for example, a typical nearest-neighbor strength), $\kappa^{-1}\equiv d_0$ is the characteristic spatial decay length of order the lattice spacing, and $f(\Delta\omega_{ij})$ is a dimensionless correction factor encoding any residual frequency dependence (which is weak in our parameter regime, so $f \approx 1$).

This equivalence clarifies why the nodal-chain product-over-detunings picture (Appendices~\ref{matrixinversion} and~\ref{nodalanalysis}) and the Green's-function evanescence picture (Appendix~\ref{greenfunction}) produce the same exponential distance dependence: virtual-process suppression through detuned intermediate modes and evanescent electromagnetic fields below cutoff both lead to couplings proportional to the screened two-dimensional Green’s function $K_0(\kappa d_{ij})$. The connection between the two approaches becomes clear when we recognize that the nodal-analysis effective coupling corresponds to
the cavity-mediated terms in the Green’s-function decomposition. Specifically, for a chain of $M$ intermediate cavity modes, the Green's function approach yields
\begin{equation}
\label{eq:Z_chain_kernel}
Z_{ij}^{\text{chain}}(\omega)\ \propto\
\sum_{\text{paths}}\ \prod_{n=1}^{M}\frac{g_{n-1,n}}{\omega_{c_n}^{2}-\omega^{2}}\,.
\end{equation}
which, in the far off-resonant limit ($|\omega-\omega_{c_n}|\gg g$), reduces to the familiar product-over-links divided by product-over-detunings, reproducing the mediator-chain scaling $J_{\mathrm{eff}}\propto \mleft(\prod g\mright)/\mleft(\prod \Delta\mright)$. This equivalence explains why the nodal (circuit) and Green's-function (enclosure) analyses yield consistent distance laws and the same dispersive detuning dependence.

\section{Quantum Lattice Crosstalk}
\label{sec:models_for_zz_couplings}
In this section, we consider the effect of cross couplings on quantum-mechanical properties of the transmon lattice. In particular, the couplings cause residual ZZ couplings, which have the observable effect that the transmons' energies become dependent on the states of neighboring transmons. This state-dependent effect is in itself sometimes referred to as quantum-mechanical crosstalk. Here, we give analytical expressions through perturbation theory, relating the ZZ coupling strengths between nearest-neighbor and next-nearest-neighbor transmons to the cross couplings. These analytical expressions are the basis from which we in Section~\ref{sec:results} experimentally infer the cross couplings from ZZ-coupling measurements.  

\subsection{ZZ couplings for nearest- and next-nearest neighbor transmons}
Beginning from the Hamiltonian model in Eq.~\eqref{eq:Hamiltonian_model_2}, we relate the model parameters, in particular, the coupling strengths $J_{ij}$, to ZZ couplings between the transmons. The ZZ couplings can be approximated using perturbation theory, and we give them below for nearest-neighbor and next-nearest-neighbor transmons. If needed, the ZZ couplings beyond next-nearest neighbor can also be analytically approximated with the same methods. We neglect spectator effects, i.e., exclude transmons which are not directly coupled to the two transmons considered for the ZZ coupling; see Appendix~\ref{app:spectator} for further details. The ZZ coupling for two nearest-neighbor transmons labeled 1 and 2 is approximated to \cite{transmon, solgun2019simple, solgun2022direct}:
\begin{equation}
    \zeta_{12} = \frac{2 (\alpha_1+\alpha_2) J_{12}^2}{(\Delta_{12} + \alpha_1)(\Delta_{12} - \alpha_2)} + \mathcal{O}(J_{12}^4/\Delta_{12}^3).
    \label{eq:ZZ_12}
\end{equation}
In the case of next-nearest-neighbor transmons labeled 1 and 3 with an intermediate transmon labeled 2, the ZZ coupling is approximated to \cite{chu2021coupler,fors2024comprehensive}:
\begin{equation}
\begin{aligned}
    \zeta_{13} =~ &\frac{2 (\alpha_1+\alpha_3)}{(\Delta_{13} + \alpha_1)(\Delta_{13} - \alpha_3)} 
    \mleft( J'_{13} - \Lambda \frac{\alpha_1 + \alpha_3}{2} \mright)^2 + \\
    &\frac{4 (\alpha_1 - \alpha_3)}{(\Delta_{13} + \alpha_1)(\Delta_{13} - \alpha_3)} 
    \mleft( \Delta_{13} + \frac{\alpha_1 + \alpha_3}{2} \mright) \Lambda J'_{13}
    + \\ &\mleft[ 2\frac{ \Delta_{13}^2 - \mleft[ \mleft( \alpha_1 + \alpha_3 \mright) / 2 \mright]^2 }{(\Delta_{13} + \alpha_1)(\Delta_{13} - \alpha_3)}  
    (\alpha_1+\alpha_3) + \mright. \\
    &   \mleft. 8\frac{ (\Delta_{12} - \Delta_{23}) }{\Delta_{12} - \Delta_{23} - \alpha_2} \alpha_2 \mright] \Lambda^2 + \mathcal{O}(J^5/\Delta^4),
\end{aligned}
\label{eq:ZZ_13}
\end{equation}
where $J'_{13} = J_{13} + \frac{1}{2}J_{12}J_{23}\mleft( \frac{1}{\Delta_{12}} - \frac{1}{\Delta_{23}} \mright) = J_{13} + \Lambda (\Delta_{12} - \Delta_{23})$, $\Lambda = -J_{12} J_{23} / (2 \Delta_{12} \Delta_{23})$, and $\mathcal{O}(J^5/\Delta^4)$ denotes higher-order corrections. $J'_{13}$ is occasionally called the effective coupling between transmons 1 and 3; it has a contribution from the direct coupling $J_{13}$ and a contribution $\frac{1}{2}J_{12}J_{23}\mleft( \frac{1}{\Delta_{12}} - \frac{1}{\Delta_{23}} \mright)$ mediated by transmon 2. The effective coupling is found from performing a Schrieffer--Wolff transformation that eliminates the intermediate transmon. We note that $\Lambda$ quantifies how $\Lambda$-like the first-excitation subspace of the system is, where smaller $0 < \Lambda \ll 1$ implies a more $\Lambda$-like system. A $\Lambda$ system is a three-level system where two of the states couple to a third intermediate state, which has a significantly higher or lower energy. 

\subsection{Modeling ZZ couplings with the scaling model}
\begin{figure}
\centering
\includegraphics[width=1.0\linewidth]{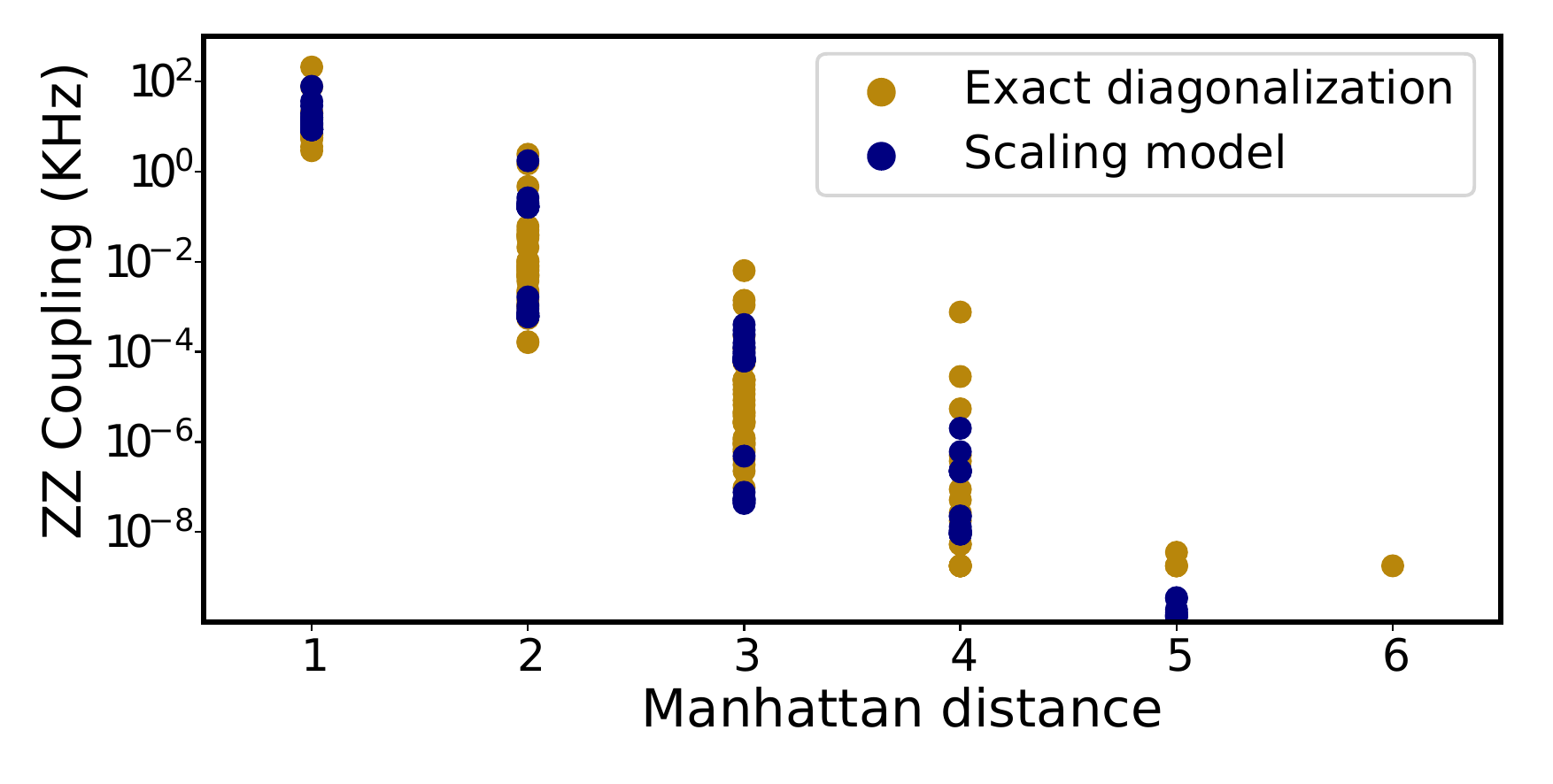}
   \caption{Simulated ZZ couplings in the $4 \times 4$ transmon lattice. The yellow markers shows the result from exact diagonalization of the model Hamiltonian in Eq.~\eqref{eq:Hamiltonian_model_2}, while the purple markers give the results of the scaling model fitted to the exact-diagonalization ZZ couplings. Note that the yellow markers saturate at $\sim \qty{e-8}{\kilo\hertz}$ beyond Manhattan distance 4. We attribute this saturation to the algorithmic precision of the exact-diagonalization routine.}
   \label{fig:simulated_zz_coupling}
\end{figure}

It is possible to continue with perturbation theory to compute the ZZ couplings between even further neighbors. However, as seen from going from nearest- to next-nearest neighbor ZZ couplings, these expressions quickly grow in complexity. Instead, we opt for a more practical approach by combining the scaling model in Eq.~\eqref{eq:J_ij} with the nearest-neighbor expression in Eq.~\eqref{eq:ZZ_12}. The idea, here, is that the scaling model estimates the effective coupling between any pair of transmons, while ignoring the other transmons in the lattice, effectively modeling each pair of transmons as independent directly coupled pairs. As such, the nearest-neighbor expression is applicable within the scaling model, which yields the approximate model:
\begin{equation}
    \zeta_{ij} = \frac{2 (\alpha_i+\alpha_j) J_{ij}^2}{(\Delta_{ij} + \alpha_i)(\Delta_{ij} - \alpha_j)} \mathrm{e}^{-2d_{ij} / d_0} ,
    \label{eq:ZZ_ij}
\end{equation}
where $i \neq j$ is for any pair of transmons in the lattice, and we recall the dispersion relation $J_{ij} \propto \sqrt{\omega_i \omega_j}$.

In Fig.~\ref{fig:simulated_zz_coupling}, we compare the scaling model in Eq.~\eqref{eq:ZZ_ij} with numerical simulations of the ZZ couplings in the lattice. The simulations compute the ZZ couplings with exact diagonalization using the model Hamiltonian in Eq.~\eqref{eq:Hamiltonian_model_2}, and the system parameters reported in Ref.~\cite{alghadeer2025characterization}, which is for the same device as characterized in Section~\ref{sec:results}. In particular, we use only the experimentally measured nearest-neighbor couplings. The scaling model is fitted to the simulated ZZ couplings with respect to the mean ZZ couplings for the different Manhattan distances. We note that the trend of the numerically simulated ZZ couplings is exponentially decaying with respect to Manhattan distance, as expected from the scaling model.

\section{Experimental Crosstalk Characterization}
\label{sec:results}
We now present experimental results to validate the theoretical crosstalk framework developed in this work. Our test platform consists of 16 fixed-frequency transmon qubits arranged in a square lattice with designed nearest-neighbor (NN) capacitive links. Fabrication details are provided in Refs.~\cite{alghadeer2025low, alghadeer2025characterization}, and device parameters are detailed in Appendix~\ref{app:device_parameters}. Each qubit has frequency $\omega_i / 2 \pi$ and anharmonicity $\alpha_i/2\pi \approx \qty{-196}{\mega\hertz}$, with qubits designed in two interleaved frequency bands (4.8 and \qty{4.9}{\giga\hertz} nominal) to maintain qubit addressability. The lattice incorporates off-chip inductive shunt pillars (see Appendices~\ref{app:device_parameters} and~\ref{app:simulations}) that raise the electromagnetic cutoff frequency from $\sim \qty{11}{\giga\hertz}$ (bare cavity) to $\sim \qty{34}{\giga\hertz}$ (shunted cavity), placing the device firmly in the evanescent regime where enclosure-mediated coupling is exponentially suppressed.

\begin{figure*}[ht]
  \centering
   \includegraphics[width=1.0\textwidth]{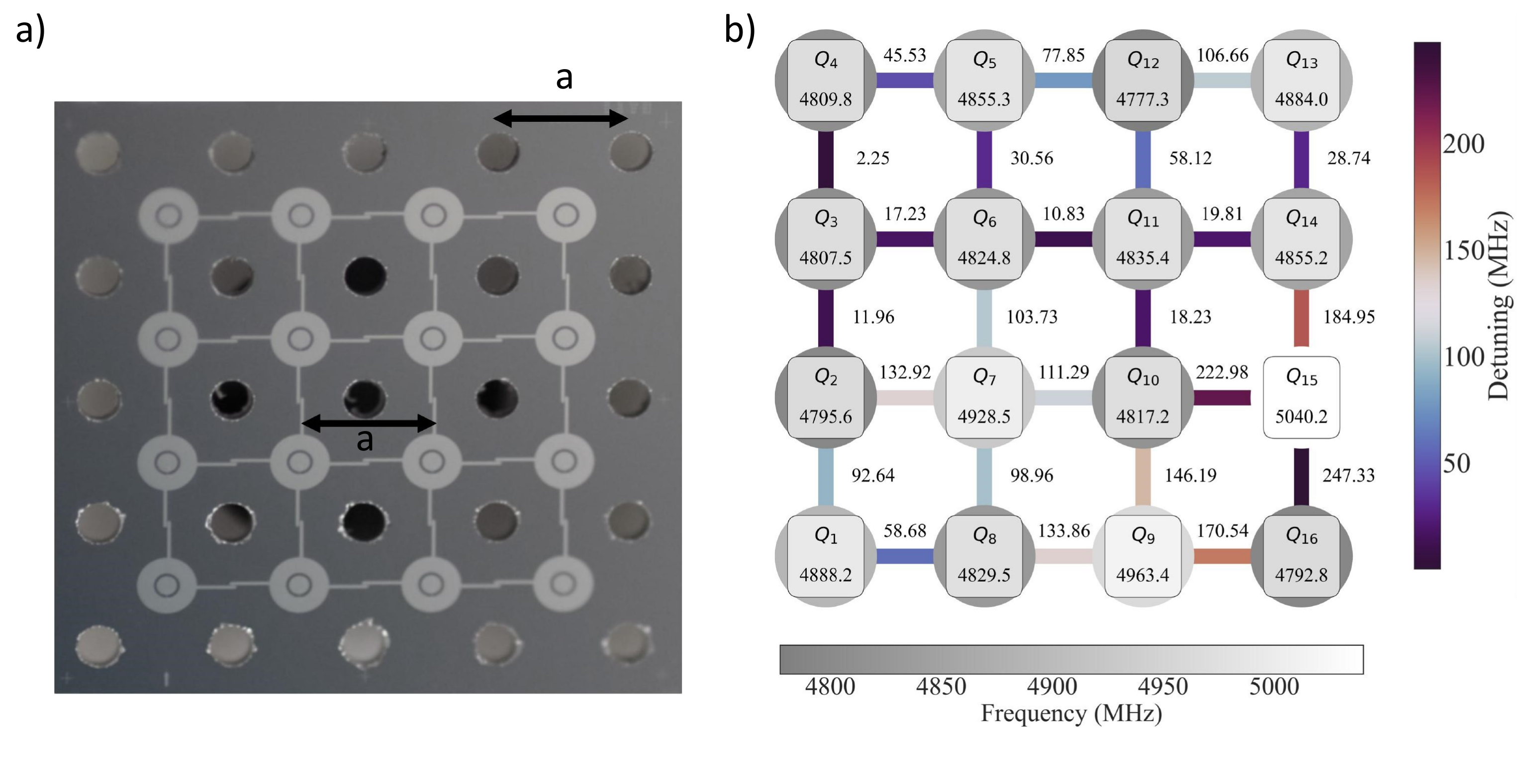}
      \caption{Device architecture and frequency allocation.
      (a) Optical micrograph of the fabricated $4\times4$ transmon lattice showing qubit pads (concentric light grey circles) connected by nearest-neighbor capacitive links, with inductive shunt pillar holes interleaved between qubit sites. Both arrays share the same square lattice pitch ($a = \qty{2}{\milli\metre}$). Inductive shunt pillars (Appendix~\ref{greenfunction}) suppress enclosure-mediated coupling by raising the cavity cutoff above the qubit band.
      (b) Measured qubit frequencies form two interleaved bands at \qtyrange[range-phrase = --, range-units = single]{4.8}{4.9}{\giga\hertz}, with most detunings satisfying $|\omega_i-\omega_j| / 2 \pi < \langle \alpha/2 \pi \rangle_{16} = \qty{-196.4}{\mega\hertz}$ (straddling regime).}
  \label{F11}
\end{figure*}

To capture both intended NN couplings and unintended long-range crosstalk, we distinguish these couplings from Eq.~(\ref{eq:Hamiltonian_model_2}):
\begin{align}
H_\mathrm{model}= \ & 
\sum_{i=0}^{15} \mleft( \omega_i \hat{a}_i^\dagger \hat{a}_i + \frac{\alpha_i}{2} \hat{a}_i^\dagger \hat{a}_i^\dagger \hat{a}_i \hat{a}_i \mright)  \nonumber \\
& + \sum_{\substack{i < j \\ j \in \mathcal{N}_i}} J_{ij} \mleft( \hat{a}_i^\dagger \hat{a}_{j} + \hat{a}_i \hat{a}_{j}^\dagger \mright) \nonumber \\
& + \sum_{\substack{i < j \\ j \notin \mathcal{N}_i}} \widetilde{J}_{ij} \mleft( \hat{a}_i^\dagger \hat{a}_j + \hat{a}_i \hat{a}_j^\dagger \mright),
\label{eq.MA_1}
\end{align}
where $\mathcal{N}_i$ denotes the four NN qubits of site $i$, $J_{ij}$ are the designed capacitive exchange couplings, and $\widetilde{J}_{ij}$ represent effective non-nearest-neighbor (NNN) interactions arising from circuit-mediated virtual processes (Appendix~\ref{nodalanalysis}) and enclosure-mediated electromagnetic coupling (Appendix~\ref{greenfunction}). Our goal is to extract both $J_{ij}$ and $\widetilde{J}_{ij}$ from spectroscopic measurements and validate their spatial and spectral scaling against theoretical predictions. Figure~\ref{F11} shows the measured idle frequencies for all 16 qubits, spanning \qtyrange[range-phrase = --, range-units = single]{4.78}{5.04}{\giga\hertz} with two clear bands centered near the design targets. The frequency allocation achieves relative spreads of \qty{0.5}{\percent} (low band) and \qty{1.5}{\percent} (high band), with most pairwise detunings $|\omega_i-\omega_j|$ below the average anharmonicity $\langle \alpha/2 \pi \rangle_{16}= \qty{-196.4}{\mega\hertz}$. Operation in this band maintains most pairwise detunings $|\omega_i-\omega_j|$ below the array-averaged anharmonicity, placing most pairs in the straddling regime~\cite{transmon}.

\subsection{ZZ spectroscopy of nearest-neighbor and long-range interactions}
We extract pairwise interaction strengths through Ramsey-based ZZ spectroscopy~\cite{Mundada2019ZZ, Cao2025AutoLabAI}, measuring the frequency shift of a target qubit when a control qubit is prepared in either $|0\rangle$ or $|1\rangle$. The resulting static ZZ interaction $\zeta_{ij}$ directly reflects the underlying exchange coupling $J_{ij}$ through the dispersive relation in Eq.~\eqref{eq:ZZ_12}. This standard formula assumes spatially and spectrally uniform coupling --- an assumption we will test and refine based on the first term of Eq.~\eqref{eq:ZZ_12}.

Constructing a high-confidence crosstalk matrix requires careful attention to measurement resolution and statistical significance. To establish the noise floor, we first characterized intrinsic frequency stability by performing approximately 400 sequential Ramsey experiments per qubit over a 12-hour period, extracting the standard deviation of each qubit's idle frequency. The overall device average fluctuation is \qty{0.88}{\kilo\hertz}, with most individual qubits exhibiting fluctuations below approximately \qty{1.0}{\kilo\hertz} (detailed results in Appendix~\ref{app:device_parameters}, Fig.~\ref{F4-Sup}). This measurement establishes resolvable ZZ shift and ensures that reported couplings are well above the intrinsic noise floor.

Based on this characterization, we apply additional filtering procedures. First, we retain only ZZ values exceeding \qty{1.0}{\kilo\hertz}, ensuring adequate signal-to-noise ratio for all reported interactions. Second, we assess the statistical significance of the remaining interactions by choosing \qty{95}{\percent} confidence intervals over repeated measurements for each element of the ZZ coupling matrix. Interactions whose confidence intervals include zero are classified as statistically insignificant and removed from the data set. The vast majority of non-nearest-neighbor interactions have measured shifts near or below the \qty{1.0}{\kilo\hertz} threshold, confirming the intended suppression of residual long-range couplings across the lattice as expected for a device with effective crosstalk suppression~\cite{alghadeer2025low}.

Finally, two pairs, $Q_9$-$Q_{16}$ and $Q_{14}$-$Q_{15}$, exhibit anomalously large ZZ shifts of \qty{75.8}{\kilo\hertz} and \qty{209.1}{\kilo\hertz} respectively. These outliers arise because both pairs operate near higher-level transitions and close to the edge of the straddling regime. Specifically, $Q_{15}$ has frequency \qty{5.04}{\giga\hertz}, significantly above the nominal \qty{4.9}{\giga\hertz} band, placing several pair combinations in a regime where the detunings approach or exceed the anharmonicity. We treat the pairs above as outliers and exclude them from the subsequent analysis. The final data set contains high-confidence nearest-neighbor pairs and statistically significant non-nearest-neighbor pairs, providing sufficient spatial coverage to extract robust scaling parameters while maintaining high measurement fidelity throughout.

\begin{figure*}[ht]
  \centering
   \includegraphics[width=1.0\textwidth]{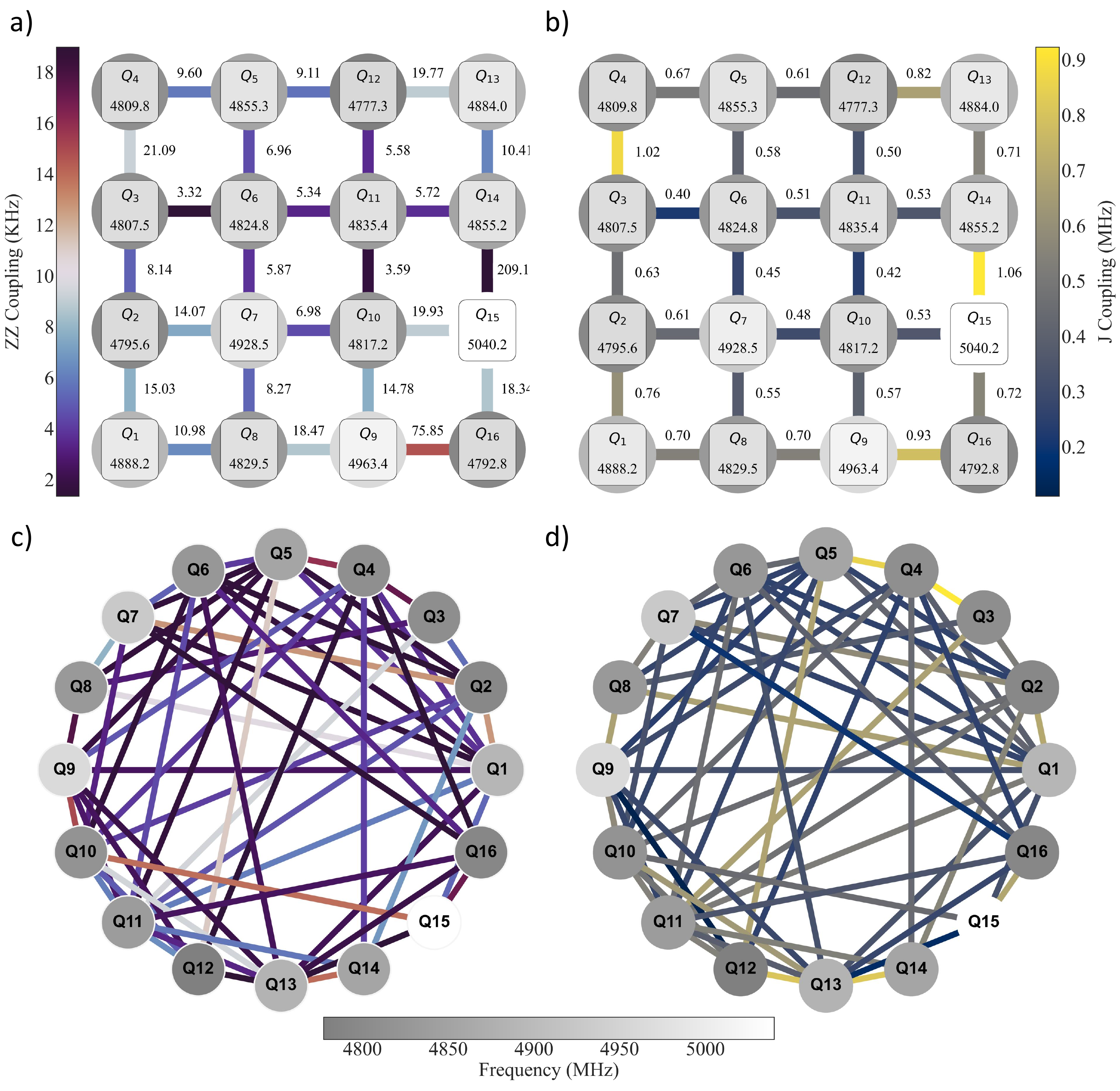}
      \caption{Measured ZZ interactions and extracted $J$ couplings.
      (a) Measured ZZ interactions between nearest-neighbor qubit pairs were extracted via Ramsey spectroscopy by preparing each control qubit in either the ground or excited state and observing the resulting state-dependent frequency shift in the target qubit. To assess temporal frequency stability, repeated Ramsey measurements ($\sim 400$ per qubit) were performed over a 12-hour window, yielding an average standard deviation of \qty{0.88}{\kilo\hertz} across the device --- below the smallest measured ZZ value reported in this work.
      (b) Estimated $J$ couplings by taking the first term in the analytical expression described in Eq.~\eqref{eq:ZZ_12} and applying it to the results in (a). 
      Interaction maps of the measured ZZ interactions in (c) and calculated $J$ couplings in (d) are shown across both nearest-neighbor and longer-range qubit pairs in the lattice.}
  \label{F2}
\end{figure*}

Figure~\ref{F2} summarizes the measured crosstalk landscape for the ZZ rates in panel (a) and NN couplings in panel (b). Across all measured NN pairs, the estimated exchange couplings [the first term in the analytical expression in Eq.~\eqref{eq:ZZ_12}] span from a minimum of \qty{401}{\kilo\hertz} to a maximum of \qty{1.064}{\mega\hertz}. The mean coupling strength is $\mu = \qty{623}{\kilo\hertz}$ with a standard deviation of $\sigma = \qty{173}{\kilo\hertz}$, corresponding to a relative spread $\sigma/\mu \approx 0.269$. This moderate variation reflects small but measurable non-uniformities in $J$ across the lattice. Panels (c) and (d) extend the interaction maps to include NNN pairs. These maps include all considered pairs in our analysis and indicate the increased complexity of the interaction landscape when long-range couplings are taken into account. Most long-range ZZ rates lie near the \qty{1.0}{\kilo\hertz} threshold, confirming effective crosstalk suppression. However, a few NNN pairs show measurable ZZ shifts, indicating that higher-order virtual processes and residual enclosure coupling are not completely eliminated. The spatial structure of these residual interactions---with next-nearest neighbors showing systematically larger shifts than more distant pairs---provides the key signature of exponential decay that we next analyze quantitatively. 

\subsection{Spatial scaling: exponential decay and bound-state length}

\begin{figure*}[ht]
  \centering
   \includegraphics[width=1.0\textwidth]{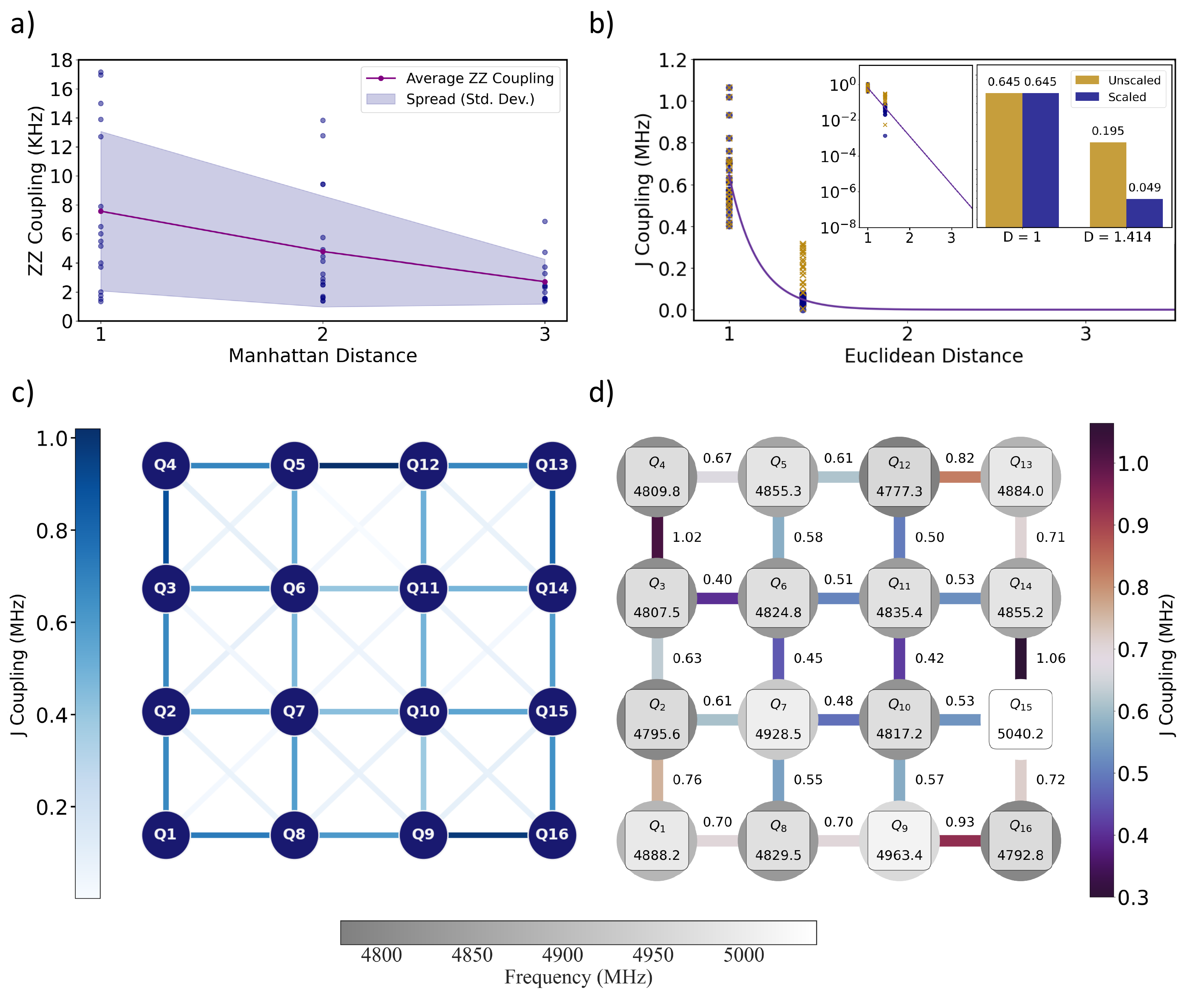}
    \caption{Distance dependence for measured and extracted ZZ and $J$ couplings.
    (a) Measured ZZ interactions (blue markers) between qubit pairs across the lattice as a function of Manhattan distance up to $D=3$. Mean ZZ interactions (purple markers connected by a line) are plotted against Manhattan distance, with the shaded region indicating the standard deviation.
    (b) Estimated exchange couplings $J$ obtained by taking the leading term of the analytical expression in Eq.~\eqref{eq:ZZ_12} (yellow markers), and by applying the exponential scaling model in Eq.~\eqref{eq:ZZ_ij} (blue markers), as a function of Euclidean distance. Predictions beyond nearest neighbors diverge significantly from measured interactions, illustrating an overestimation of long-range coupling strengths not supported by direct avoided-crossing measurements~\cite{alghadeer2025low}. The left inset shows the extracted $J$ coupling strength (in MHz) plotted on a logarithmic scale as a function of Euclidean distance, highlighting the exponential decay beyond next-nearest neighbors. The right inset compares the average measured $J$ coupling obtained from both methods, with and without the applied global scaling factor.
    (c) Qualitative grid diagram of the calculated coupling $J$ after applying a global scaling factor to account for model overestimation. Data include both nearest-neighbor and diagonal qubit pairs across the lattice.
    (d) Quantitative spatial distribution of the corrected $J$ values for all nearest-neighbor qubit pairs. }
  \label{F3}
\end{figure*}

We now test the central prediction that crosstalk should decay exponentially with spatial separation due to capacitance matrix localization and evanescent electromagnetic fields. For quantitative analysis, we characterize qubit separations using both Manhattan distance $D$ (sum of grid steps) and Euclidean distance $d_{ij}$ (straight-line separation), sometimes referred to as $L^1$ and $L^2$ distance, respectively. Lattice coordinates for qubit $i$ are defined as
\begin{equation}
\text{position}(i) = \mleft(r_i,\,c_i\mright)
= \mleft( \mleft\lfloor \frac{i}{4} \mright\rfloor,\, i \bmod 4 \mright),
\label{eq:lattice_position}
\end{equation}
with Euclidean distance $d_{ij} = a\sqrt{(r_i - r_j)^2 + (c_i - c_j)^2}$ where the lattice constant $a = \qty{2}{\milli\metre}$, and $r_i$ and $c_i$ denote row $i$ and column $i$, respectively.

Figure~\figpanelNoPrefix{F3}{a} shows mean ZZ as a function of Manhattan distance, averaging over all pairs with the same topological separation. A pronounced suppression with distance is observed in the detectable subset: the mean ZZ coupling decreases from $\sim\qty{22}{\kilo\hertz}$ ($D=1$)—or $\sim\qty{8}{\kilo\hertz}$ after removing outliers as shown in panel~(a)—to $\sim\qty{5}{\kilo\hertz}$ ($D=2$) and $\sim\qty{2.5}{\kilo\hertz}$ ($D=3$). Since many longer-range couplings fall below the Ramsey sensitivity, these values should be interpreted as conditional on detection, and then more as upper bounds than mean values; within this subset, the typical ZZ scale is reduced by approximately a factor of four from $D=1$ to $D=3$.
Fitting to $\langle\zeta\rangle \propto \exp(-D/D_0)$ yields $D_0 \approx 0.83$ sites (\qty{8.3}{\milli\metre} characteristic decay length), indicating that long-range interactions are heavily suppressed within 1--2 lattice spacings. This rapid falloff directly validates the combined suppression from circuit localization and enclosure engineering. Figure~\figpanelNoPrefix{F3}{b} replots the same data versus Euclidean distance $d_{ij}$, now converting ZZ to exchange coupling $J$ by taking the first term in the analytical expression in Eq.~\eqref{eq:ZZ_12} (yellow points). This naive model---which assumes spatially and spectrally uniform $J$---overestimates long-range interactions, predicting NNN couplings of \qtyrange[range-phrase = --, range-units = single]{0.1}{0.4}{\mega\hertz} that are contradicted by direct avoided-crossing measurements~\cite{alghadeer2025low}. The overestimation arises because taking the first term in the analytical expression in Eq.~\eqref{eq:ZZ_12} treats $J$ as distance-independent, ignoring the exponential spatial suppression established theoretically.

To correct this systematic error, we introduced a unified scaling model that incorporates both spatial decay and frequency-dependent suppression in Eq.~\eqref{eq:ZZ_ij}. Fitting Eq.~\eqref{eq:ZZ_ij} to the full data set yields spatial and frequency suppression parameters consistent with the theoretical framework. The spatial scaling exhibits the expected combination of exponential decay (from capacitance localization, Appendix~\ref{matrixinversion}) and geometric modulation (from 2D evanescent fields, Appendix~\ref{greenfunction}), while the characteristic frequency scale aligns with typical qubit detunings in our two-band architecture. With these fitted parameters, the scaling-corrected $J$ values [blue points in \figpanel{F3}{b}] collapse onto a physically consistent picture: NN couplings remain near \qty{0.6}{\mega\hertz} as expected from capacitance design, while NNN couplings are suppressed, consistent with avoided-crossing measurements that showed negligible long-range exchange~\cite{alghadeer2025low}. The naive model (yellow points) systematically overestimates NNN interactions by factors of 5--10$\times$, while the unified scaling model resolves this discrepancy across the full lattice. 

The extracted exchange couplings agree closely with the predictions of the unified scaling model, which accurately reproduces all nearest-neighbor interactions while exponentially suppressing longer-range terms. Diagonal next-nearest couplings are reduced by approximately 70\%. Detailed comparisons between measured and modeled $J_{ij}$ values are provided in Appendix~\ref{app:device_parameters}. Figure~\figpanelNoPrefix{F3}{c} and \figpanelNoPrefix{F3}{d} present the final adjusted coupling map using Eq.~\eqref{eq:ZZ_ij}. Figure~\figpanelNoPrefix{F3}{c} shows the qualitative spatial distribution of corrected $J$ values across the lattice, revealing the intended NN-dominated structure with dramatically suppressed long-range interactions. Figure~\figpanelNoPrefix{F3}{d} shows the quantitative NN coupling map.

\section{Conclusion}
\label{sec:discussion_conclusion}
We analyzed static crosstalk in a fixed-frequency transmon array and observed that neglecting spatial and spectral structure can overestimate non-nearest-neighbor interactions. The measured dependencies of coupling (crosstalk) strength on distance and detuning are consistent with three complementary mechanisms: exponential localization in the inverse capacitance matrix, suppression through virtual intermediate modes, and evanescent electromagnetic coupling below the enclosure cutoff. These observations suggest that lattice geometry and enclosure design act as control parameters for shaping residual interactions relevant to error-correcting architectures. The pillar-shunted enclosure represents 
one approach to mitigating enclosure-mediated exchange while maintaining modularity. Taken together, these results provide a starting point for device-level studies aimed at translating qualitative trends into quantitative design criteria for suppression of correlated errors.

From an architectural perspective, error-correcting codes benefit when spatially separated qubits remain effectively independent. Residual static couplings can seed correlations during syndrome extraction. The measured suppression with distance indicates that lattice geometry and enclosure design may contribute to code-level independence targets; the specifics are likely application-dependent. In the near term, crosstalk maps can inform scheduling and compilation to avoid simultaneous operations on strongly interacting pairs, and motivate targeted decoupling sequences where appropriate. 

Looking ahead, several extensions appear natural. Systematic sweeps of enclosure dimensions, pillar spacing, and lattice pitch across multiple devices could clarify how general the spatial trends are. Incorporating distance-aware penalties into placement and routing may help assess whether crosstalk-aware compilation improves two-qubit performance. Extending the static treatment to include time-dependent effects during gates would link static couplings to dynamical error accumulation. Similar combined analytical–experimental methods could address other correlated mechanisms---drive line coupling, flux crosstalk in tunable designs, and packaging-induced interference---balancing model complexity against predictive utility.

\begin{acknowledgments}
M.B.~acknowledges support from EPSRC QT Fellowship grant EP/W027992/1, and EP/Z53318X/1. S.P.F.~and A.F.K.~acknowledge support from the Knut and Alice Wallenberg Foundation through the Wallenberg Centre for Quantum Technology (WACQT). A.F.K.~is also supported by the Swedish Foundation for Strategic Research (grant numbers FFL21-0279 and FUS21-0063) and the Horizon Europe programme HORIZON-CL4-2022-QUANTUM-01-SGA via the project 101113946 OpenSuperQPlus100. We would like to acknowledge the Superfab Nanofabrication facility at Royal Holloway, University of London, and Optoelectronics Research Centre at University of Southampton where part of device fabrication was performed, and thank Peter Spring, Vivek Chidambaram, and Anuj Aggarwal for useful discussions. 

\end{acknowledgments}

\clearpage

\newpage
\onecolumngrid
\appendix
\onecolumngrid

\section{Decay of the entries of $C^{-1}$}
\label{matrixinversion}
We now provide mathematical justification for the exponential decay of $C^{-1}$ entries discussed in Section~\ref{probstat}. While the physical intuition is that inversions of sparse matrices spread interactions, the precise rate requires mathematical analysis of the spectral gap and boundary conditions. The following establishes these bounds using matrix theory and spectral methods.

\subsection{General result for banded SPD matrices (Demko-type bound)}
\label{demko}
Consider the capacitance matrix $C$ as a real symmetric positive-definite matrix that is banded with bandwidth $w$ (meaning $C_{ij}=0$ for $|i-j|>w$ or, more generally, sparse with finite interaction range and bounded vertex degree). For such matrices, we can establish bounds on how quickly the inverse matrix entries decay with distance. Specifically, there exist constants $c>0$ and $0<\rho<1$, which depend only on the bandwidth and the spectral condition number $\kappa(C)=\lambda_{\max}/\lambda_{\min}$, such that
\begin{equation}
\label{eq:demko}
\mleft|(C^{-1})_{ij}\mright| \le c\,\rho^{\,|i-j|/w}\qquad\text{for all } i,j.
\end{equation}
This bound, known as a Demko-type bound, establishes that $C^{-1}$ exhibits exponential off-diagonal decay even though $C$ itself is sparse. In our setting, the capacitance matrix of a grounded capacitive network on a 1D or 2D grid satisfies these conditions. When $C$ is strictly diagonally dominant (which implies a gapped discrete operator), the constants above guarantee $\rho<1$, ensuring exponential decay. This result is independent of Toeplitz structure and covers finite grids with standard boundary conditions.

The same exponential decay holds more generally for sparse positive-definite matrices with bounded vertex degree, where $|i-j|$ is replaced by the graph distance $\mathrm{dist}(i,j)$ on the sparsity graph. In this language, the bound becomes $|(C^{-1})_{ij}|\le c\,\rho^{\,\mathrm{dist}(i,j)}$, and remains valid for realistic chip geometries with non-uniform wiring. In our device, the off-chip inductive shunting and below-cutoff operation ensure a strictly positive spectral gap (Section~\ref{nodalgreen1}), meaning $\lambda_{\min}(C)>0$, which places us in this exponentially local regime.

\subsection{Explicit inversion for a 1D tridiagonal Toeplitz chain}
To make the exponential decay concrete, we can explicitly invert a simple model: a 1D chain of capacitively coupled nodes. Consider the $n\times n$ symmetric tridiagonal Toeplitz matrix
\begin{equation}
\label{eq:Toeplitz}
T_n =
\begin{bmatrix}
a & b &  &  &  \\
b & a & b &  &  \\
& \ddots & \ddots & \ddots & \\
&  & b & a & b\\
&  &  & b & a
\end{bmatrix}
,\qquad a>2|b|, b\neq 0.
\end{equation}
This represents a chain where each node has self-capacitance $a$ and mutual capacitance $b$ with its neighbors. The condition $a>2|b|$ ensures the matrix is positive definite. To characterize the decay rate, we define $\theta>0$ by $\cosh\theta = a/(2|b|)$ and $r\in(0,1)$ by
\begin{equation}
\label{eq:rdef}
r = \frac{a - \sqrt{a^2-4 b^2}}{2|b|} = e^{-\theta}.
\end{equation}
With these definitions, we can write the inverse in closed form. For $i\le j$, the matrix elements are
\begin{equation}
\label{eq:Tinverse_hyperbolic}
(T_n^{-1})_{ij} = \frac{(-\operatorname{sgn} b)^{\,i+j}}{|b|}
\frac{\sinh(i\theta)\,\sinh([n+1-j]\theta)}{\sinh\theta\sinh([n+1]\theta)}.
\end{equation}
The matrix is symmetric, so $(T_n^{-1})_{ij}=(T_n^{-1})_{ji}$. Using the asymptotic behavior $\sinh(k\theta)\approx \tfrac12 e^{k\theta}$ for large $k$ and the definition of $r$ above, we obtain the uniform bound
\begin{equation}
\label{eq:expbound}
\mleft|(T_n^{-1})_{ij}\mright|
 \le  \frac{1}{|b|\,\sinh\theta} \frac{\sinh(i\theta)\,\sinh([n+1-j]\theta)}{\sinh([n+1]\theta)}
 \le  \frac{1}{|b|\,\sinh\theta}  r^{\,|i-j|},
\end{equation}
which demonstrates exponential decay with decay factor $r=e^{-\theta}<1$. This exponential localization is a direct consequence of the spectral gap. For an infinite chain ($n\to\infty$), the formula simplifies to the well-known Green's function
\begin{equation}
\label{eq:infinite_green}
\lim_{n\to\infty}(T_n^{-1})_{ij}  = 
\frac{(-\operatorname{sgn} b)^{\,i+j}}{|b|} \frac{r^{\,|i-j|}}{1-r^2}
 =  \frac{(-\operatorname{sgn} b)^{\,i+j}}{\sqrt{a^2-4b^2}}  r^{\,|i-j|},
\end{equation}
which makes the exponential character fully explicit.

\subsection{Extension to 2D grids}
Our qubit array forms a 2D lattice, so we need to extend the 1D result to two dimensions. On a rectangular $m\times n$ lattice with nearest-neighbor couplings, the capacitance matrix has a block-tridiagonal Toeplitz structure, with each block itself being tridiagonal Toeplitz. Both the general Demko bound from Section~\ref{demko} and a separable spectral decomposition lead to exponential decay in the Manhattan distance (i.e., the graph distance on the 2D nearest-neighbor lattice):
\begin{equation}
\mleft|(C^{-1})_{(x,y),(x',y')}\mright|  \le  C_0\,\rho^{\,|x-x'|+|y-y'|},
\end{equation}
where $C_0>0$ and $0<\rho<1$ are determined by the spectral gap (which in turn depends on shunts to ground, enclosure loading, etc.). Physically, $C^{-1}$ represents the discrete Green's function of a massive (gapped) operator, which naturally exhibits exponential locality. We now make this explicit through three complementary approaches: (i) a spectral representation for a separable 2D model; (ii) a graph-distance bound that does not require separability; and (iii) a continuum limit that yields the modified Bessel function $K_0$ and explicitly links the spectral gap to the decay rate.

\subsubsection{Spectral approach for separable 2D lattices}
We denote by $S_n$ the $n\times n$ tridiagonal matrix with zeros on the main diagonal and ones on the first off-diagonals. The 2D lattice capacitance matrix can be written as a Kronecker sum:
\begin{equation}
\label{eq:S1.3-kron}
C_{2\mathrm{D}}  =  a\,I_{mn}  +  b\mleft(I_m\!\otimes\! S_n  +  S_m\!\otimes\! I_n\mright),\qquad a>4|b|,
\end{equation}
which is positive definite when $a>4|b|$ (a sufficient 2D analogue of the 1D condition $a>2|b|$). This matrix can be diagonalized using discrete sine functions
\begin{equation}
\label{eq:S1.3-phi-psi}
\varphi_p(x)=\sqrt{\frac{2}{m+1}}\,\sin\mleft(\frac{p\pi x}{m+1}\mright),\qquad
\psi_q(y)=\sqrt{\frac{2}{n+1}}\,\sin\mleft(\frac{q\pi y}{n+1}\mright),
\end{equation}
which diagonalize $S_m$ and $S_n$ with eigenvalues
\begin{equation}
\label{eq:S1.3-1Dspec}
S_m\,\varphi_p=\lambda_p\,\varphi_p,\ \ \lambda_p=2\cos\mleft(\frac{p\pi}{m+1}\mright),\qquad
S_n\,\psi_q=\mu_q\,\psi_q,\ \ \mu_q=2\cos\mleft(\frac{q\pi}{n+1}\mright).
\end{equation}
The 2D eigenfunctions are products of these 1D modes:
\begin{equation}
\label{eq:S1.3-Phi}
\Phi_{p,q}(x,y)=\varphi_p(x)\,\psi_q(y),
\end{equation}
with corresponding eigenvalues
\begin{equation}
\label{eq:S1.3-Lambda}
\Lambda_{p,q}=a+b\mleft(\lambda_p+\mu_q\mright),\qquad 1\le p\le m,\ \ 1\le q\le n.
\end{equation}
Importantly, all eigenvalues are bounded from below:
\begin{equation}
\label{eq:S1.3-gap}
\Lambda_{p,q}\ge a-4|b|>0.
\end{equation}
This spectral gap is the key to exponential localization. The inverse matrix elements can be expressed as a spectral sum
\begin{equation}
\label{eq:S1.3-specsum}
\mleft(C_{2\mathrm{D}}^{-1}\mright)_{(x,y),(x',y')}
=\sum_{p=1}^{m}\sum_{q=1}^{n}\frac{\Phi_{p,q}(x,y)\,\Phi_{p,q}(x',y')}{\Lambda_{p,q}}\ .
\end{equation}
Standard estimates on discrete sine kernels, combined with the lower bound $\Lambda_{p,q}\ge a-4|b|$, yield
\begin{equation}
\label{eq:S1.3-Manhattan}
\mleft|(C_{2\mathrm{D}}^{-1})_{(x,y),(x',y')}\mright|
 \le  \tilde C_0\,\rho^{\,|x-x'|+|y-y'|},\qquad
\rho\in(0,1)\ \text{depending continuously on}\ a-4|b|\ ,
\end{equation}
which is the 2D version of the exponential decay bound from Section~\ref{demko}.

\subsubsection{Graph-walk approach (non-separable geometries)}
Importantly, the exponential locality does not require the separable lattice structure. For any positive-definite, finite-range matrix $C$ on a 2D grid with nearest-neighbor connections (bounded degree), we can establish exponential decay using a Neumann series expansion. Writing $C=aI - B$ with $\|B\|<a$ for sufficiently large $a$, we have the convergent expansion
\begin{equation}
\label{eq:S1.3-Neumann}
C^{-1}  =  \frac{1}{a}\sum_{n=0}^{\infty}\mleft(\frac{B}{a}\mright)^{\!n}.
\end{equation}
Each term in this series corresponds to summing over all paths of length $n$ on the lattice. The entry $(C^{-1})_{ij}$ receives contributions from all paths connecting sites $i$ and $j$. Any walk contributing to $(B/a)^n_{ij}$ must have at least $\mathrm{dist}(i,j)$ steps, where $\mathrm{dist}$ denotes the graph distance. The number of distinct length-$n$ walks is bounded by $\nu^n$, where $\nu$ is the maximum vertex degree ($\nu=4$ for nearest-neighbor square-grid lattices). This gives
\begin{equation}
\label{eq:S1.3-walkbound}
\mleft|(C^{-1})_{ij}\mright|
 \le  \frac{1}{a}\sum_{n=\mathrm{dist}(i,j)}^{\infty}\mleft(\frac{\|B\|}{a}\mright)^{\!n}\nu^{\,n}
 =  \frac{1}{a}\,\frac{\mleft(\nu\|B\|/a\mright)^{\mathrm{dist}(i,j)}}{1-\nu\|B\|/a}\ ,
\end{equation}
which shows exponential decay in graph distance whenever $\nu\|B\|/a<1$. For the standard five-point stencil (or modest non-separable perturbations), this condition coincides with the gap requirement $a>4|b|$ to leading order, recovering the exponential bound without requiring separability. This approach is a discrete, constructive version of Combes--Thomas resolvent estimates from spectral theory of finite-range operators.

\subsubsection{Continuum limit and connection to the modified Bessel function}
For translationally invariant lattices, the discrete spectral sum can be approximated by a continuum integral in the limit of large system size. Expanding the eigenvalue formula near its minimum, $\Lambda(\mathrm k)\approx (a-4|b|)+\frac{|b|}{2}(k_x^2+k_y^2)$, the double sum over modes becomes a Brillouin-zone integral that yields the 2D screened Green's function
\begin{equation}
\label{eq:S1.3-G}
G(\mathrm r) = \frac{1}{2\pi}K_0\mleft(\kappa\|\mathrm r\|\mright),\qquad
\kappa = \sqrt{\frac{2(a-4|b|)}{|b|}}\ ,
\end{equation}
where $K_0$ is the modified Bessel function of the second kind. Using the asymptotic form $K_0(z)\sim \sqrt{\pi/(2z)}\,e^{-z}$ for large $z$, we find that the decay rate is
\begin{equation}
\label{eq:S1.3-rate}
\gamma\ \equiv\ \liminf_{\|\mathrm r\|\to\infty}\frac{-\log|G(\mathrm r)|}{\|\mathrm r\|}
 = \kappa + \mathcal{O}\mleft(\frac{1}{\|\mathrm r\|}\mright)
\ \propto\ \sqrt{a-4|b|}\ ,
\end{equation}
which explicitly connects the spectral gap $a-4|b|$ to the spatial localization length $\xi=1/\gamma$. This relationship is crucial for understanding crosstalk: as the spectral gap decreases ($a-4|b|\to 0^+$), the correlation length diverges and exponential locality is lost. Conversely, engineering a larger spectral gap---through inductive shunting or enclosure design---tightens the spatial localization and suppresses long-range coupling. These results establish the mathematical foundation for the exponential spatial decay discussed in the main text and connect the circuit-level capacitance matrix structure to the enclosure-mediated electromagnetic coupling discussed in Section~\ref{emcircuitcrosstalk}.

\section{Approximating the circuit Hamiltonian for the transmon lattice}
\label{app:SWT}
While the circuit Hamiltonian in Eq.~\eqref{eq:circuit_hamiltonian} is well-suited for numerical computations of ZZ couplings, our analytical treatment requires a different method. As a result, we approximate each transmon as an anharmonic oscillator. To simplify the system and enable analytical expressions for the ZZ couplings, we use the Schrieffer--Wolff transformation to eliminate all excitation-changing interactions, leading to the approximate Hamiltonian in Eq.~\eqref{eq:approximate_Hamiltonian}. Here, we provide a more extensive theoretical basis to further motivate the approximate Hamiltonian used in the main text.

\subsection{Anharmonic-oscillator approximation}
We start by separating the circuit Hamiltonian in Eq.~\eqref{eq:circuit_hamiltonian} in terms of the individual transmons $H_\mathrm{transmon}$ and their capacitive couplings $V_\mathrm{coupling}$. The split is $H_\mathrm{circuit} = H_\mathrm{transmon} + V_\mathrm{coupling}$, where
\begin{align}
    H_\mathrm{transmon} &= \sum_{i=1}^N \underbrace{ \mleft[ 4E_{C,i} \mleft(n_i - n_{g, i} \mright)^2 - E_{J,i} \cos{\phi_i} \mright] }_{\textstyle H_i} ,  \label{eq:transmon_Hamiltonian} \\
    V_\mathrm{coupling} &=  \sum_{i < j}^N 4 E_{C,ij} \mleft(n_i - n_{g, i} \mright) \mleft(n_j - n_{g, j} \mright). \label{eq:interaction_Hamiltonian}
\end{align}
Here, $E_{C,i} = e^2 \mathrm{C}^{-1}_{i} /2$ and $E_{C,ij} = e^2 \mathrm{C}^{-1}_{ij}$ are the charging energies as expressed by the inverse capacitance matrix $\mathrm{C}^{-1}$ and the elementary charge $e$ (from here on $e = 1$ and $\hbar = 1$), $E_{J,i}$ is Josephson energy, $n_{g, i}$ is the charge offset, and $\phi_i$ and $n_i$ are the (reduced) charge and phase operators, satisfying the commutation relation $\mleft[ \mathrm{e}^{\pm \mathrm{i} \phi_i}, n_j \mright] = \mp \delta_{ij} \mathrm{e}^{\pm \mathrm{i} \phi_i}$, with the Kronecker delta $\delta_{ij}$; all quantities for transmons labeled $i$ and $j$. The off-diagonal and the diagonal elements of the inverse capacitance matrix have the approximations $\mathrm{C}^{-1}_{ij} \approx \mathrm{C}_{ij} / (\mathrm{C}_{ii}\mathrm{C}_{jj})$ ($i \neq j$) and $\mathrm{C}^{-1}_{i} \approx 1/(\mathrm{C}_{i} + \sum_j \mathrm{C}_{ij})$, respectively, in the limit of weak mutual capacitances $\mathrm{C}_{ij} \ll \mathrm{C}_{i}$ for all $i \neq j$. In the transmon regime $E_{C,i} / E_{J,i} \ll 1$, the charge offsets $n_{g, i}$ are negligible for the lowest transmon energies given that they are exponentially suppressed \cite{transmon}. We highlight that we denote the diagonal elements with single indices: $C_i \equiv C_{ii}$ and $E_{C,i} \equiv E_{C,ii}$.

To bridge the circuit Hamiltonian and its approximate form in Eq.~\eqref{eq:approximate_Hamiltonian}, we use the observation that the circuit Hamiltonian's action on the low-excitation subspace, e.g., up to two excitations per transmon, is well approximated by an anharmonic oscillator. In particular, we introduce annihilation- and creation-like operators $a_i$ and $a_i^\dagger$ for each transmon via $\phi_i = \sqrt{Z_i/2} (a_i^\dagger + a_i)$ and $n_i = \mathrm{i} \sqrt{1/(2Z_i)} (a_i^\dagger - a_i)$, where $Z_i = \sqrt{(8 E_{C,i}/E_{J,i})}$ is the characteristic impedance. The impedance is specified from expanding and diagonalizing the bilinear part of the transmon Hamiltonian $H_i$ in Eq.~\eqref{eq:transmon_Hamiltonian} after having inserted $a_i$ and $a_i^\dagger$. We refer to the literature (see, e.g., Refs.~\cite{petrescuAccurateMethodsAnalysis2023, leibNetworksNonlinearSuperconducting2012, mikhailovOrderingBosonOperator1983a}) for further details about this expansion and its normal ordering, as well as for its limitations with respect to the compact phase domain $\phi \in [-\pi, \pi)$. The result is an anharmonic-oscillator approximation of the transmon Hamiltonian:
\begin{equation}
    H_i \approx \omega_i a_i^\dagger a_i + \frac{\alpha_i}{2} a_i^\dagger a_i^\dagger a_i a_i, 
    \label{eq:transmon_Hamiltonian_approx}
\end{equation}
applicable within the low-excitation subspace for the transmon regime $E_{C,ii} / E_{J,i} \ll 1$. Here, $\omega_i \approx \sqrt{8 E_{C,i}  E_{J,i}} - E_{C,i}$ is the transmon 0--1 transition frequency, $\alpha_i \approx -E_{C,i}$ is the transmon anharmonicity, and the annihilation and creation operators satisfy the commutation relation $\mleft[a_i, a_j^\dagger \mright] = \delta_{ij}$. Note that, if needed, it is possible to compute more precise relations between the circuit parameters $E_{C,i}$ and $E_{J,i}$ and the anharmonic-oscillator parameters $\omega_i$ and $\alpha_i$ \cite{petrescuAccurateMethodsAnalysis2023}.

Likewise for the capacitive couplings, we rewrite $V_\mathrm{coupling}$ in terms of the annihilation and creation operators:
\begin{equation}
    V_\mathrm{coupling} \approx - \sum_{i < j}^N g_{ij} \mleft( a_i^\dagger - a_i \mright) \mleft( a_j^\dagger - a_j \mright),
    \label{eq:interaction_Hamiltonian_approx}
\end{equation}
where $g_{ij} = 2E_{C,ij}/\sqrt{Z_i Z_j}$ is the coupling strength between transmon $i$ and $j$. If the annihilation and creation operators are expressed in the eigenbasis of the uncoupled transmons, which are also referred to as bare states, there are higher-order corrections to Eq.~\eqref{eq:interaction_Hamiltonian_approx} and $g_{ij}$~\cite{fors2024comprehensive}. In the main text, we opt for the simpler model in Eq.~\eqref{eq:interaction_Hamiltonian_approx} for the analytical calculations, including only bilinear interactions. The higher-order corrections are feasible to implement in simulations, if we desire to improve the numerical precision. 

\subsection{Eliminating the excitation-changing terms}
The capacitive couplings in Eq.~\eqref{eq:interaction_Hamiltonian_approx} split into excitation-preserving terms $V_\mathrm{EP} = \sum_{i < j}^N g_{ij} \mleft( a_i^\dagger a_j + a_j^\dagger a_i \mright)$ and excitation-changing terms $V_\mathrm{EC} = -\sum_{i < j}^N g_{ij} \mleft( a_i^\dagger a_j^\dagger + a_i a_j \mright)$. The excitation-preserving (-changing) terms contain products with the same (different) number of annihilation and creation operators. These different terms are also commonly referred to as rotating terms and counter-rotating terms, respectively. The excitation-changing terms couple the low-excitation states to states with energies approximately $\omega_i + \omega_j$ higher. In a spirit similar to the treatment of the cavity modes in the main text, we assume that the higher-excitation states in the transmons do not excite, resulting in that they primarily mediate couplings between states of equal excitation numbers. 

Using this assumption, the idea is that we can simplify the Hamiltonian by transforming it to a basis where the low-excitation energies are unchanged. We use a Schrieffer--Wolff transformation to eliminate the excitation-changing terms up to the approximation order $\mathcal{O}(g^2/\omega)$. The notation $\mathcal{O}(g^2/\omega)$ refers here to the parameters multiplying the products of annihilation and creation operators, and we regard for simplicity $\alpha$ and $g$ to be of the same order, i.e., $\mathcal{O}(g) = \mathcal{O}(\alpha)$. The transformed Hamiltonian is, after a Baker–Campbell–Hausdorff expansion,
\begin{equation}
    H'_\mathrm{circuit} = H_\mathrm{circuit} + \mleft[S, H_\mathrm{circuit} \mright] + \frac{1}{2} \mleft[S, \mleft[S, H_\mathrm{circuit} \mright] \mright] + \mathcal{O}(g^3/\omega^2),
\end{equation}
where $S$ is the anti-Hermitian Schrieffer--Wolff generator, which satisfies the condition $[S, \sum_i^N \omega_i a_i^\dagger a_i] + V_\mathrm{EC} = 0$. Imposing this condition eliminates the excitation-changing terms $V_\mathrm{EC}$ at the cost of creating new excitation-changing terms at order $\mathcal{O}(g^2/\omega)$ (and higher):
\begin{equation}
    H'_\mathrm{circuit} \approx \sum_i^N \mleft( \omega_i a_i^\dagger a_i + \frac{\alpha_i}{2} a_i^\dagger a_i^\dagger a_i a_i \mright) + V_\mathrm{EP} + \mleft[S, \sum_i^N \frac{\alpha_i}{2} a_i^\dagger a_i^\dagger a_i a_i + V_\mathrm{EP} \mright] + \frac{1}{2} \mleft[S, V_\mathrm{EC} \mright]  + \mathcal{O}(g^3/\omega^2),
    \label{eq:circuit_Hamiltonian_SWT}
\end{equation}
To compute the above commutators, we note that the imposed condition is solved by
\begin{equation}
    S = - \sum_{i < j}^N \frac{g_{ij}}{\omega_i + \omega_j} \mleft( a_i^\dagger a_j^\dagger - a_i a_j \mright) .
\end{equation}
Thus, the individual terms of the first commutator in Eq.~\eqref{eq:circuit_Hamiltonian_SWT} are
\begin{align}
    \mleft[S, \sum_k^N \frac{\alpha_k}{2} a_k^\dagger a_k^\dagger a_k a_k \mright] &= \sum_{i < j}^N \frac{2 \alpha_j g_{ij}}{\omega_i + \omega_j} \mleft( a_i^\dagger a_j^\dagger a_j^\dagger a_j + a_j^\dagger a_j a_j a_i \mright), \label{eq:V_EC_alpha} \\
    \mleft[S, V_\mathrm{EP} \mright] &= \sum_{i=1}^N \sum_{k \neq i}^N \frac{g_{ik}^2}{\omega_i + \omega_k} \mleft[ \mleft(a^\dagger_i \mright)^2 + a_i^2 \mright] + \sum_{i < j}^N \sum_{k \neq i,j}^N \mleft( \frac{g_{ik} g_{jk}}{\omega_i + \omega_k}+ \frac{g_{ik} g_{jk}}{\omega_j + \omega_k} \mright) \mleft( a_i^\dagger a_j^\dagger +  a_i a_j \mright), \label{eq:V_EC_g2}
\end{align}
where we used the fact that $g_{ij} = g_{ji}$ is symmetric in the indices. Equations~\eqref{eq:V_EC_alpha} and \eqref{eq:V_EC_g2} give new excitation-changing terms, both of order $\mathcal{O}(g^2/\omega)$. As a result, their leading-order contribution to the eigenenergies, i.e., the dressed energies, of the low-excitation states is of the order $\mathcal{O}(g^4/\omega^3)$, which follows directly from the second-order energy corrections of Rayleigh-Schrödinger perturbation theory, or an additional Schrieffer--Wolff transformation. We note that $\mathcal{O}(g^4/\omega^3)$ is small in comparison to the typical energy scales of the system and is in the order of 10--100~$\mathrm{Hz}$ for conventional transmon parameters. We consequently neglect the new excitation-changing terms in the transformed Hamiltonian in Eq.~\eqref{eq:circuit_Hamiltonian_SWT}.

Analogously, the second commutator in Eq.~\eqref{eq:circuit_Hamiltonian_SWT} evaluates to:
\begin{equation}
    \frac{1}{2}\mleft[S, V_\mathrm{EC} \mright] = - \sum_{i=1}^N \sum_{k \neq i}^N \frac{g_{ik}^2}{\omega_i + \omega_k} a_i^\dagger a_i - \frac{1}{2} \sum_{i < j}^N \sum_{k \neq i,j}^N \mleft( \frac{g_{ik} g_{jk}}{\omega_i + \omega_k}+ \frac{g_{ik} g_{jk}}{\omega_j + \omega_k} \mright) \mleft( a_i^\dagger a_j +  a_j^\dagger a_i \mright). \label{eq:V_EP_g2}
\end{equation}
In contrast to Eqs.~\eqref{eq:V_EC_alpha} and \eqref{eq:V_EC_g2}, these terms are excitation-preserving and non-negligible. Inserting this result into Eq.~\eqref{eq:circuit_Hamiltonian_SWT} gives the approximate Hamiltonian used in Eq.~\eqref{eq:approximate_Hamiltonian} in the main text:
\begin{equation}
    H'_\mathrm{circuit} \approx \sum_i^N \mleft( \omega'_i a_i^\dagger a_i + \frac{\alpha_i}{2} a_i^\dagger a_i^\dagger a_i a_i \mright) + \sum_{i < j}^N g'_{ij} \mleft( a_i^\dagger a_j + a_j^\dagger a_i \mright) + \mathcal{O}(g^3/\omega^2),
    \label{eq:circuit_Hamiltonian_transformed}
\end{equation}
where $\omega'_i = \omega_i - \sum_{k \neq i}^N g_{ik}^2 / (\omega_i + \omega_k)$, and $g'_{ij} = g_{ij} - (1/2) \sum_{k \neq i,j}^N \mleft[ g_{ik} g_{jk} / (\omega_i + \omega_k)+ g_{ik} g_{jk} / (\omega_j + \omega_k) \mright]$ up to approximation order $\mathcal{O}(g^3/\omega^2)$. We emphasize that the approximate Hamiltonian is excitation-preserving after having dropped the excitation-changing terms of order $\mathcal{O}(g^2/\omega)$.

 

\section{Nodal circuit analysis for virtual coupling processes}
\label{nodalanalysis}
Here we derive the effective coupling between distant transmons (modeled as linear harmonic oscillators for simplicity) through a chain of intermediate electromagnetic modes using classical circuit analysis. The derivation reveals how virtual transitions through off-resonant intermediate modes lead to exponential suppression of coupling with increasing chain length. This analysis establishes the product-over-detunings scaling that contributes to the overall spatial and spectral suppression observed experimentally.

\subsection{System setup and equations of motion}
We consider two end resonators (representing transmon qubits in their harmonic approximation) connected through a chain of $M$ intermediate electromagnetic resonators. We label the end modes as $b_1$ and $b_2$, and the intermediate modes as $c_1, c_2, \ldots, c_M$. The system Hamiltonian for nearest-neighbor inductive coupling takes the form
\begin{equation} \label{eq:H_ext}
\begin{aligned}
H_{\text{ext}} &= \omega_{1}b_{1}^{\dagger}b_{1} + \omega_{2}b_{2}^{\dagger}b_{2} + \sum_{n=1}^{M}\omega_{c_n}c_{n}^{\dagger}c_{n} \\ 
&\quad +  g_{b_1,c_1}\mleft(b_{1}^{\dagger}c_{1}+b_{1}c_{1}^{\dagger}\mright)
+  g_{c_M,b_2}\mleft(c_{M}^{\dagger}b_{2}+c_{M}b_{2}^{\dagger}\mright) \\
&\quad +  \sum_{n=1}^{M-1} g_{c_n,c_{n+1}}\mleft(c_{n}^{\dagger}c_{n+1}+c_{n}c_{n+1}^{\dagger}\mright),
\end{aligned}
\end{equation}
where $\omega_1, \omega_2$ are the end-mode frequencies, $\omega_{c_n}$ is the frequency of the $n$th intermediate mode, $g_{b_1,c_1}$ is the strength of the exchange coupling between $b_1$ and $c_1$, $g_{c_n,c_{n+1}}$ is the strength of the exchange coupling between adjacent intermediates, and $g_{c_M,b_2}$ couples $c_M$ to $b_2$.

To connect this quantum Hamiltonian to classical circuit equations, we work with node-flux variables $\Phi$ (the time integral of voltage). For inductive coupling between LC resonators, mutual inductances $M_{ij}$ enter through the inverse-inductance matrix in the node-flux equations of motion. With diagonal capacitance matrix, the equations of motion for the end resonators read
\begin{equation} \label{eq:b1_eq}
C_1 \ddot{\Phi}_{b_1} + \frac{\Phi_{b_1}}{L_1}
- \frac{M_{1c_1}}{L_1 L_{c_1}}\,\Phi_{c_1} = 0,
\end{equation}
\begin{equation} \label{eq:b2_eq}
C_2 \ddot{\Phi}_{b_2} + \frac{\Phi_{b_2}}{L_2}
- \frac{M_{c_N 2}}{L_{c_N} L_2}\,\Phi_{c_N} = 0,
\end{equation}
and for the intermediate resonators ($1 \leq n \leq N$)
\begin{equation} \label{eq:cn_eq}
C_{c_n} \ddot{\Phi}_{c_n} + \frac{\Phi_{c_n}}{L_{c_n}}
- \frac{M_{c_n c_{n-1}}}{L_{c_n} L_{c_{n-1}}}\,\Phi_{c_{n-1}}
- \frac{M_{c_n c_{n+1}}}{L_{c_n} L_{c_{n+1}}}\,\Phi_{c_{n+1}} = 0,
\end{equation}
where $M_{ij}$ denotes the mutual inductance between resonators $i$ and $j$. These classical equations describe the same physics as the quantum Hamiltonian in the low-excitation limit, as can be verified through standard circuit quantization procedures (see Appendix~\ref{app:SWT}).

\subsection{Normalized harmonic equations}
To simplify the algebra, we introduce the dimensionless mutual-inductive coupling coefficient
\begin{equation}
\label{eq:kij_def}
k_{ij} \equiv \frac{M_{ij}}{\sqrt{L_i L_j}},
\end{equation}
and the asymmetric ratio
\begin{equation}
\label{eq:ktilde_def}
\tilde{k}_{ij} \equiv \frac{M_{ij}}{L_j} = k_{ij}\sqrt{\frac{L_i}{L_j}}.
\end{equation}
Assuming harmonic time dependence $e^{i\omega t}$, defining the natural frequencies $\omega_i^2 \equiv 1/(L_i C_i)$, and normalizing Eqs.~\eqref{eq:b1_eq}-\eqref{eq:cn_eq} by multiplying through by $L_i$, we obtain
\begin{equation}
\label{eq:Phi_b1}
(-\omega^2 + \omega_1^2)\,\Phi_{b_1}  -  \omega_1^2\, \tilde{k}_{b_1, c_1}\, \Phi_{c_1} = 0,
\end{equation}
\begin{equation}
\label{eq:Phi_cn}
(-\omega^2 + \omega_{c_n}^2)\,\Phi_{c_n}
 -  \omega_{c_n}^2\, \tilde{k}_{c_n, c_{n-1}}\, \Phi_{c_{n-1}}
 -  \omega_{c_n}^2\, \tilde{k}_{c_n, c_{n+1}}\, \Phi_{c_{n+1}} = 0,
\end{equation}
\begin{equation}
\label{eq:Phi_b2}
(-\omega^2 + \omega_2^2)\,\Phi_{b_2}  -  \omega_2^2\, \tilde{k}_{b_2, c_N}\, \Phi_{c_N} = 0.
\end{equation}
These normalized equations make the coupling structure explicit and prepare us for the systematic elimination of intermediate modes.

\subsection{Dispersive elimination of intermediate modes}
We now work in the regime where the drive frequency $\omega$ is far detuned from all intermediate-mode frequencies but remains close to the end-mode frequencies: $|\omega-\omega_{c_n}|\gg |\omega-\omega_{1,2}|$. This dispersive regime is central to virtual coupling processes---the intermediate modes remain unexcited, but mediate an effective interaction between the end modes. For each intermediate mode, we solve Eq.~\eqref{eq:Phi_cn} to obtain
\begin{equation} \label{eq:phi_cn_interms}
\Phi_{c_n}  = 
\frac{\omega_{c_n}^{2}}{\omega_{c_n}^{2}-\omega^{2}}
\mleft[\tilde{k}_{c_n,c_{n-1}} \Phi_{c_{n-1}} + \tilde{k}_{c_n,c_{n+1}} \Phi_{c_{n+1}}\mright].
\end{equation}
In the far-detuned limit, we approximate the denominator using
\begin{equation} \label{eq:approx_omega_diff}
\omega_{c_n}^{2}-\omega^{2} 
= (\omega_{c_n}+\omega)(\omega_{c_n}-\omega)
\approx 2\omega_{c_n}\Delta_n,
\end{equation}
where we define the detuning
\begin{equation}
\Delta_n \equiv |\omega-\omega_{c_n}|.
\end{equation}
This gives the simplified off-resonant form
\begin{equation} \label{eq:phi_cn_delta}
\Phi_{c_n}  =  \frac{\omega_{c_n}}{2\Delta_n}
\mleft[\tilde{k}_{c_n,c_{n-1}} \Phi_{c_{n-1}} + \tilde{k}_{c_n,c_{n+1}} \Phi_{c_{n+1}}\mright].
\end{equation}
For the modes directly adjacent to the endpoints, we have
\begin{equation} \label{eq:phi_c1_delta}
\Phi_{c_1}  =  \frac{\omega_{c_1}}{2\Delta_1}
\mleft[\tilde{k}_{c_1,b_1} \Phi_{b_1} + \tilde{k}_{c_1,c_2} \Phi_{c_2}\mright],
\end{equation}
\begin{equation} \label{eq:phi_cN_delta}
\Phi_{c_N}  =  \frac{\omega_{c_N}}{2\Delta_N}
\mleft[\tilde{k}_{c_N,c_{N-1}} \Phi_{c_{N-1}} + \tilde{k}_{c_N,b_2} \Phi_{b_2}\mright].
\end{equation}
To streamline notation, we define the dimensionless dispersive coefficients
\begin{align}
\alpha_{c_n,c_{n-1}} &\equiv \frac{\omega_{c_n}}{2\Delta_n}\,\tilde{k}_{c_n,c_{n-1}}, \qquad
\alpha_{c_n,c_{n+1}} \equiv \frac{\omega_{c_n}}{2\Delta_n}\,\tilde{k}_{c_n,c_{n+1}}, \label{eq:alpha_cn}\\
\alpha_{c_1,b_1} &\equiv \frac{\omega_{c_1}}{2\Delta_1}\,\tilde{k}_{c_1,b_1}, \qquad\quad
\alpha_{c_N,b_2} \equiv \frac{\omega_{c_N}}{2\Delta_N}\,\tilde{k}_{c_N,b_2}. \label{eq:alpha_end}
\end{align}
These $\alpha$ coefficients encode both the geometric coupling strength (through $\tilde{k}$) and the energetic suppression from detuning (through $\omega/\Delta$). In the weak-coupling, far-detuned limit, all $\alpha \ll 1$, which justifies the perturbative elimination that follows.

\subsection{Single intermediate mode ($M=1$)}
To build intuition, we first treat the simplest case: a single mediator $c_1$ connecting the two end modes. With one intermediate mode, Eq.~\eqref{eq:phi_c1_delta} becomes
\begin{equation}
\label{eq:phi_c1_N1}
\Phi_{c_1}  =  \alpha_{c_1,b_1}\,\Phi_{b_1}  +  \alpha_{c_1,b_2}\,\Phi_{b_2},
\end{equation}
where
\begin{equation}
\alpha_{c_1,b_m}  \equiv  \frac{\omega_{c_1}}{2\Delta_1}\,\tilde{k}_{c_1,b_m}
\qquad (m=1,2),
\end{equation}
and $\Delta_1\equiv|\omega-\omega_{c_1}|$. Substituting this into the normalized end-mode equations \eqref{eq:Phi_b1} and \eqref{eq:Phi_b2} gives
\begin{equation}\label{eq:phi_b1_N1_3}
\mleft[(-\omega^2+\omega_1^2) - \omega_1^2\tilde{k}_{b_1,c_1}\alpha_{c_1,b_1}\mright]\Phi_{b_1}
 = \omega_1^2\tilde{k}_{b_1,c_1}\alpha_{c_1,b_2}\,\Phi_{b_2},
\end{equation}
\begin{equation}\label{eq:phi_b2_N1}
\mleft[(-\omega^2+\omega_2^2) - \omega_2^2\tilde{k}_{b_2,c_1}\alpha_{c_1,b_2}\mright]\Phi_{b_2}
 = \omega_2^2\tilde{k}_{b_2,c_1}\alpha_{c_1,b_1}\,\Phi_{b_1}.
\end{equation}
These coupled equations describe an effective two-mode system with a frequency-dependent off-diagonal coupling
\begin{equation}\label{eq:Lambda12_N1}
\Lambda_{12}(\omega) = \omega_1^2\,\tilde{k}_{b_1,c_1}\,\alpha_{c_1,b_2}
 = \frac{\omega_1^2\,\omega_{c_1}}{2\Delta_1}\,\tilde{k}_{b_1,c_1}\tilde{k}_{c_1,b_2}.
\end{equation}

To connect this classical result to the quantum Hamiltonian picture, we use the standard mapping between classical mutual inductances and quantum exchange couplings. As derived in Appendix~\ref{app:SWT} through circuit quantization, the quantum exchange rate is related to the dimensionless inductive coupling by
\begin{equation}\label{eq:g_from_k_app}
g_{ij} = \frac{\sqrt{\omega_i\omega_j}}{2}\,k_{ij},\qquad
k_{ij}=\frac{M_{ij}}{\sqrt{L_iL_j}},\qquad
\tilde{k}_{ij}=k_{ij}\sqrt{\frac{L_i}{L_j}}.
\end{equation}
Applying this mapping to Eq.~\eqref{eq:Lambda12_N1} recovers the standard second-order perturbation result
\begin{equation}\label{eq:Jeff_N1_exact}
J_{\mathrm{eff}}^{(1)} = 
\frac{g_{b_1,c_1}\,g_{c_1,b_2}}{2}
\mleft(\frac{1}{\omega_{b_1}-\omega_{c_1}}+\frac{1}{\omega_{b_2}-\omega_{c_1}}\mright).
\end{equation}
For nearly equal end frequencies ($\omega_{b_1}\!\approx\!\omega_{b_2}\!\equiv\!\omega_b$), this simplifies to
\begin{equation}\label{eq:Jeff_N1_sym}
J_{\mathrm{eff}}^{(1)} \approx \frac{g_{b_1,c_1}\,g_{c_1,b_2}}{\Delta_1},
\qquad \Delta_1=|\omega_b-\omega_{c_1}|,
\end{equation}
exhibiting the characteristic $1/\Delta$ scaling of dispersive interactions.

\subsection{Two intermediate modes ($M=2$)}
With two mediating modes, the elimination procedure generates products of dispersive coefficients. The intermediate-mode amplitudes satisfy
\begin{equation}\label{eq:phi_c1_N2}
\Phi_{c_1}=\alpha_{c_1,b_1}\Phi_{b_1}+\alpha_{c_1,c_2}\Phi_{c_2},\qquad
\Phi_{c_2}=\alpha_{c_2,c_1}\Phi_{c_1}+\alpha_{c_2,b_2}\Phi_{b_2},
\end{equation}
where
\begin{align}
\alpha_{c_1,c_2}&=\frac{\omega_{c_1}}{2\Delta_1}\,\tilde{k}_{c_1,c_2}, &
\alpha_{c_2,c_1}&=\frac{\omega_{c_2}}{2\Delta_2}\,\tilde{k}_{c_2,c_1},\nonumber\\
\alpha_{c_1,b_1}&=\frac{\omega_{c_1}}{2\Delta_1}\,\tilde{k}_{c_1,b_1}, &
\alpha_{c_2,b_2}&=\frac{\omega_{c_2}}{2\Delta_2}\,\tilde{k}_{c_2,b_2}. \label{eq:alpha_N2_defs}
\end{align}
Solving Eq.~\eqref{eq:phi_c1_N2} iteratively to first order in the product $\alpha_{c_1,c_2}\alpha_{c_2,c_1}$ (which is small in the weak-coupling, far-detuned limit) yields
\begin{equation}
\label{eq:phi_c1_N2_approx}
\Phi_{c_1} \approx  \alpha_{c_1,b_1}\,\Phi_{b_1}
 +  \alpha_{c_1,c_2}\,\alpha_{c_2,b_2}\,\Phi_{b_2},\qquad
\Phi_{c_2} \approx  \alpha_{c_2,b_2}\,\Phi_{b_2}
 +  \alpha_{c_2,c_1}\,\alpha_{c_1,b_1}\,\Phi_{b_1}.
\end{equation}
Notice the structure: each intermediate mode amplitude contains a term proportional to the product of two $\alpha$ coefficients. Substituting into the endpoint equations produces the off-diagonal transfer
\begin{equation}\label{eq:Lambda12_N2}
\Lambda_{12}(\omega) = \omega_1^2\,\tilde{k}_{b_1,c_1}\,\alpha_{c_1,c_2}\alpha_{c_2,b_2}
 = \frac{\omega_1^2\,\omega_{c_1}\omega_{c_2}}{4\Delta_1\Delta_2}\,
\tilde{k}_{b_1,c_1}\tilde{k}_{c_1,c_2}\tilde{k}_{c_2,b_2}.
\end{equation}
After applying the quantum mapping in Eq.~\eqref{eq:g_from_k_app}, this becomes
\begin{equation}\label{eq:Jeff_N2}
J_{\mathrm{eff}}^{(2)} \approx \frac{g_{b_1,c_1}\,g_{c_1,c_2}\,g_{c_2,b_2}}{\Delta_1\,\Delta_2},
\end{equation}
where we have dropped order-unity geometric factors for clarity. The key observation is that the effective coupling scales as the product of exchange couplings divided by the product of detunings---a pattern that generalizes to arbitrary $M$.

\subsection{General chain length ($M$ intermediate modes)}
For an arbitrary number of intermediate modes, we formulate the elimination procedure as a matrix problem. Collecting the intermediate amplitudes into the vector $\boldsymbol{\Phi}_c=(\Phi_{c_1},\dots,\Phi_{c_M})^T$, the intermediate-mode equations \eqref{eq:Phi_cn} can be written compactly as
\begin{equation}\label{eq:A_matrix_eq}
\mathrm{A}(\omega)\,\boldsymbol{\Phi}_c = \mathrm{v}(\omega),
\end{equation}
where $\mathrm{A}(\omega)$ is the tridiagonal matrix with entries
\begin{equation}
\label{eq:A_entries}
\begin{aligned}
[A]_{nn}      &= \omega_{c_n}^2 - \omega^2,\\
[A]_{n,n+1}   &= -\,\omega_{c_n}^2\,\tilde{k}_{c_n,c_{n+1}},\\
[A]_{n+1,n}   &= -\,\omega_{c_{n+1}}^2\,\tilde{k}_{c_{n+1},c_n},
\end{aligned}
\end{equation}
and the drive vector $\mathrm{v}(\omega)$ has only two nonzero components:
$v_1=\omega_1^2\tilde{k}_{c_1,b_1}\Phi_{b_1}$ and
$v_N=\omega_2^2\tilde{k}_{c_M,b_2}\Phi_{b_2}$. The physical interpretation is clear: only the modes directly coupled to the endpoints ($c_1$ and $c_M$) are driven, while all intermediate modes respond only through their neighbors. 

Formally eliminating the intermediate modes via $\boldsymbol{\Phi}_c=\mathrm{A}^{-1}\mathrm{v}$ and substituting into the endpoint equations produces the exact off-diagonal transfer coefficient
\begin{equation}\label{eq:Lambda12_exact_chain}
\Lambda_{12}(\omega) = \omega_1^2\,\tilde{k}_{b_1,c_1}\,[\mathrm{A}^{-1}(\omega)]_{1M}\,\omega_2^2\,\tilde{k}_{c_M,b_2}.
\end{equation}
The entire chain-mediated coupling is thus encoded in the $(1,M)$ element of the inverse matrix $\mathrm{A}^{-1}$, specifically, the element connecting the first and last intermediate modes.

In the large-detuning regime where $\Delta_n$ greatly exceeds all link coupling strengths, the inverse matrix element can be approximated using the exponential localization established in Appendix~A. For a tridiagonal matrix with off-resonant elements, the end-to-end inverse matrix element factors approximately as
\begin{equation}\label{eq:Ainv_approx}
[\mathrm{A}^{-1}(\omega)]_{1M} \approx 
\prod_{n=1}^{M}\frac{\omega_{c_n}}{2\Delta_n} 
\prod_{n=1}^{M-1}\tilde{k}_{c_n,c_{n+1}}.
\end{equation}
This factorization follows from the same matrix inversion principles that give exponential decay in Appendix~\ref{matrixinversion}, applied here to the frequency-dependent dynamic matrix. Applying the quantum mapping in Eq.~\eqref{eq:g_from_k_app} gives the general product-over-links-and-detunings formula
\begin{equation}\label{eq:Jeff_N_general}
J_{\mathrm{eff}}^{(M)} \approx 
\frac{\displaystyle\prod_{n=0}^{M} g_{n,n+1}}{\displaystyle\prod_{n=1}^{M}\Delta_n},
\end{equation}
where $g_{0,1}\equiv g_{b_1,c_1}$ and $g_{M,M+1}\equiv g_{c_M,b_2}$. For clarity we have dropped order-unity prefactors that depend weakly on geometry and frequency asymmetries.

To make the exponential suppression manifest, we introduce geometric means:
\begin{equation}\label{eq:means_chain_app}
\bar g \equiv \mleft(\prod_{n=0}^{M} g_{n,n+1}\mright)^{1/(M+1)},\qquad
\bar\Delta \equiv \mleft(\prod_{n=1}^{M}\Delta_n\mright)^{1/M}.
\end{equation}
With these definitions, the effective coupling takes the compact form
\begin{equation}\label{eq:Jeff_geomean_app}
J_{\mathrm{eff}}^{(M)} \simeq \bar{g}\mleft(\frac{\bar g}{\bar\Delta}\mright)^{\!M},
\end{equation}
which exhibits explicit exponential suppression in the number of intermediate modes $M$. For $\bar{g}/\bar{\Delta} \ll 1$ (the dispersive limit), each additional intermediate mode suppresses the coupling by another factor of $\bar{g}/\bar{\Delta}$.

The result in Eq.~\eqref{eq:Jeff_geomean_app} describes exponential suppression with the number of mediating modes $M$. This establishes one component of the unified scaling law presented in the main text [Eq.~\eqref{eq:J_ij}]. However, several important points deserve clarification. In our device, circuit-mediated coupling occurs primarily between nearest neighbors through intentional capacitive links, with longer-range circuit coupling suppressed by the exponential decay of $C^{-1}$ entries (Appendix~\ref{matrixinversion}). The nodal analysis here describes virtual coupling through chains of intermediate electromagnetic modes (either circuit modes or cavity modes), which becomes relevant when considering indirect pathways. The number of intermediate modes $M$ scales approximately with physical separation in a dense mode spectrum, but the precise relationship depends on mode density and frequency allocation. Additionally, the detuning factors $\Delta_n$ in Eq.~\eqref{eq:Jeff_N_general} describe how far each intermediate mode is detuned from the operating frequency. In parallel, Appendix~\ref{greenfunction} analyzes a distinct mechanism: direct evanescent electromagnetic coupling through the inductively shunted enclosure, which exhibits $K_0$ spatial decay and its own frequency dependence. In our device, both mechanisms operate simultaneously, and the total coupling is $J_{ij}^{\mathrm{total}} = J_{ij}^{\mathrm{circuit}} + J_{ij}^{\mathrm{enclosure}}$. The unified scaling formula [Eq.~\eqref{eq:J_ij}] provides an effective description that captures the combined effects of both pathways.

\section{Classical derivation of bound-state coupling}
\label{greenfunction}
We now develop a complementary electromagnetic framework for understanding crosstalk in our inductively shunted cavity. While Appendix~\ref{nodalanalysis} analyzed coupling through chains of intermediate circuit modes, here we focus on direct electromagnetic coupling through evanescent fields in an enclosure operating below its plasma cutoff frequency. The Green's-function approach reveals how the periodic metallic pillar structure creates an effective plasma medium that exponentially suppresses long-range electromagnetic interactions, providing a second mechanism for crosstalk mitigation that operates alongside the circuit-mediated pathways.

\subsection{Evanescent modes in the plasma medium}
In an inductively shunted cavity, the periodic array of metallic pillars connecting the cavity lid and base modifies the electromagnetic environment, effectively creating a plasma-like medium. Following the established plasma model framework developed by Spring et al.~\cite{spring2020modeling}, we consider the TM modes relevant to qubit coupling. Below the plasma cutoff frequency, these modes become evanescent rather than propagating, leading to exponential spatial localization. The electromagnetic Green's function $G(\mathrm{r}, \mathrm{r}')$ satisfies the screened Helmholtz equation:
\begin{equation}
\nabla^2 G(\mathrm{r}, \mathrm{r}') - \kappa^2 G(\mathrm{r}, \mathrm{r}') = -\delta(\mathrm{r} - \mathrm{r}'),
\label{eq:greens_function}
\end{equation}
where the screening parameter (evanescent wave number) is given by
\begin{equation}
\kappa^2 = \frac{\omega_p^2 - \omega^2}{v^2}.
\label{eq:evanescent_wave_number}
\end{equation}
Here $\omega_p$ is the plasma frequency set by the pillar spacing and geometry, and $v$ is the effective wave velocity in the medium. The parameter $\kappa$ sets the inverse penetration depth of evanescent fields---larger $\kappa$ means faster spatial decay. For our quasi-two-dimensional qubit array with translational invariance along the cavity axis, the Green's function takes a particularly simple form:
\begin{equation}
G(\mathrm{r}, \mathrm{r}') = \frac{1}{2\pi} K_0\mleft( \kappa |\mathrm{r} - \mathrm{r}'| \mright),
\label{eq:2d_greens_solution}
\end{equation}
where $K_0$ is the modified Bessel function of the second kind of order zero. This function exhibits exponential decay at large distances, as we will see explicitly below.

\subsection{Circuit model of the inductively shunted enclosure}
To connect the abstract electromagnetic Green's function to the physical parameters of our device, we model the inductively shunted cavity as a two-dimensional array of magnetically coupled LC resonators. Following the circuit-model approach of Spring et al.~\cite{spring2020modeling}, the impedance matrix $Z_{2D}$ for an $n \times m$ array of resonators satisfies the relation
\begin{equation}
Z_{2D} \mathrm{I} = \mathrm{V},
\label{eq:impedance_matrix}
\end{equation}
where $\mathrm{I}$ and $\mathrm{V}$ are the current and voltage vectors. As demonstrated by Spring et al., this matrix can be decomposed using a Kronecker sum:
\begin{equation}
Z_{2D} = Z_{1D}^{(n)} \oplus Z_{1D}^{(m)} - Z_0 I_{nm \times nm},
\label{eq:matrix_decomposition}
\end{equation}
where $\oplus$ denotes the Kronecker sum operation, $Z_0 = i\omega L_0 - i/(\omega C_0)$ is the impedance of each uncoupled resonator, and $Z_{1D}^{(n)}$ is the impedance matrix of a one-dimensional chain of $n$ coupled resonators:
\begin{equation}
Z_{1D}^{(n)} = \begin{bmatrix}
Z_1 & -Z_g & 0 & \cdots & 0 \\
-Z_g & Z_2 & -Z_g & \cdots & 0 \\
\vdots & \ddots & \ddots & \ddots & \vdots \\
0 & \cdots & -Z_g & Z_2 & -Z_g \\
0 & \cdots & 0 & -Z_g & Z_1
\end{bmatrix}.
\label{eq:z1d}
\end{equation}
Here $Z_g = i\omega L_g$ is the coupling impedance between adjacent resonators (determined by the mutual inductance of neighboring pillars), $Z_1 = Z_0 + Z_g + Z_b$ accounts for boundary effects, and $Z_2 = Z_0 + 2Z_g$ describes resonators in the bulk of the array.

The eigenvalue structure of this 2D system reveals how the discrete resonator array approximates the continuum electromagnetic medium. The eigenvalues of $Z_{2D}$ are related to those of the 1D chains by
\begin{equation}
\lambda(Z_{2D})_{ij} = \lambda(Z_{1D}^{(n)})_i + \lambda(Z_{1D}^{(m)})_j - Z_0.
\label{eq:eigenvalue_relation}
\end{equation}
Resonant mode frequencies occur when $\lambda(Z_{2D})_{ij} = 0$. For open boundary conditions ($L_b = 0$), the eigenvalues of the 1D chain are
\begin{equation}
\lambda(Z_{1D}^{(n)})_i = Z_0 + 2Z_g \cos\mleft(\frac{i\pi}{n}\mright),
\label{eq:1d_eigenvalues}
\end{equation}
which leads to the mode-frequency condition
\begin{equation}
\omega^2 = \frac{1}{L_0 C_0\mleft[1 + 2\beta\mleft(\cos\mleft(\frac{i\pi}{n}\mright) + \cos\mleft(\frac{j\pi}{m}\mright)\mright)\mright]},
\label{eq:omega_squared_beta}
\end{equation}
where $\beta = L_g/L_0$ is the dimensionless coupling parameter. This yields the discrete mode frequencies
\begin{equation}
\omega_{ij} = \omega_0 \frac{1}{\sqrt{1 + 2\beta\mleft(\cos\mleft(\frac{i\pi}{n}\mright) + \cos\mleft(\frac{j\pi}{m}\mright)\mright)}},
\label{eq:mode_frequencies}
\end{equation}
where $\omega_0 = 1/\sqrt{L_0 C_0}$ is the natural frequency of an isolated resonator.

\subsection{Continuum limit and dispersion relation}
In the continuum limit, where the number of resonators becomes large ($n, m \to \infty$) and the lattice constant $a$ remains finite, we can replace the discrete mode indices with continuous wavevectors. Substituting $i\pi/n \to k_x a$ and $j\pi/m \to k_y a$, the discrete mode frequencies become a continuous dispersion relation:
\begin{equation}
\omega(k_x, k_y) = \omega_0 \frac{1}{\sqrt{1 + 2\beta(\cos(k_x a) + \cos(k_y a))}}.
\label{eq:dispersion_relation}
\end{equation}
Near the band minimum (small wavevector limit), we can expand the cosines to quadratic order:
\begin{equation}
\cos(k_x a) + \cos(k_y a) \approx 2 - \frac{1}{2}a^2(k_x^2 + k_y^2).
\label{eq:small_k_expansion}
\end{equation}
This gives the quadratic dispersion relation near the cutoff frequency $\omega_c = \omega_0/\sqrt{1 + 4\beta}$:
\begin{equation}
\omega(k) \approx \omega_c \mleft(1 + \frac{1}{2}\frac{k^2}{k_0^2}\mright),
\label{eq:quadratic_dispersion}
\end{equation}
where $k_0^2 = 1/(\beta a^2)$ sets the characteristic wavevector scale, and the effective wave velocity is $v^2 = \frac{a^2}{4\beta}(1 + 4\beta)$. For the TM-like band edge of the shunted enclosure we identify $\omega_c \equiv \omega_p$ and subsequently use $\omega_c$ to denote the cutoff. Comparing this dispersion relation with the continuum description in Eq.~\eqref{eq:evanescent_wave_number}, we see that the circuit model directly yields the plasma-like behavior with the same effective velocity $v$.

\subsection{Coupling through transimpedance}
Having established the mode structure, we now turn to calculating the coupling between qubits. The coupling rate between two qubits $i$ and $j$ can be expressed through network theory in terms of the transimpedance $Z_{ij}(\omega)$ between their circuit nodes:
\begin{equation}
\label{eq:Jij_impedance}
J_{ij}
= -\,\frac{1}{4}\sqrt{\frac{\omega_i\,\omega_j}{L_i L_j}} 
\mathrm{Im}\mleft[\frac{Z_{ij}(\omega_i)}{\omega_i}+\frac{Z_{ij}(\omega_j)}{\omega_j}\mright],
\end{equation}
where $\omega_i$ and $\omega_j$ are the qubit frequencies, and $L_i$ and $L_j$ are their effective inductances. The transimpedance itself can be expressed as an overlap integral involving the electromagnetic Green's function and the current distributions $\mathrm{J}_i(\mathrm{r})$ and $\mathrm{J}_j(\mathrm{r})$ associated with each qubit:
\begin{equation}
Z_{ij}(\omega) = i\omega \mu_0 \iint \mathrm{J}_i(\mathrm{r}) \cdot \mathrm{G}(\mathrm{r}, \mathrm{r'}; \omega) \cdot \mathrm{J}_j(\mathrm{r'}) \, d^3\mathrm{r} \, d^3\mathrm{r'}.
\label{eq:Zij}
\end{equation}
By reciprocity in the passive enclosure, $Z_{ij} = Z_{ji}$. This formula shows that the coupling is determined by how the current distributions of the two qubits overlap through the intermediary of the electromagnetic Green's function-essentially, how well the field from one qubit reaches the other.

For a cavity or waveguide system with discrete modes, the Green's function can be expanded using the eigenmode functions $\mathrm{f}_{\omega'}(\mathrm{r})$ and the density of states $\rho(\omega)$:
\begin{equation}
G(\mathrm{r}, \mathrm{r'}; \omega) = \int_{\omega_c}^{\infty} \frac{\rho(\omega')\mathrm{f}_{\omega'}(\mathrm{r})\mathrm{f}_{\omega'}^*(\mathrm{r'})}{(\omega')^2 - \omega^2} \, d\omega'.
\label{eq:green_omega2}
\end{equation}
For large-area devices, the sum over discrete modes is well-approximated by this continuum integral near the band edge. In our two-dimensional geometry with dispersion relation $\omega \approx \omega_c + \frac{v^2 k^2}{2\omega_c}$ near cutoff [from Eq.~\eqref{eq:quadratic_dispersion}], the density of states is
\begin{equation}
\rho(\omega) = \frac{A \omega_c}{2\pi v^2},
\label{eq:dos_final}
\end{equation}
where $A$ is the in-plane area of the device.

\subsection{Resolvent representation and the modified Bessel function}
When a qubit with frequency $\omega_q < \omega_c$ (below cutoff) is introduced into this system, it couples to the continuum of cavity modes above the cutoff frequency. To evaluate this coupling, it is convenient to work with the momentum-space resolvent. Using the dispersion relation $\omega_{\mathrm{k}}^2 - \omega^2 \approx v^2 (k^2 + \kappa^2)$ near the band edge, we can write
\begin{equation}
G(\mathrm{r},\mathrm{r}';\omega) = \int \frac{d^2\mathrm{k}}{(2\pi)^2} \,
\frac{e^{i \mathrm{k} \cdot (\mathrm{r} - \mathrm{r}')}}{k^2 + \kappa^2(\omega)},
\label{eq:resolvent_representation}
\end{equation}
where $\kappa^2 = (\omega_c^2 - \omega^2)/v^2$ is the screening parameter from Eq.~\eqref{eq:evanescent_wave_number}, and $v$ is the effective wave velocity derived from the dispersion relation. The denominator $k^2 + \kappa^2$ shows that the poles of the Green's function lie off the real axis, corresponding to evanescent rather than propagating modes.

Performing the angular integration in polar coordinates yields
\begin{equation}
G(R;\omega) = \frac{1}{2\pi} \int_0^\infty \frac{k \, J_0(kR)}{k^2 + \kappa^2} \, dk
= \frac{1}{2\pi} K_0(\kappa R),
\label{eq:G_after_angular}
\end{equation}
where we have used a standard Bessel-function identity.
This confirms the modified Bessel-function form obtained earlier in Eq.~\eqref{eq:2d_greens_solution}, now derived from the momentum-space formulation, up to overall normalization absorbed in $Z_{ij}$.

For completeness, we note that at frequencies above cutoff ($\omega > \omega_c$), we substitute $\kappa \to iq$, where $q(\omega) = \sqrt{\omega^2 - \omega_c^2}/v$ is now real. Using the identity $K_0(iq R) = \frac{\pi i}{2} H_0^{(1)}(q R)$, where $H_0^{(1)}$ is the Hankel function of the first kind, we obtain
\begin{equation}
G(R; \omega \gtrsim \omega_c) = \frac{i}{4} H_0^{(1)}\mleft( q(\omega) R \mright),
\label{eq:above_cutoff_hankel}
\end{equation}
which exhibits oscillatory behavior $|G| \sim 1/\sqrt{R} \times e^{iqR}$ characteristic of propagating waves. This confirms that operating below the cutoff frequency is essential for achieving exponential spatial localization of electromagnetic fields.

\subsection{Bound-state length and spatial decay}
Returning to the below-cutoff regime relevant to our device, the Green's function also satisfies the 2D Helmholtz equation
\begin{equation}
\mleft[\nabla^2 - \kappa^2\mright]G(\mathrm{r}, \mathrm{r}', \omega) = -\delta(\mathrm{r} - \mathrm{r}'),
\label{eq:bound_state_greens}
\end{equation}
whose solution is
\begin{equation}
G(\mathrm{r}, \mathrm{r}', \omega) = \frac{1}{2\pi}K_0(\kappa|\mathrm{r} - \mathrm{r}'|).
\label{eq:2d_bound_state_solution}
\end{equation}
The characteristic length scale for spatial decay is the bound-state length, defined as the inverse of the screening parameter:
\begin{equation}
\delta_b = \frac{1}{\kappa} = \frac{v}{\sqrt{\omega_c^2 - \omega_q^2}} \approx \frac{v}{\sqrt{2\omega_c(\omega_c - \omega_q)}}.
\label{eq:bound_state_length}
\end{equation}
Qubits separated by distances much larger than $\delta_b$ experience exponentially suppressed enclosure-mediated coupling. From the circuit model parameters derived earlier, this bound-state length becomes
\begin{equation}
\delta_b = a\sqrt{\frac{1 + 4\beta}{4\beta}} \cdot \frac{1}{\sqrt{2(\omega_c - \omega_q)/\omega_c}},
\label{eq:bound_state_length_circuit}
\end{equation}
which shows explicitly how the pillar spacing $a$ and coupling parameter $\beta$ control the spatial reach of evanescent electromagnetic fields. This formula provides direct design guidance: smaller pillar spacing $a$ or larger coupling parameter $\beta$ (achieved through stronger mutual inductance between pillars) reduces $\delta_b$ and tightens the spatial localization.

The asymptotic behavior of the modified Bessel function for large argument,
\begin{equation}
K_0(\kappa R) \sim \sqrt{\frac{\pi}{2\kappa R}} \, e^{-\kappa R}
\quad (\kappa R \gg 1),
\label{eq:K0_asympt}
\end{equation}
makes the exponential spatial decay explicit. Finite losses ($\kappa \to \kappa + i\eta$) add further suppression and do not alter the $K_0$ envelope. Any transimpedance built from the overlap integral
\begin{equation}
Z_{ij}(\omega) \propto i \omega \iint \mathrm{J}_i \cdot G(\omega) \cdot \mathrm{J}_j
\label{eq:transimpedance_overlap}
\end{equation}
inherits this $e^{-\kappa R}$ scaling at large distances.

\subsection{Coupling for localized qubit distributions}
For compact qubit structures, where the current distributions $\mathrm{J}_i$ and $\mathrm{J}_j$ are well-localized compared to the qubit separation $d_{ij}$, we can approximate the current overlap by localized weights. In this limit, the transimpedance simplifies to
\begin{equation}
Z_{ij}(\omega) \propto i \, \omega \, G(d_{ij};\omega) 
= \frac{i\omega}{2\pi} \, K_0\mleft[ \kappa(\omega) \, d_{ij} \mright],
\label{eq:Zij_localized}
\end{equation}
where geometric form factors have been absorbed into an overall prefactor. Using the coupling formula from Eq.~\eqref{eq:Jij_impedance} and noting that $K_0$ is real and positive for real $\kappa$, while $Z_{ij} \propto i\omega \times (\text{real})$, we obtain
\begin{equation}
J_{ij} = J_0 \mleft[ \omega_i^2 K_0\mleft( \kappa(\omega_i) d_{ij} \mright) 
+ \omega_j^2 K_0\mleft( \kappa(\omega_j) d_{ij} \mright) \mright],
\label{eq:Jij_K0}
\end{equation}
where the prefactor is
\begin{equation}
J_0 = \frac{1}{8\pi} \sqrt{\frac{C_i C_j}{L_i L_j}} \times \mleft( \mu_0 \mathcal{G}_{ij} \mright),
\label{eq:J0_definition}
\end{equation}
with $\mathcal{G}_{ij}$ encoding geometric overlap factors specific to the qubit geometries.

\subsection{Frequency-dependent coupling}
We now address the important case of two qubits with different frequencies $\omega_i \neq \omega_j$. For large qubit separations where $d_{ij} \gg \delta_b$, we can use the asymptotic form of $K_0$ from Eq.~\eqref{eq:K0_asympt}:
\begin{equation}
\label{eq:Jij_large_distance}
J_{ij} \approx \tilde J_0\mleft[
\frac{e^{-\kappa(\omega_i) d_{ij}}}{\sqrt{\kappa(\omega_i) d_{ij}}}
+\frac{e^{-\kappa(\omega_j) d_{ij}}}{\sqrt{\kappa(\omega_j) d_{ij}}}
\mright],
\end{equation}
where $\tilde J_0$ absorbs geometry-dependent prefactors. To analyze how the coupling depends on the frequency difference, we define the average frequency $\bar\omega=(\omega_i+\omega_j)/2$ and frequency difference $\Delta\omega = \omega_i - \omega_j$. A first-order Taylor expansion gives
\begin{equation}
\kappa(\omega_{i/j})\approx\bar\kappa\ \pm\ \frac{\kappa'(\bar\omega)}{2},\Delta\omega,
\qquad
\bar\kappa\equiv \kappa(\bar\omega),\quad
\kappa'(\omega)= -\frac{\omega}{v\sqrt{\omega_c^2-\omega^2}}.
\label{eq:kappa_taylor}
\end{equation}
Inserting Eq.~\eqref{eq:kappa_taylor} into Eq.~\eqref{eq:Jij_large_distance} and treating the slowly varying prefactors $1/\sqrt{\kappa d}$ as approximately constant yields 
\begin{equation}
\label{eq:Jij_cosh_dependence}
J_{ij} \propto e^{-\bar\kappa d_{ij}}
\mleft[e^{-\kappa'(\bar\omega)(\Delta\omega/2)\,d_{ij}}
+e^{+\kappa'(\bar\omega)(\Delta\omega/2)\,d_{ij}}\mright]
=\,2\,e^{-\bar\kappa d_{ij}}\,
\cosh\mleft[-\,\kappa'(\bar\omega)\,\frac{\Delta\omega}{2}\,d_{ij}\mright],
\end{equation}
which shows that the dependence on the detuning enters through a hyperbolic cosine function.

For two qubits with different frequencies, the complete analysis of the impedance integral (accounting for both the spatial evanescence and the frequency-dependent denominators in the mode expansion) reveals that the transimpedance has the form
\begin{equation}
\label{eq:transimpedance_2D}
Z_{ij}(\omega) \propto i\omega\, K_0 \mleft(\kappa(\omega)\,d_{ij}\mright),
\qquad
\kappa(\omega) = \frac{\sqrt{\omega_c^2 - \omega^2}}{v}.
\end{equation}
For large separations ($\kappa d_{ij} \gg 1$),
\begin{equation}
\label{eq:transimpedance_asymptotic}
Z_{ij}(\omega) \sim i\omega\,\frac{e^{-\kappa(\omega) d_{ij}}}{\sqrt{\kappa(\omega) d_{ij}}}.
\end{equation}
For two frequencies $\omega_i$ and $\omega_j$ with average $\bar\omega=(\omega_i+\omega_j)/2$, expanding $\kappa(\omega_{i,j})$ around $\bar\omega$ and combining the spatial 
decay terms yields
\begin{equation}
\label{eq:enclosure_coupling}
J_{ij} \propto \frac{e^{-\bar\kappa\,d_{ij}}}{\sqrt{\bar\kappa\,d_{ij}}}
 \cosh\mleft[\frac{1}{2}\,|\kappa'(\bar\omega)|\,|\omega_i - \omega_j|\,d_{ij}\mright],
\end{equation}

For the device parameters relevant to this work---qubit frequencies $\omega_q/2\pi \sim \qty{5}{\giga\hertz}$, cavity cutoff $\omega_c/2\pi \sim \qty{33}{\giga\hertz}$, and typical separations $d_{ij} \sim \qty{2}{\milli\metre}$---the argument of the hyperbolic cosine is small. Consequently, frequency-dependent variations in the evanescent decay rate $\kappa(\omega)$ contribute negligibly to the electromagnetic coupling strength in this device architecture. 

\subsection{Unified coupling formula}
Collecting these results, we can write a compact expression that captures both the spatial and spectral suppression:
\begin{equation}
\label{eq:complete_coupling_green}
J_{ij}^{\mathrm{enclosure}} = J_0\,K_0 \mleft(\bar\kappa\,d_{ij}\mright)
 \times \cosh\mleft[\frac{|\kappa'(\bar\omega)|\,|\omega_i-\omega_j|\,d_{ij}}{2}\mright].
\end{equation}
For large separations where $d_{ij}\gg\delta_{b,\mathrm{eff}}$, using the asymptotic form $K_0(x)\approx \sqrt{\pi/(2x)}\,e^{-x}$, this becomes
\begin{equation}
\label{eq:asymptotic_enclosure}
J_{ij}^{\mathrm{enclosure}}  \propto  \frac{e^{-\bar\kappa d_{ij}}}{\sqrt{\bar\kappa d_{ij}}}
= \frac{e^{-d_{ij}/\delta_b}}{\sqrt{d_{ij}/\delta_b}}.
\end{equation}
The spatial decay is exponential with characteristic length $\delta_b$, modulated by a weak $1/\sqrt{d_{ij}}$ power law arising from the two-dimensional geometry of the enclosure. In our device, both circuit-mediated coupling (through the capacitive/inductive network, 
with exponential decay of $C^{-1}$ entries established in Appendix~\ref{matrixinversion} and virtual-process suppression from Appendix~\ref{nodalanalysis}) and enclosure-mediated coupling (through evanescent electromagnetic fields analyzed here) operate simultaneously. The total coupling is
\begin{equation}
J_{ij}^{\mathrm{total}} = J_{ij}^{\mathrm{circuit}} + J_{ij}^{\mathrm{enclosure}}.
\end{equation}

Both mechanisms predict exponential spatial decay through distinct physical pathways. The circuit-mediated contribution exhibits frequency-dependent suppression through virtual processes when intermediate modes are off-resonant (Appendix~\ref{nodalanalysis}), while the enclosure-mediated contribution shows weak frequency dependence within the parameter regime of this device. 

\section{Spectator effects on ZZ couplings for nearest-neighbor transmons}
\label{app:spectator}
In this appendix, we estimate the error of neglecting spectator effects in Eq.~\eqref{eq:ZZ_12}, which gives the ZZ coupling $\zeta_{12}^\mathrm{spectator}$ for a pair of nearest-neighbor transmons. Here, spectator effects refer to the fact that the pair of transmons are capacitively coupled to additional transmons. We suppose that the additional transmons are in their ground state. We label the transmons in the pair with 1 and 2, and the spectator qubit with 3. 
Then, by relabeling $2 \leftrightarrow 3$ in Eq.~\eqref{eq:ZZ_13}, the ZZ coupling $\zeta_{12}^\mathrm{spectator}$ including the spectator effect of transmon 3 directly follows:
\begin{equation}
\begin{aligned}
    \zeta_{12}^\mathrm{spectator} =~ &\frac{2 (\alpha_1+\alpha_2)}{(\Delta_{12} + \alpha_1)(\Delta_{12} - \alpha_2)} 
    \mleft( J'_{12} - \Lambda' \frac{\alpha_1 + \alpha_2}{2} \mright)^2 +
    \frac{4 (\alpha_1 - \alpha_2)}{(\Delta_{12} + \alpha_1)(\Delta_{12} - \alpha_2)} 
    \mleft( \Delta_{12} + \frac{\alpha_1 + \alpha_2}{2} \mright) \Lambda' J'_{12}
    + \\ &\mleft[ 2\frac{ \Delta_{12}^2 - \mleft[ \mleft( \alpha_1 + \alpha_2 \mright) / 2 \mright]^2 }{(\Delta_{12} + \alpha_1)(\Delta_{12} - \alpha_2)}  
    (\alpha_1+\alpha_2) + 8\frac{ (\Delta_{13} + \Delta_{23}) }{\Delta_{13} + \Delta_{23} - \alpha_3} \alpha_3 \mright] \Lambda'^2 + \mathcal{O}(J^4/\Delta^3),
\end{aligned}
\label{eq:ZZ_12_with_spectator}
\end{equation}
where $J'_{12} = J_{12} + \frac{1}{2}J_{13}J_{23}\mleft( \frac{1}{\Delta_{13}} + \frac{1}{\Delta_{23}} \mright) = J_{12} + \Lambda' (\Delta_{13} + \Delta_{23})$, and $\Lambda' = J_{13} J_{23} / (2 \Delta_{13} \Delta_{23})$. We remark that $\mathcal{O}(J^4/\Delta^3)$ denotes terms proportional to, e.g., $J_{12}^4/\Delta_{12}^3$ and $J_{12}^2 J_{23}^2/(\Delta_{12} \Delta_{23}^2)$. Assuming $J_{12} \approx J_{23}$, these terms are negligible in comparison to the leading-order contribution $\zeta_{12}$ in Eq.~\eqref{eq:ZZ_12}.

To estimate the spectator error $\Delta \zeta_{12} = \zeta_{12} - \zeta_{12}^\mathrm{spectator}$, we draw attention to the facts that $|J_{13}| \ll |J_{12}|, |J_{23}| $ and $|J_{ij}| \ll |\Delta_{ij}|$ for all $i \neq j$; see Table~\ref{basic_device_parameters} in Appendix~\ref{app:device_parameters}. Thus $|\Lambda'| \ll 1$, which warrants an expansion of Eq.~\eqref{eq:ZZ_12_with_spectator} to first order in $\Lambda'$. We also note that $|\alpha_1 - \alpha_2| \ll |\alpha_1 + \alpha_2|$. With these observations, we find that the second and third terms in Eq.~\eqref{eq:ZZ_12_with_spectator} are negligible, and that the spectator error is
\begin{equation}
    \Delta \zeta_{12} = \frac{2 (\alpha_1+\alpha_2)}{(\Delta_{12} + \alpha_1)(\Delta_{12} - \alpha_2)} J_{12} \mleft[ 2(\Delta_{12} + \Delta_{23} ) - (\alpha_1 + \alpha_2) \mright] \Lambda'  + \mathcal{O} \mleft( \alpha\Lambda'^2, J_{12}^4/\Delta_{12}^3 \mright) .
    \label{eq:spectator_error}
\end{equation}
Further comparing the spectator error with $\zeta_{12}$ in Eq.~\eqref{eq:ZZ_12} reveals
\begin{equation}
    \mleft| \frac{\Delta \zeta_{12}}{\zeta_{12}} \mright| = \mleft| \frac{2(\Delta_{12} + \Delta_{23} ) - (\alpha_1 + \alpha_2)}{J_{12}} \Lambda' \mright| 
    \approx \mleft| \frac{\Delta_{12} + \Delta_{23}  - (\alpha_1 + \alpha_2) / 2}{\Delta_{23}} \mright| \mleft|\frac{J_{13}}{\Delta_{13}} \mright|,
    \label{eq:relative_spectator_error}
\end{equation}
where we in the last approximate equality have assumed $J_{12} \approx J_{23}$. Viewing Table~\ref{basic_device_parameters}, we note that $\Delta_{23}$ is on a similar scale as $\Delta_{12}$ and $(\alpha_1 + \alpha_2) / 2$, implying that the first factor in Eq.~\eqref{eq:dispersion_relation} is of scale $\sim 1$. Therefore, we conclude that
\begin{equation}
    \mleft| \frac{\Delta \zeta_{12}}{\zeta_{12}} \mright| \ll 1,
\end{equation}
and that the spectator effects are negligible for the nearest-neighbor ZZ couplings.

\section{Additional device parameters and measurements}
\label{app:device_parameters}

This appendix provides comprehensive device characterization supporting the crosstalk analysis in Section~\ref{sec:results}. We present three key validation results: (i) exponential spatial decay of ZZ couplings consistent with the bound-state length $\delta_b$ predicted in Appendix~\ref{greenfunction}, (ii) frequency-dependent suppression following dispersive scaling, demonstrating that naive extraction of $J$ from ZZ measurements overestimates long-range interactions, and (iii) frequency stability below \qty{1}{\kilo\hertz} across all qubits, confirming that slow drift does not limit our crosstalk characterization. Together with fabrication details and enclosure transmission measurements, these results establish the experimental foundation for the crosstalk model presented in the main text.

\subsection{Spatial and spectral scaling of ZZ interactions}
\begin{figure}[ht]
  \centering
   \includegraphics[width=0.86\textwidth]{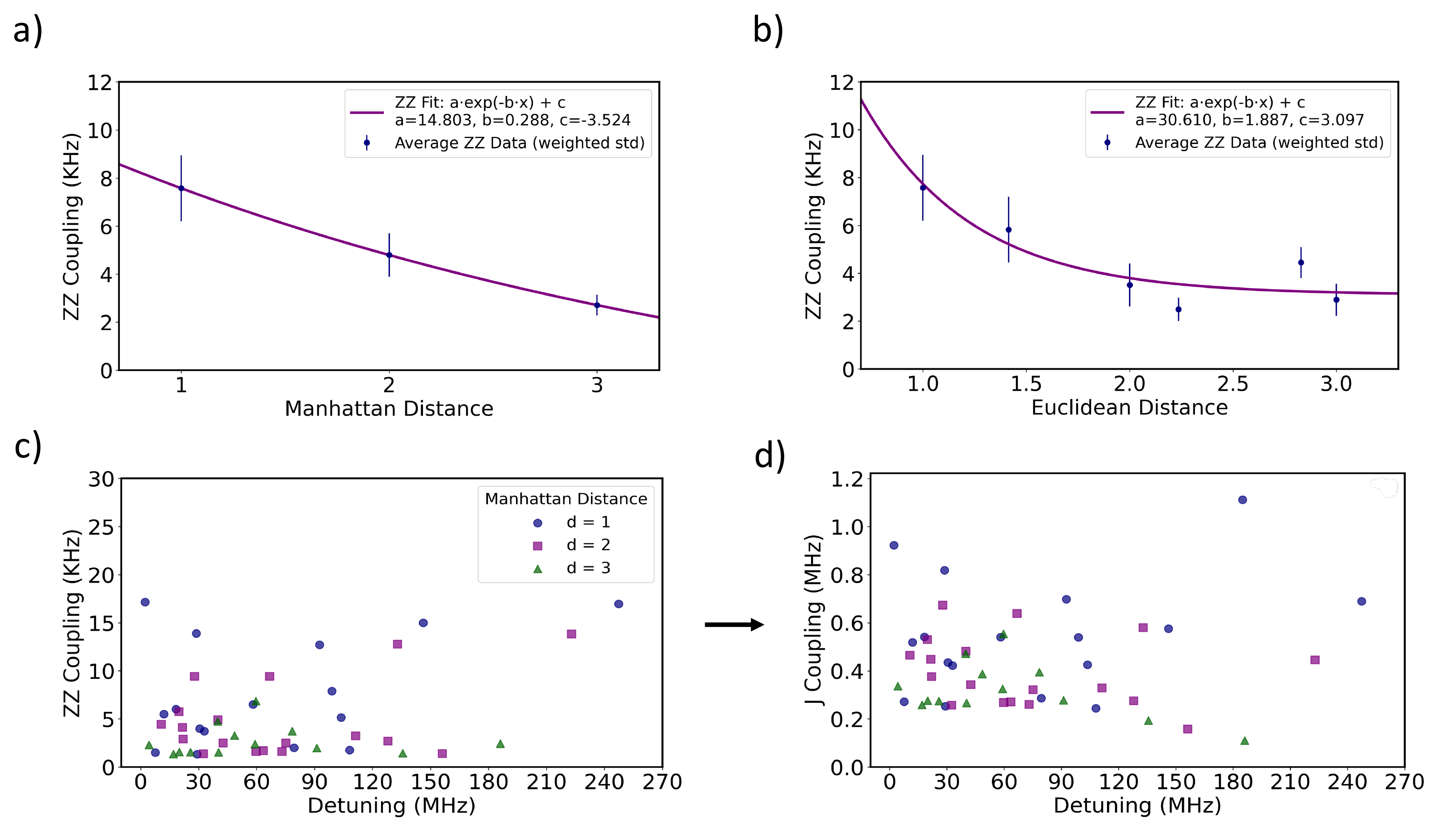}
      \caption{Spatial and spectral scaling of static ZZ interactions. (a) Detected ZZ couplings as a function of Manhattan distance $D$ across the $4\times4$ lattice. An increasing fraction of qubit pairs fall below the Ramsey sensitivity at larger separations, indicating strong suppression beyond nearest neighbours; averages at $D\ge2$ should therefore be interpreted as conditional on detection. (b) The same data replotted as a function of Euclidean distance, showing consistency with a short-range interaction length scale comparable to the enclosure bound-state prediction ($\delta_b \sim \qty{2}{\milli\metre}$), although the functional form of the decay is not uniquely determined by the detectable subset. (c) Measured ZZ couplings as a function of qubit–qubit detuning $|\omega_i-\omega_j|$ for representative pairs, illustrating the role of spectral separation in suppressing virtual exchange. (d) Effective exchange couplings $J$ extracted from the ZZ measurements using the standard dispersive relation. Neglecting spatial and spectral scaling leads to systematic overestimation of long-range couplings, while incorporating both dependencies yields a nearest-neighbour-dominated interaction picture.}
  \label{F3-Sup}
\end{figure}

Figure~\ref{F3-Sup} summarizes statistical trends in the measured static ZZ interactions across the $4\times4$ lattice. Figure~\figpanelNoPrefix{F3-Sup}{a} shows ZZ couplings as a function of Manhattan distance $D$ for all qubit pairs, where values indistinguishable from the independently characterized Ramsey frequency fluctuations (Appendix~\ref{ramseynoise}) are treated as non-detections. While the number of detectable couplings decreases rapidly with increasing separation, the remaining detected values indicate strong suppression beyond nearest neighbours. Because an increasing fraction of couplings at $D\ge2$ fall below the experimental sensitivity, averages at larger distances should be interpreted as conditional on detection and primarily provide upper bounds on typical ZZ magnitudes rather than defining a unique decay law. Figure~\figpanelNoPrefix{F3-Sup}{b} replots the same data as a function of Euclidean distance, showing that the detectable couplings are consistent with a short-range interaction length scale, in agreement with expectations from capacitance-matrix localization (Appendix~\ref{matrixinversion}) and evanescent enclosure-mediated fields (Appendix~\ref{greenfunction}). Given the limited dynamic range set by measurement resolution,
the data do not uniquely distinguish between functional forms of decay. 

Figures~\figpanelNoPrefix{F3-Sup}{c} and~\figpanelNoPrefix{F3-Sup}{d} address the role of qubit frequency detuning. Figure~\figpanelNoPrefix{F3-Sup}{c} shows measured ZZ couplings as a function of detuning $|\omega_i-\omega_j|$ for representative qubit pairs. Significant scatter is observed, and no simple monotonic dependence on detuning
is expected for ZZ interactions, which can become enhanced near the edges of the straddling regime. 
In Fig.~\figpanelNoPrefix{F3-Sup}{d}, the same data are converted to effective exchange couplings $J$ using the standard dispersive relation~\cite{solgun2022direct}. In this representation, an overall suppression of $J$ with increasing detuning is observed, consistent with virtual-exchange-mediated coupling. 
If spatial and spectral dependencies are neglected, the extracted $J$ values substantially overestimate long-range interactions. Only after incorporating both spatial separation $d_{ij}$ and detuning $|\omega_i-\omega_j|$ via Eqs.~\eqref{eq:J_ij} and~\eqref{eq:ZZ_ij} do the inferred couplings collapse onto a consistent, nearest-neighbour-dominated picture. This demonstrates that joint spatial and spectral scaling is essential for quantitatively accurate crosstalk modeling in multi-qubit architectures. A direct comparison between scaled and unscaled extraction of $J$ from measured ZZ shifts is shown in Fig.~\figpanelNoPrefix{F3}{b}, where neglecting spatial and spectral scaling is seen to systematically overestimate long-range couplings.

\subsection{Frequency-stability validation}
\label{ramseynoise}
\begin{figure}[H]
  \centering
   \includegraphics[width=.9\textwidth]{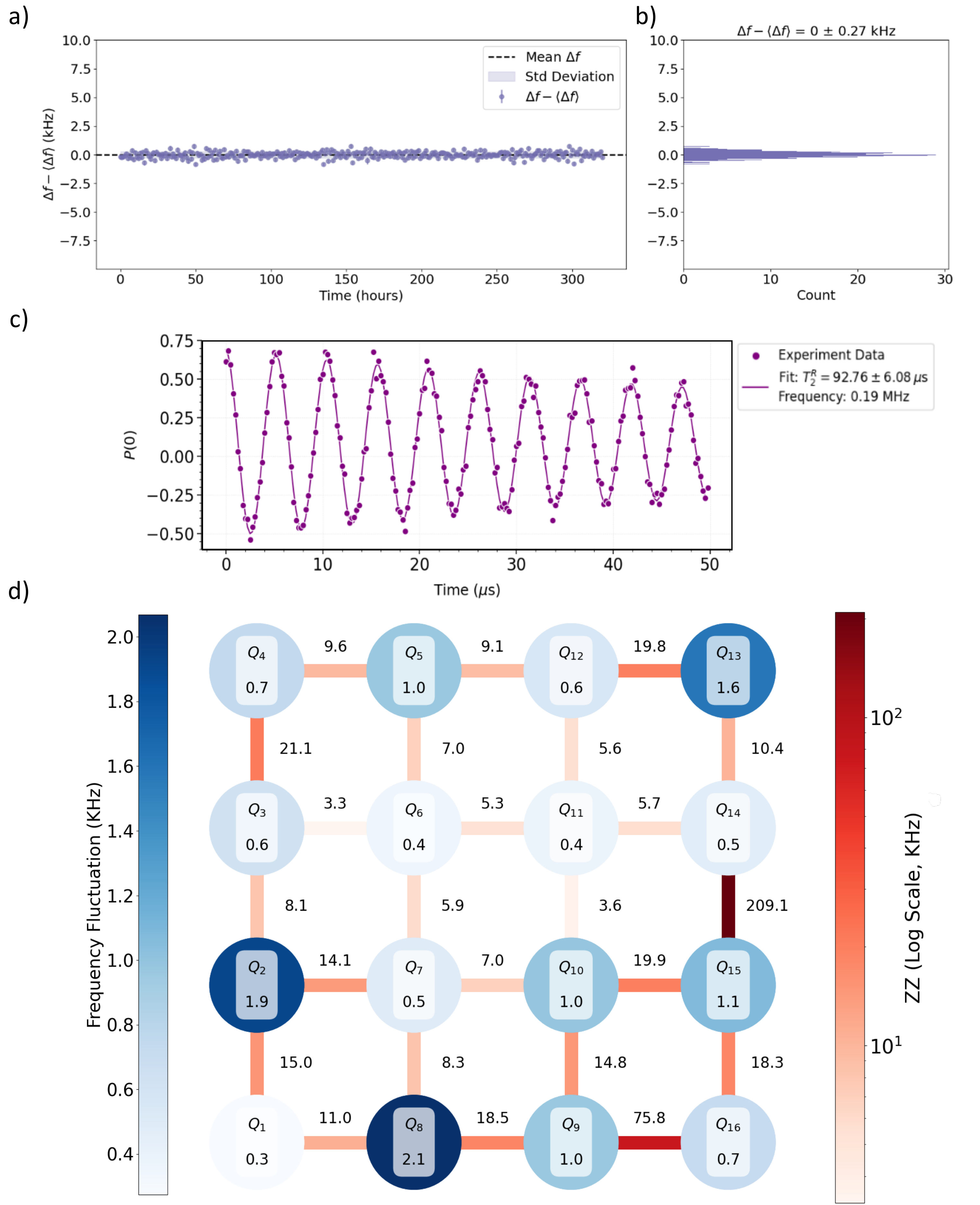}
      \caption{Frequency-stability characterization. (a) Continuous Ramsey monitoring of qubit $Q_1$ over 12~hours shows slow frequency fluctuations with $\qty{0.88}{\kilo\hertz}$ standard deviation. (b) Distribution of frequency fluctuations across the monitoring period, yielding $\Delta f - \langle \Delta f \rangle = \qty{0}{\kilo\hertz} \pm \qty{0.27}{\kilo\hertz}$. (c) Representative Ramsey data for $Q_{1}$ with sinusoidal fit used to extract the instantaneous qubit frequency. (d) Frequency fluctuation levels across the lattice (numbers inside circles) compared to nearest-neighbour ZZ rates (colored links). The average fluctuation of $\sim\qty{0.88}{\kilo\hertz}$ lies below the weakest measured ZZ coupling, validating the effective resolution of the Ramsey-based ZZ measurements.}
  \label{F4-Sup}
\end{figure}

Extracting static ZZ couplings from spectroscopy requires that qubit frequency fluctuations remain small compared to the measured interaction strengths. Figure~\ref{F4-Sup} demonstrates that our device satisfies this criterion. Figure~\figpanelNoPrefix{F4-Sup}{a} shows continuous Ramsey monitoring of $Q_1$ over approximately 12 hours ($\sim 400$ sequential measurements). The extracted frequency fluctuates slowly with a standard deviation of \qty{0.88}{\kilo\hertz}. Figure~\figpanelNoPrefix{F4-Sup}{b} extends this analysis across the full $4\times4$ lattice. Frequency fluctuation levels for each qubit appear as numbers inside the circles (lattice positions), while nearest-neighbor ZZ interaction strengths are shown as colored connecting lines. The device-wide average fluctuation is \qty{0.88}{\kilo\hertz}, which lies below even the weakest ZZ coupling included in our analysis ($\sim \qty{3}{\kilo\hertz}$ for the most detuned nearest-neighbor pair). This margin ensures that slow frequency drift introduces negligible systematic error in the extracted ZZ rates. The lack of spatial correlation between fluctuation levels and lattice position further confirms that environmental noise rather than crosstalk drives the observed frequency instability.

\subsection{Comprehensive qubit characterization}

Table~\ref{basic_device_parameters} provides full microwave characterization for all 16 qubits. Several trends emerge from the statistics. First, qubit frequencies span \qtyrange[range-phrase = --, range-units = single]{4.78}{5.04}{\giga\hertz}
and are arranged into two interleaved bands near \qty{4.8}{\giga\hertz} and \qty{4.9}{\giga\hertz}, confirming successful implementation of the two-band frequency allocation. This $\sim \qty{250}{\mega\hertz}$ frequency range provides sufficient detuning to suppress nearest-neighbor exchange couplings while maintaining addressability. Anharmonicities cluster tightly around $\alpha/2\pi = -196.4 \pm 1.1$~MHz, indicating highly uniform Josephson junction fabrication across the chip. The small \qty{0.6}{\percent} variation in $\alpha$ validates our assumption of identical qubit nonlinearity in the theoretical model.

Coherence times, however, show significant device-to-device variation. $T_1$ ranges from \qty{24}{\micro\second} to \qty{126}{\micro\second} (mean \qty{71}{\micro\second}), while $T_{2E}$ spans \qtyrange[range-phrase = --, range-units = single]{33}{124}{\micro\second} (mean \qty{78}{\micro\second}). This $\sim 5 \times$ spread is typical for fixed-frequency transmon arrays and arises primarily from variation in TLS defect density rather than crosstalk, as evidenced by the lack of spatial correlation between coherence and lattice position. Qubits with short $T_1$ do not cluster near high-connectivity regions, nor do they correlate with large ZZ rates. The readout resonator internal quality factors $Q_i$ vary by nearly an order of magnitude (3.1--$24.6 \times 10^4$), likely reflecting variations in surface participation ratio and material loss. Dispersive shifts $\chi$ range from 175 to \qty{250}{\kilo\hertz}, chosen to provide fast high-fidelity readout while maintaining $|\chi| \ll \alpha$ for the rotating-wave approximation. All parameters fall within the ranges expected for state-of-the-art fixed-frequency transmon devices.

\begin{table}[H]
    \centering
    \caption{Comprehensive microwave characterization for all 16 qubits. Frequencies $\omega_r/2\pi$ (readout) and $\omega_q/2\pi$ (qubit), internal quality factor $Q_i$, external coupling rate $\kappa_{\text{ext}}$, dispersive shift $\chi$, anharmonicity $\alpha$, and coherence times $T_1$, $T_{2R}$, $T_{2E}$ (averaged over 400 measurements). 
    }
    \resizebox{0.81\textwidth}{!}{
        \begin{tabular}{|c|c|c|c|c|c|c|c|c|c|}
            \hline
            Parameters & \( \omega_{r}/2\pi \) & \( \omega_{q}/2\pi \) & \( Q_{i} \) & \( \kappa_{\text{ext}} \) & \( \chi/2\pi \) & \( \alpha/2\pi \)  & \( \langle T_{1} \rangle \) & \( \langle T_{2R} \rangle \) & \( \langle T_{2E} \rangle \) \\
            \hline
            Qubits & MHz & MHz & \( 10^{4} \) & MHz & kHz  & MHz & \( \mu \)s & \( \mu \)s & \( \mu \)s \\
            \hline
            \( Q_{1} \)  & 9997.4 & 4888.2 & 11.7 & 2.6 & -200.0 & -196.6 & 126 ± 18 & 107 ± 12 & 124 ± 23 \\
            \( Q_{2} \)  & 9386.0 & 4795.6 & 11.8 & 1.3 & -225.0 & -197.2 & 89 ± 13  & 56 ± 15  & 86 ± 12 \\
            \( Q_{3} \)  & 9299.2 & 4807.5 & 6.5  & 1.8 & -200.0 & -196.2 & 61 ± 6   & 44 ± 5   & 102 ± 18 \\
            \( Q_{4} \)  & 8649.5 & 4809.8 & 5.3  & 2.8 & -200.0 & -198.6 & 54 ± 9   & 39 ± 14  & 97 ± 20 \\
            \( Q_{5} \)  & 8755.6 & 4855.3 & 3.1  & 0.8 & -225.0 & -196.4 & 68 ± 10  & 38 ± 4   & 45 ± 10 \\
            \( Q_{6} \)  & 9220.6 & 4824.8 & 6.4  & 0.5 & -225.0 & -194.0 & 63 ± 7   & 49 ± 4   & 68 ± 10 \\
            \( Q_{7} \)  & 9474.2 & 4928.5 & 10.8 & 1.7 & -175.0 & -195.6 & 77 ± 12  & 45 ± 12  & 82 ± 15 \\
            \( Q_{8} \)  & 9908.6 & 4829.5 & 5.8  & 2.0 & -175.0 & -197.2 & 63 ± 8   & 32 ± 7   & 71 ± 8 \\
            \( Q_{9} \)  & 9802.3 & 4963.4 & 7.8  & 4.4 & -250.0 & -195.0 & 24 ± 7   & 24 ± 5   & 33 ± 9 \\
            \( Q_{10} \) & 9535.4 & 4817.2 & 7.3  & 3.2 & -200.0 & -196.9 & 51 ± 10  & 35 ± 9   & 63 ± 17 \\
            \( Q_{11} \) & 9112.8 & 4835.4 & 16.5 & 1.4 & -175.0 & -196.1 & 74 ± 7   & 49 ± 3   & 78 ± 13 \\
            \( Q_{12} \) & 8851.8 & 4777.3 & 11.4 & 0.9 & -250.0 & -196.4 & 92 ± 24  & 57 ± 11  & 76 ± 15 \\
            \( Q_{13} \) & 8943.2 & 4884.0 & 16.7 & 1.6 & -175.0 & -195.3 & 102 ± 13 & 63 ± 7   & 107 ± 22 \\
            \( Q_{14} \) & 9025.3 & 4855.2 & 9.3  & 1.4 & -175.0 & -197.0 & 56 ± 12  & 56 ± 10  & 60 ± 15 \\
            \( Q_{15} \) & 9645.5 & 5040.2 & 7.3  & 1.7 & -225.0 & -196.1 & 55 ± 7   & 58 ± 9   & 65 ± 9 \\
            \( Q_{16} \) & 9728.7 & 4792.8 & 24.6 & 2.7 & -175.0 & -197.5 & 60 ± 6   & 46 ± 4   & 65 ± 9 \\
            \hline
            \multicolumn{10}{|c|}{\textbf{Summary Statistics}} \\
            \hline
            Max & 9997.4 & 5040.2 & 24.6 & 4.4 & -175.0 & -194.0 & 126 & 107 & 124 \\
            Min & 8649.5 & 4777.3 & 3.1  & 0.5 & -250.0 & -198.6 & 24 & 24 & 33 \\
            Mean (\(\mu\)) & 9333.5 & 4856.5 & 10.1 & 1.9 & -203.1 & -196.4 & 71 & 51 & 78 \\
            Std.\ Dev.\ (\(\sigma\))   & 407.1 & 73.0 & 5.3 & 1.0  & 26.3 & 1.1 & 21 & 17 & 20 \\
            \( \sigma/\sqrt{N} \) (\( N=16 \)) & 101.8 & 18.3 & 1.3 & 0.2 & 6.6 & 0.3 & 5 & 4 & 5 \\
            \hline
        \end{tabular}
    }
    \label{basic_device_parameters}
\end{table}

\subsection{Effect of scaling corrections on nearest-neighbor couplings}

Figure~\ref{F3} summarizes the comparison of nearest-neighbor exchange couplings in the $4\times4$ lattice obtained from two different treatments of the same experimental data by taking the first term in the analytical expression described in Eq.~\eqref{eq:ZZ_12}, and by using the scaling model in Eq.~\eqref{eq:ZZ_ij} incorporating exponential scaling as a function of Euclidean distance. For each of the 24 nearest-neighbor pairs ($D=1$), the $J$ values extracted from measured ZZ shifts are compared against the corresponding values obtained after applying the distance--frequency scaling model. The corrections for $D=1$ pairs are effectively zero, as expected, confirming that the scaling function correctly preserves nearest-neighbor couplings. The large base coupling strengths ($J/2\pi\sim \qtyrange{0.4}{1.1}{\mega\hertz}$) and small relative detunings ($|\omega_i - \omega_j|/2\pi\sim \qtyrange{30}{100}{\mega\hertz}$, well below the characteristic suppression scale $\Delta_0/2\pi\sim \qty{200}{\mega\hertz}$) ensure that the exponential frequency term remains perturbative.

Table~\ref{couplings_Dsqrt2} extends this comparison to the first non-nearest pairs ($D=\sqrt{2}$), corresponding to diagonal couplings across unit cells. In this regime, the scaling corrections become significant: exchange couplings extracted from the measured ZZ shifts span \qtyrange{5}{316}{\kilo\hertz}, while the corresponding values obtained after applying the distance-scaling model decrease to \qtyrange{2}{97}{\kilo\hertz}, yielding an average reduction of roughly \qty{70}{\percent}. The mean difference (\qty{136}{\kilo\hertz}) indicates that diagonal interactions are strongly suppressed by the exponential distance factor, consistent with the measured spatial decay trend in \figpanel{F3}{b}. This validates that the unified scaling model effectively attenuates long-range and diagonal interactions while maintaining the designed nearest-neighbor network connectivity.

\begin{table}[H]
    \centering
    \caption{Pairs at first next-nearest-neighbors (\(D = 1.414\)) comparing measured and distance-scaled calculated couplings using the exponential suppression model in Eq.~\eqref{eq:ZZ_ij}.}
    \resizebox{0.6\textwidth}{!}{
        \begin{tabular}{|c|c|c|c|}
            \hline
            Pair            & Unscaled $J/2\pi$ & Scaled \(J/2\pi\) & Difference \\
            \hline
                            & MHz            & MHz              & MHz        \\
            \hline
            $Q_{1}$--$Q_{6}$   & 0.0853 & 0.0261 & 0.0592 \\
            $Q_{2}$--$Q_{5}$   & 0.2298 & 0.0704 & 0.1594 \\
            $Q_{2}$--$Q_{7}$   & 0.2107 & 0.0645 & 0.1462 \\
            $Q_{3}$--$Q_{6}$   & 0.2744 & 0.0840 & 0.1904 \\
            $Q_{3}$--$Q_{8}$   & 0.3160 & 0.0968 & 0.2192 \\
            $Q_{4}$--$Q_{7}$   & 0.2837 & 0.0869 & 0.1968 \\
            $Q_{5}$--$Q_{10}$  & 0.2974 & 0.0911 & 0.2063 \\
            $Q_{6}$--$Q_{9}$   & 0.1320 & 0.0404 & 0.0916 \\
            $Q_{6}$--$Q_{11}$  & 0.1305 & 0.0399 & 0.0905 \\
            $Q_{7}$--$Q_{10}$  & 0.1828 & 0.0560 & 0.1268 \\
            $Q_{7}$--$Q_{12}$  & 0.0801 & 0.0245 & 0.0556 \\
            $Q_{8}$--$Q_{11}$  & 0.2399 & 0.0735 & 0.1664 \\
            $Q_{9}$--$Q_{14}$  & 0.2024 & 0.0620 & 0.1404 \\
            $Q_{10}$--$Q_{13}$ & 0.3066 & 0.0939 & 0.2127 \\
            $Q_{10}$--$Q_{15}$ & 0.1127 & 0.0345 & 0.0782 \\
            $Q_{11}$--$Q_{14}$ & 0.0055 & 0.0017 & 0.0038 \\
            $Q_{11}$--$Q_{16}$ & 0.1669 & 0.0511 & 0.1158 \\
            $Q_{12}$--$Q_{15}$ & 0.2619 & 0.0802 & 0.1817 \\
            \hline
            \multicolumn{4}{|c|}{\textbf{Summary Statistics}} \\
            \hline
            Mean (\(\mu\))      & 0.1955 & 0.0599 & 0.1356 \\
            Std.\ Dev.\ (\(\sigma\)) & 0.0872 & 0.0267 & 0.0605 \\
            Min                 & 0.0055 & 0.0017 & 0.0038 \\
            Max                 & 0.3160 & 0.0968 & 0.2192 \\
            \hline
        \end{tabular}
    }
    \label{couplings_Dsqrt2}
\end{table}

\subsection{Device fabrication and pillar architecture}

\begin{figure}[ht]
  \centering
   \includegraphics[width=1.0\textwidth]{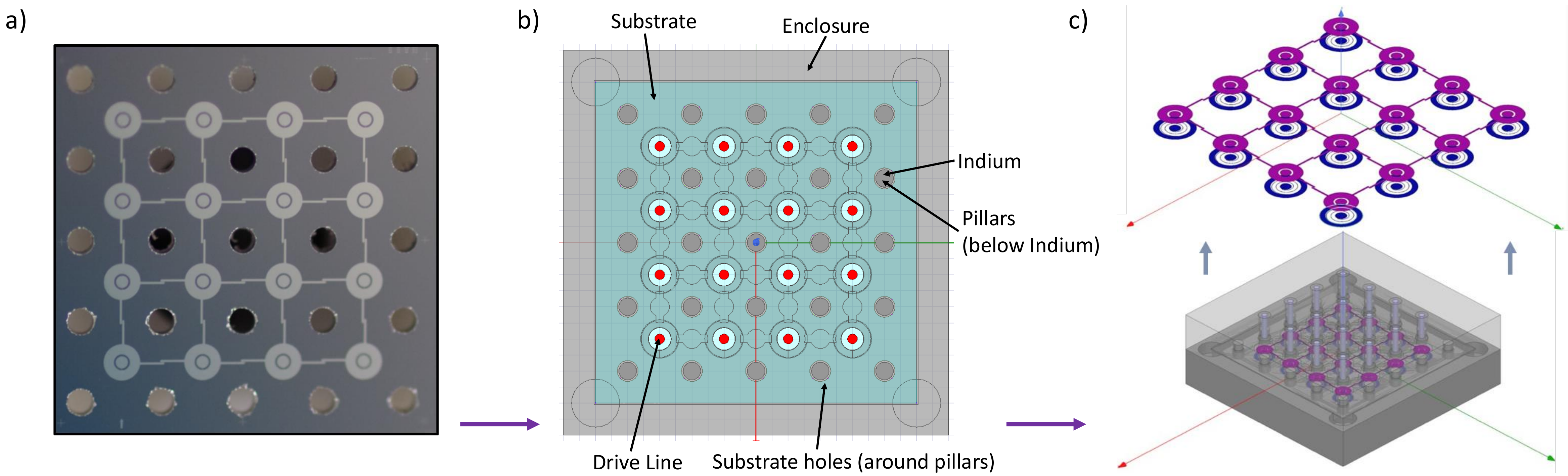}
\caption{Fabricated device and inductive shunt architecture. (a) Optical image showing 16 transmon qubits with through-silicon-via readout resonators. (b) Pillar-array schematic: aluminum posts connect lid and base, creating inductive shunts with cutoff $\omega_c \sim 1/a$, independent of enclosure size $L$. (c) Cross-section showing pillars interleaved between qubits. This off-chip approach requires no additional lithography and maintains tileability for scaling.}
  \label{F1-Sup}
\end{figure}

Figure~\ref{F1-Sup} presents the fabricated $4\times4$ superconducting qubit array and its engineered enclosure. Figure~\figpanelNoPrefix{F1-Sup}{a} shows an optical micrograph of the device. Each transmon consists of a single Josephson junction (fabricated via double-angle shadow evaporation) shunted by a capacitor. The readout resonators are patterned on the opposite chip face, physically isolating the resonator structures from the qubit plane to minimize parasitic capacitive crosstalk between circuit elements. Figure~\figpanelNoPrefix{F1-Sup}{b} and \figpanelNoPrefix{F1-Sup}{c} illustrate the inductive shunt pillar strategy developed to suppress enclosure-mediated coupling. A periodic array of cylindrical aluminum pillars directly connects the enclosure base and lid, with each pillar press-fit into precision-machined holes and bonded using indium for both mechanical stability and low-resistance electrical contact. These pillars act as distributed inductive shunts that modify the cavity electromagnetic mode structure, raising the fundamental mode frequency from $\omega_c^{\text{bare}}/2\pi \sim \qty{11}{\giga\hertz}$ (bare rectangular cavity) to $\omega_c^{\text{shunt}}/2\pi \sim \qty{34}{\giga\hertz}$ (shunted cavity), as validated by simulation in Appendix~\ref{app:spectator}.

The key design principle is that the cutoff frequency scales as $\omega_c/2\pi \propto 1/a$, where $a$ is the pillar spacing, independent of the total enclosure size $L$. This decoupling enables suppression of long-wavelength cavity modes without constraining the device footprint-essential for scaling to larger arrays. In our implementation, $a = \qty{2}{\milli\metre}$ pillar spacing yields $\omega_c / 2 \pi \approx \qty{34}{\giga\hertz}$, placing the \qtyrange[range-phrase = --, range-units = single]{4.8}{5.0}{\giga\hertz} qubit operating band firmly in the evanescent regime where $\omega_q < \omega_c$. Fields at qubit frequencies cannot propagate across the enclosure and instead decay exponentially with characteristic bound-state length $\delta_b \approx \qty{2}{\milli\metre}$ [Eq.~\eqref{eq:bound_state_length_circuit}], suppressing enclosure-mediated crosstalk by factors of $e^{-d_{ij}/\delta_b} \sim 0.5$--0.01 for qubit separations spanning \qtyrange[range-phrase = --, range-units = single]{1}{2}{\milli\metre}.

Practically, the pillar architecture requires precision-aligned through-wafer holes (\qty{500}{\micro\metre} diameter) in the silicon substrate, but critically involves no additional chip-level lithography or metallization. The pillars are entirely off-chip components, maintaining fabrication simplicity while achieving dramatic crosstalk reduction. The interleaved positioning between qubits preserves tileability: the same pillar lattice can be extended to arbitrarily large arrays without redesigning the qubit layout, making this approach directly scalable to surface-code-scale devices.

\subsection{Enclosure transmission measurements}

\begin{figure}[ht]
  \centering
   \includegraphics[width=0.85\textwidth]{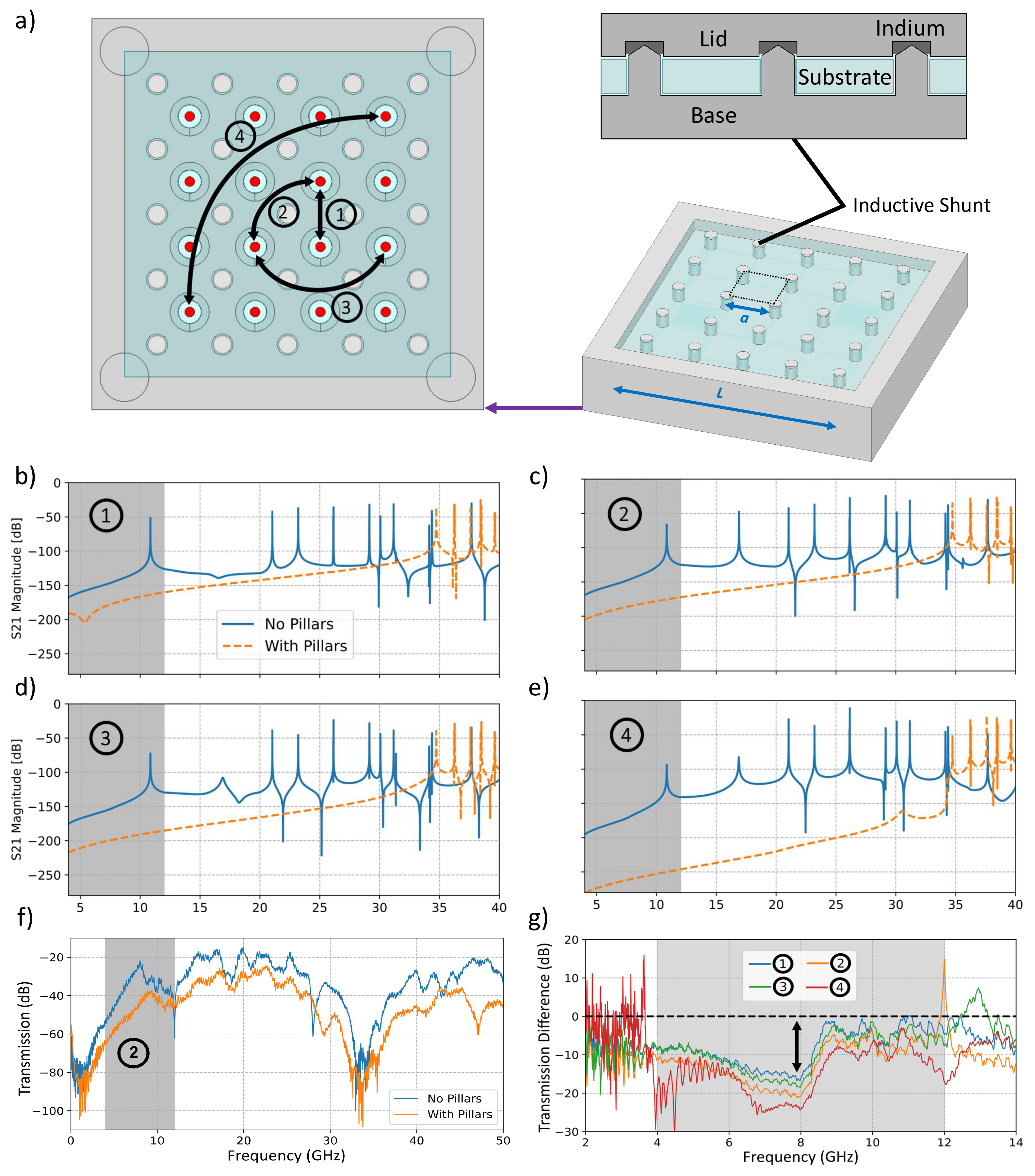}
\caption{Enclosure transmission measurements validate mode engineering.
(a) Geometry: pillar spacing $a$ sets cutoff $\omega_c \sim 1/a$, independent of size $L$.
(b--e) HFSS simulations show mode shift from $\sim \qty{6}{\giga\hertz}$ (bare) to $> \qty{13}{\giga\hertz}$ (shunted).
(f--g) Experimental S$_{21}$ at sub-K temperature confirms \qtyrange[range-phrase = --, range-units = single]{15}{25}{\decibel} isolation improvement with pillars, placing qubits firmly in the evanescent regime where $\omega_q < \omega_c$.}
  \label{F2-Sup}
\end{figure}

Figure~\ref{F2-Sup} validates the pillar-induced mode engineering through both electromagnetic simulation and cryogenic S$_{21}$ measurements. Figure~\figpanelNoPrefix{F2-Sup}{a} defines the geometry: a square enclosure of total size $L$ contains a periodic pillar array with spacing $a$, yielding cutoff frequency $\omega_c \sim 1/a$ from the circuit model (Appendix~\ref{greenfunction}). Figure~\figpanelNoPrefix{F2-Sup}{b}--\figpanelNoPrefix{F2-Sup}{e} show HFSS finite-element simulations of transmission between coaxial ports placed at representative qubit positions on the chip.

The simulations reveal a dramatic transformation of the mode structure. In the bare enclosure [Figure~\figpanelNoPrefix{F2-Sup}{b} and \figpanelNoPrefix{F2-Sup}{d}], transmission exhibits strong resonances starting near \qty{11}{\giga\hertz}. These near-resonant cavity modes provide efficient pathways for enclosure-mediated qubit–qubit coupling, limiting crosstalk suppression. With pillars inserted, the mode spectrum shifts entirely: the lowest resonances now appear above \qty{35}{\giga\hertz}, and the \qtyrange[range-phrase = --, range-units = single]{4}{10}{\giga\hertz} range shows significantly reduced transmission. This frequency shift directly confirms the plasma-model prediction [Eqs.~\eqref{eq:dispersion_relation}--\eqref{eq:quadratic_dispersion}]: the inductive shunts create an effective high-pass filter that suppresses long-wavelength electromagnetic modes.

Figure~\figpanelNoPrefix{F2-Sup}{f} and \figpanelNoPrefix{F2-Sup}{g} present experimental validation through S$_{21}$ transmission measurements performed at sub-Kelvin temperature. We compare multiple port pairs in enclosures with and without the pillar array, spanning the frequency range relevant to device operation. While individual cavity resonances are not sharply resolved---due to mode overlap, finite port coupling, and residual loss---the broadband suppression is unmistakable. Across all port pairs measured, the pillar-equipped enclosure shows \qtyrange[range-phrase = --, range-units = single]{15}{25}{\decibel} higher isolation over the \qtyrange[range-phrase = --, range-units = single]{4}{10}{\giga\hertz} range compared to the bare enclosure. This isolation improvement directly translates to exponentially suppressed qubit–qubit coupling mediated by cavity fields. This engineered electromagnetic environment, combined with the naturally localized capacitance matrix structure (Appendix~\ref{matrixinversion}), achieves the exponential spatial suppression of crosstalk observed in Section~\ref{sec:results}---without sacrificing nearest-neighbor coupling strength or requiring complex mode-matching structures.

\section{Numerical simulation: field distributions and cutoff engineering}
\label{app:simulations}
To validate the electromagnetic model developed in Appendix~\ref{greenfunction} and quantify the effect of inductive shunting on the enclosure mode structure, we performed full-wave finite-element simulations of the cavity enclosure with and without the periodic pillar array. These simulations demonstrate how the inductive shunts raise the electromagnetic cutoff frequency, pushing propagating modes above the qubit operating band and creating the evanescent regime essential for crosstalk suppression.

\subsection{Effect of inductive shunting on mode spectrum}
The key observation is that the fundamental mode frequency increases from \qty{11}{\giga\hertz} (bare enclosure) to \qty{34}{\giga\hertz} (shunted enclosure). This directly validates the plasma-model prediction from Appendix~\ref{greenfunction}: the periodic inductive shunts create an effective plasma medium with raised cutoff frequency $\omega_c$. We recall that our qubits operate at 4.8--5.0~GHz, which is below the shunted-enclosure cutoff of \qty{34}{\giga\hertz}. In the bare enclosure, the fundamental mode at \qty{11}{\giga\hertz} is only $2 \times$ above the qubit band, allowing near-resonant enclosure-mediated coupling. The inductive shunts move the device firmly into the evanescent regime where $\omega_q < \omega_c$, as required by Eq.~\eqref{eq:evanescent_wave_number}.

In addition, the shunted enclosure exhibits a higher density of modes above cutoff, with many closely spaced resonances. This is consistent with the formation of a dense photonic band structure above the plasma frequency, as predicted by the circuit model in Eqs.~\eqref{eq:dispersion_relation}--\eqref{eq:quadratic_dispersion}. From the circuit parameters extracted in Appendix~\ref{greenfunction}, the predicted cutoff is $\omega_c = \omega_0/\sqrt{1 + 4\beta}$, where $\beta = L_g/L_0$ is the coupling parameter. The observed cutoff shift implies $\beta \approx 0.35$, consistent with the pillar geometry and spacing used in the device.

\subsection{Spatial field distributions}
\begin{figure}[ht]
    \centering
    \setlength{\tabcolsep}{6pt} 

    \begin{tabular}{ccc}
        \includegraphics[width=0.30\textwidth]{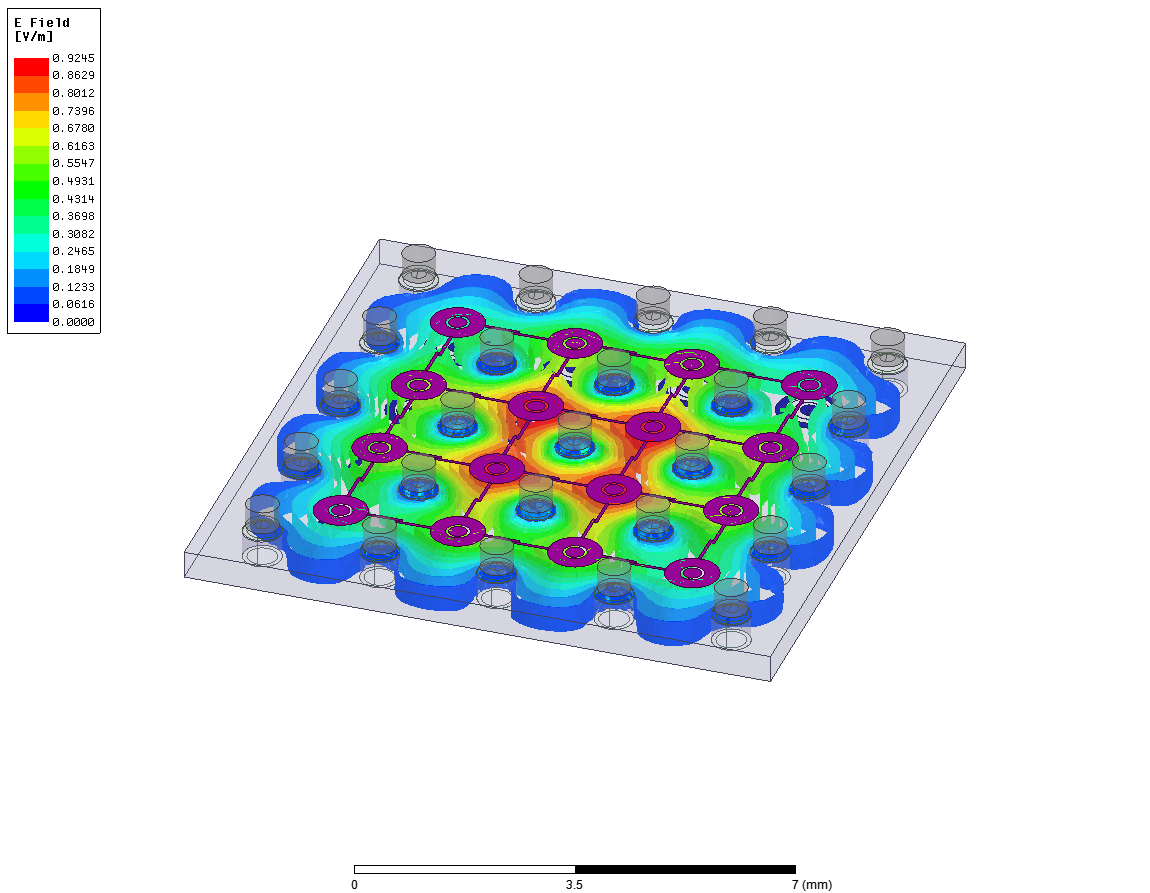} &
        \includegraphics[width=0.30\textwidth]{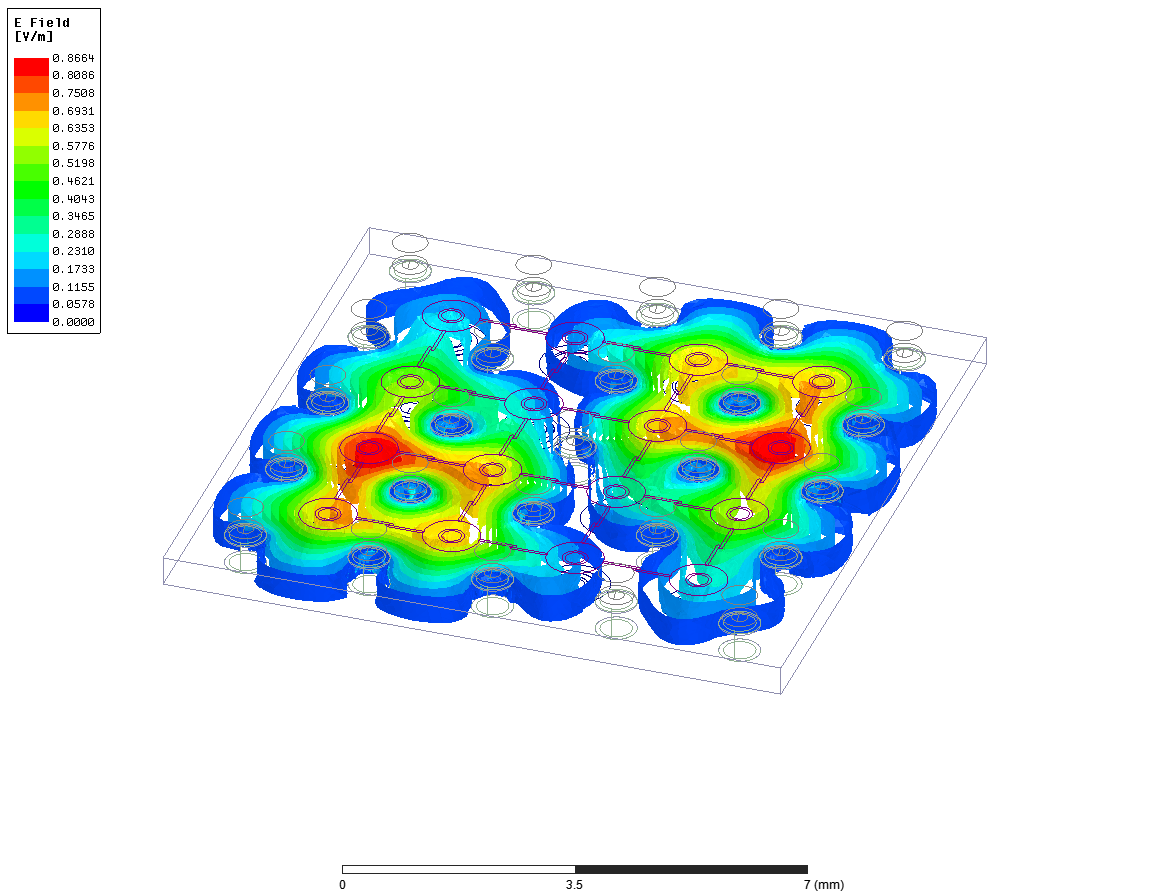} &
        \includegraphics[width=0.30\textwidth]{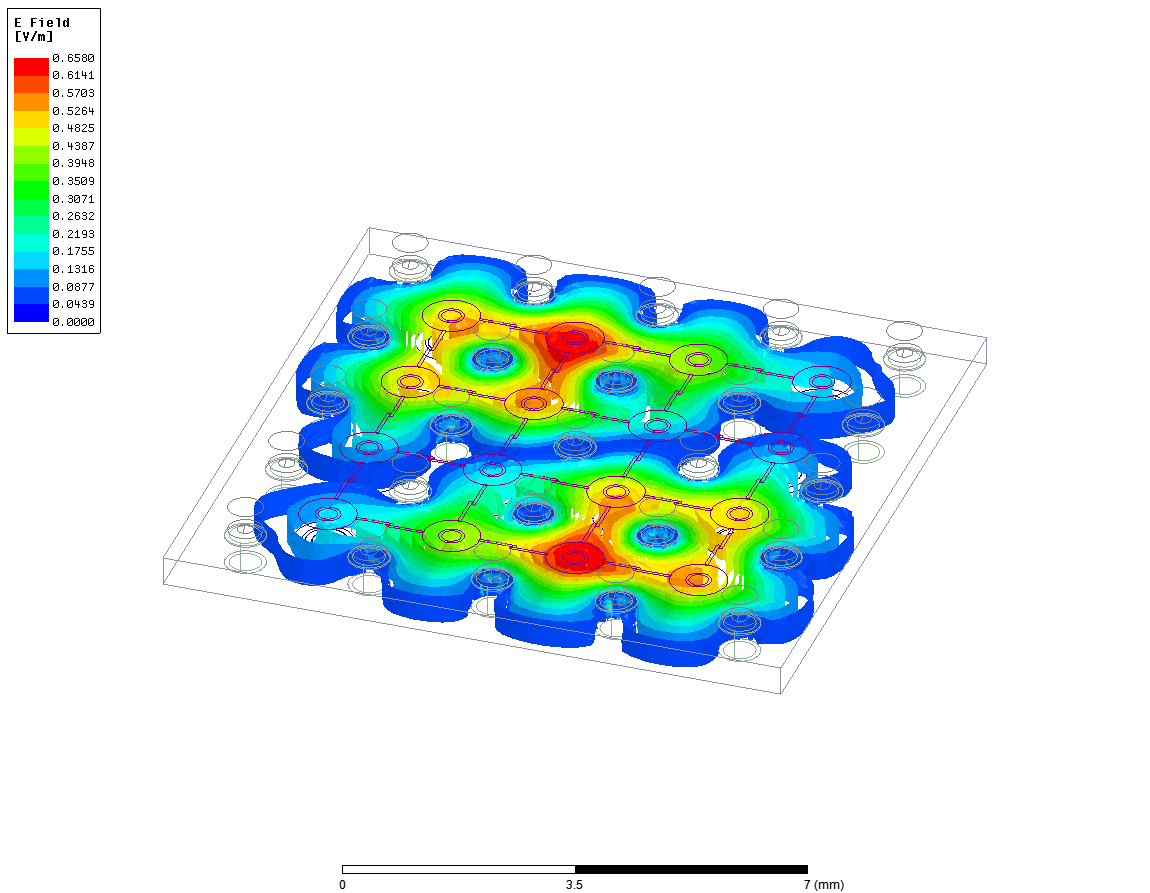} \\
        {\small (a) Mode 1 (\qty{34.08}{\giga\hertz})} &
        {\small (b) Mode 2 (\qty{35.28}{\giga\hertz})} &
        {\small (c) Mode 3 (\qty{35.28}{\giga\hertz})} \\
        \addlinespace[0.8em]
        \includegraphics[width=0.30\textwidth]{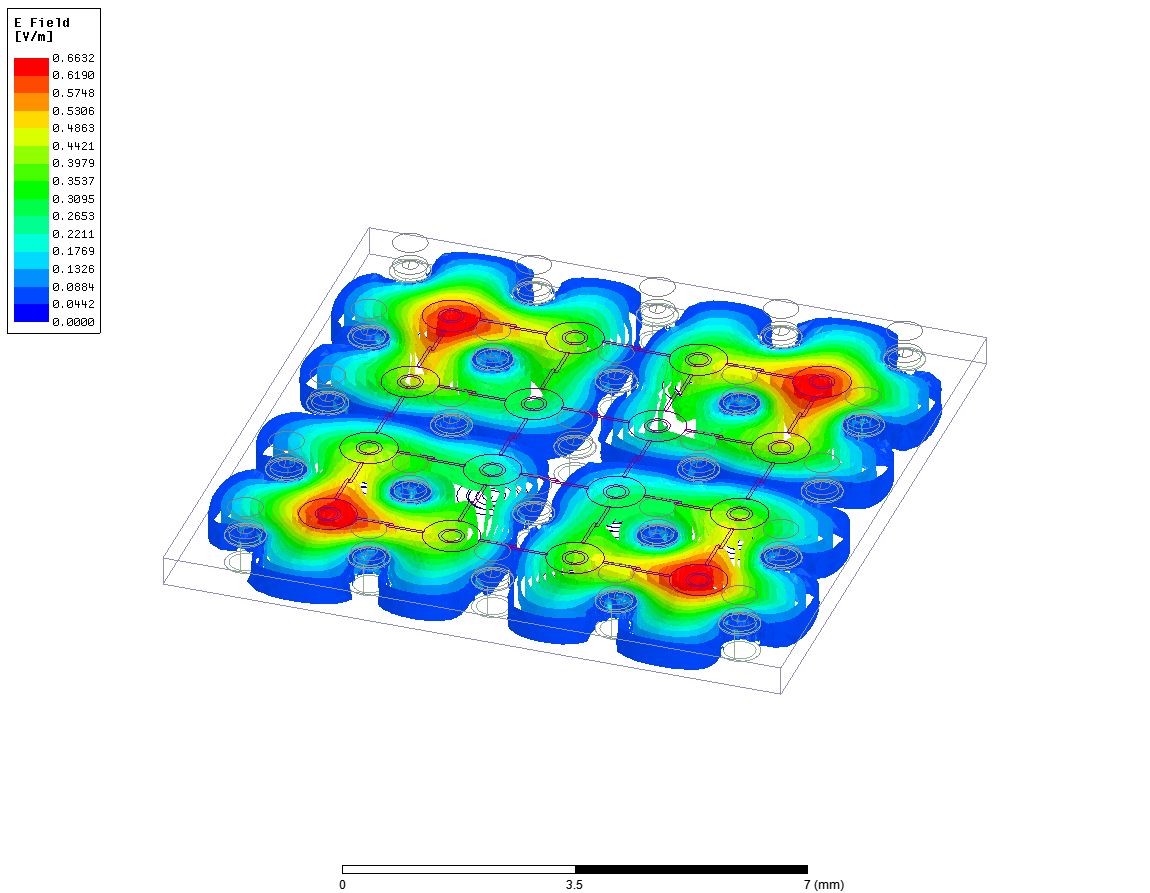} &
        \includegraphics[width=0.30\textwidth]{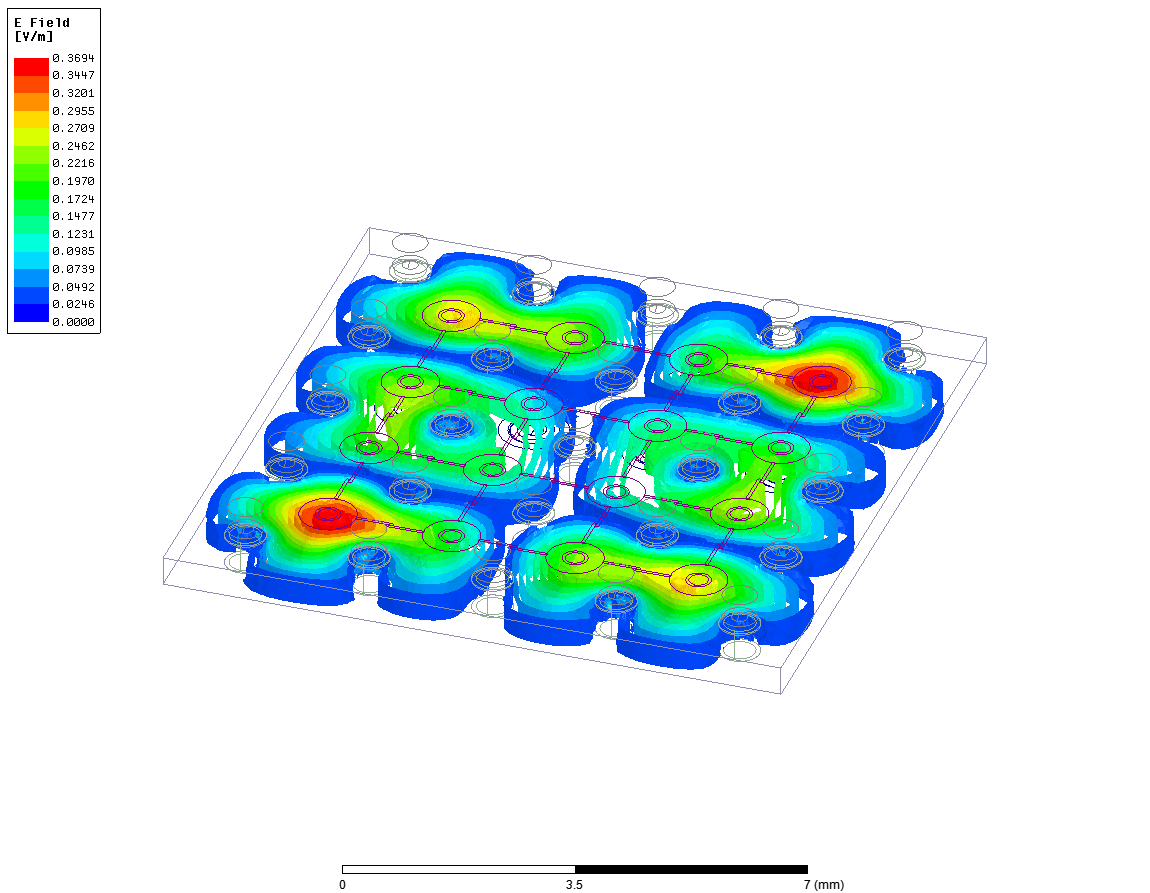} &
        \includegraphics[width=0.30\textwidth]{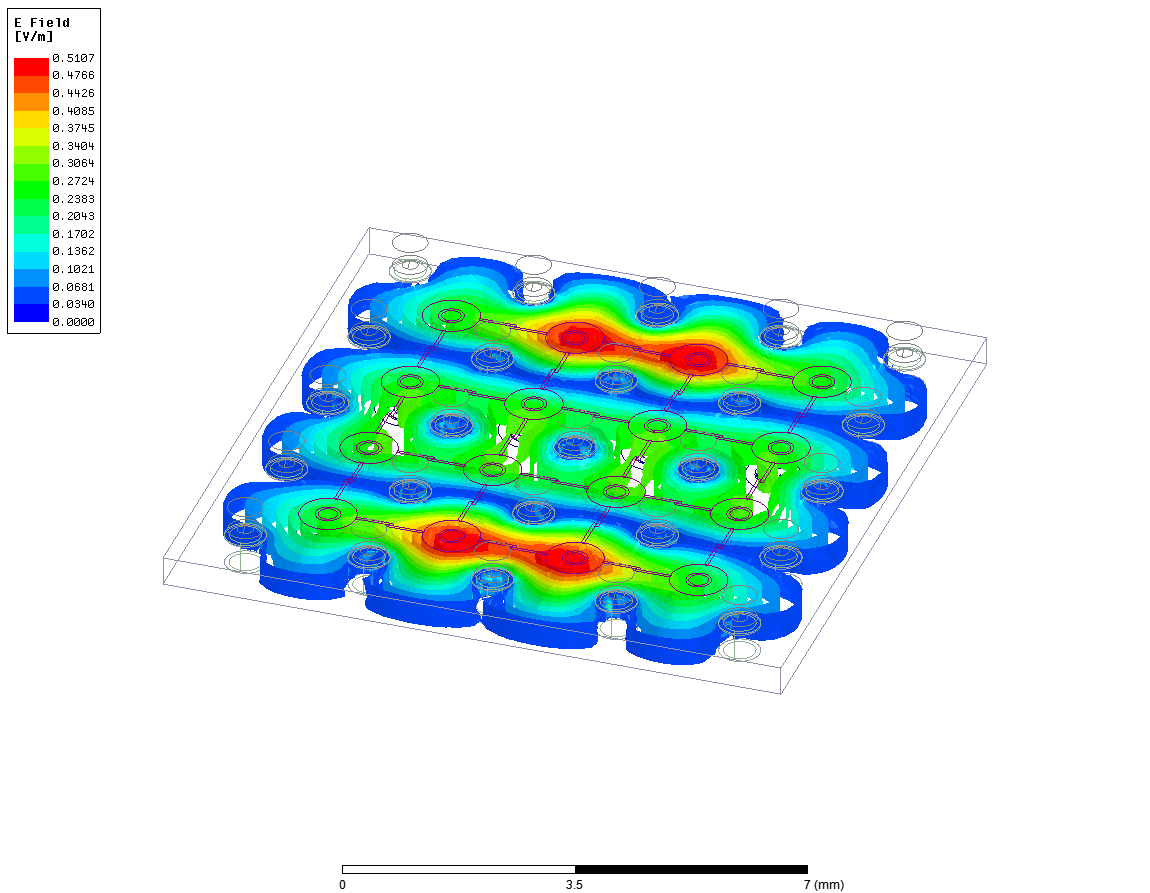} \\
        {\small (d) Mode 4 (\qty{36.43}{\giga\hertz})} &
        {\small (e) Mode 5 (\qty{36.99}{\giga\hertz})} &
        {\small (f) Mode 6 (\qty{36.99}{\giga\hertz})} \\
    \end{tabular}

    \caption{Simulated electric field distributions of the first six modes of the shunted enclosure. All mode frequencies lie well above the qubit operating band (4.8--5.0~GHz), confirming that the device operates in the evanescent regime where enclosure-mediated coupling is exponentially suppressed with distance. The enclosure geometry is omitted for clarity. Degenerate mode pairs (2--3 and 5--6) reflect the four-fold symmetry of the structure.}
    \label{fig:matrix_figures}
\end{figure}

Figure~\ref{fig:matrix_figures} shows the numerically simulated electric field distributions for the first six enclosure modes in the presence of inductive shunt pillars. The eigenmodes are plotted without the enclosing geometry for clarity, emphasizing the spatial field patterns. The simulated frequencies are 34.08, 35.28, 35.28, 36.43, 36.99, and \qty{35.99}{\giga\hertz}, corresponding to the lowest resonances of the modified enclosure---all well above the 4.8--5.0~GHz qubit band.

Several features are evident from the field patterns. Modes 2--3 and 5--6 exhibit near degeneracy, reflecting the four-fold rotational symmetry of the square enclosure and the periodic unit-cell structure of the via array. The small frequency splitting arises from weak symmetry breaking due to port locations. Further, all modes show clear standing-wave patterns with nodes and antinodes, characteristic of resonances in a bounded cavity. The wavelengths are short compared to the enclosure size ($\lambda \sim$ 1--2~cm for modes at 34--36~GHz). Importantly, at the qubit operating frequencies (4.8--5.0~GHz), which lie far below the fundamental mode at \qty{34}{\giga\hertz}, no propagating modes exist. Fields at qubit frequencies are purely evanescent, decaying exponentially from any source with the characteristic length scale $\delta_b = v/\sqrt{2\omega_c(\omega_c - \omega_q)} \approx \qty{2.1}{\centi\metre}$ (using the parameters from Appendix~\ref{greenfunction}). This evanescent nature is the key to crosstalk suppression: qubits separated by distances $d_{ij} \gg \delta_b$ experience exponentially suppressed enclosure-mediated coupling, as observed experimentally in Section~\ref{sec:results}.

\newpage
\bibliographystyle{nature}

\begin{thebibliography}{99}
\bibitem{Arute2019}
F.~Arute, K.~Arya, R.~Babbush \emph{et al.}, ``Quantum supremacy using a programmable superconducting processor,'' 
\href{https://doi.org/10.1038/s41586-019-1666-5}{\emph{Nature} \textbf{574}, 505--510 (2019)}.

\bibitem{megrant2025scaling}
A.~Megrant and Y.~Chen, ``Scaling up superconducting quantum computers,'' 
\href{https://doi.org/10.1038/s41928-025-01381-7}{\emph{Nat. Electron.} \textbf{8}, 549--551 (2025)}.

\bibitem{google2023suppressing}
R.~P.~Acharya, A.~A.~Liener, R.~A.~Liener \emph{et al.}, ``Suppressing quantum errors by scaling a surface code logical qubit,'' 
\href{https://doi.org/10.1038/s41586-023-05784-4}{\emph{Nature} \textbf{614}, 676--681 (2023)}.

\bibitem{acharya2025quantum}
R.~P.~Acharya, L.~Aghababaie-Beni, I.~Aleiner \emph{et al.}, ``Quantum error correction below the surface code threshold,'' 
\href{https://doi.org/10.1038/s41586-024-08449-y}{\emph{Nature} \textbf{638}, 920--926 (2025)}.

\bibitem{mohseni2024build}
M.~Mohseni, A.~Scherer, K.~G.~Johnson \emph{et al.},
``How to build a quantum supercomputer: Scaling challenges and opportunities,''
\href{https://arxiv.org/abs/2411.10406}
{arXiv:2411.10406 [quant-ph] (2024)}.


\bibitem{krinner2020benchmarking}
S.~Krinner, S.~Lazar, A.~Remm \emph{et al.}, ``Benchmarking coherent errors in controlled-phase gates due to spectator qubits,'' 
\href{https://doi.org/10.1103/PhysRevApplied.14.024042}{\emph{Phys. Rev. Appl.} \textbf{14}, 024042 (2020)}.

\bibitem{shor1995scheme}
P.~W.~Shor, ``Scheme for reducing decoherence in quantum computer memory,'' 
\href{https://doi.org/10.1103/PhysRevA.52.R2493}{\emph{Phys. Rev. A} \textbf{52}, R2493 (1995)}.


\bibitem{lacroix2024scaling}
N.~Lacroix, A.~Bourassa, F.~J.~H.~Heras \emph{et al.}, ``Scaling and logic in the color code on a superconducting quantum processor,'' 
\href{https://arxiv.org/abs/2412.14256}{arXiv:2412.14256 [quant-ph]} (2024).

\bibitem{eickbusch2024demonstrating}
A.~Eickbusch, M.~McEwen, V.~Sivak \emph{et al.}, ``Demonstrating dynamic surface codes,'' 
\href{https://arxiv.org/abs/2412.14360}{arXiv:2412.14360 [quant-ph]} (2024).

\bibitem{le2023scalable}
N.~H.~Le, M.~Cykiert, and E.~Ginossar, ``Scalable and robust quantum computing on qubit arrays with fixed coupling,'' 
\href{https://doi.org/10.1038/s41534-022-00668-3}{\emph{npj Quantum Inf.} \textbf{9}, 1 (2023)}.

\bibitem{transmon}
J.~Koch, T.~M.~Yu, J.~Gambetta \emph{et al.}, ``Charge-insensitive qubit design derived from the Cooper pair box,'' 
\href{https://doi.org/10.1103/PhysRevA.76.042319}{\emph{Phys. Rev. A} \textbf{76}, 042319 (2007)}.

\bibitem{alghadeer2025characterization}
M.~Alghadeer, S.~D.~Fasciati, S.~Cao \emph{et al.},
``Characterization of nanostructural imperfections in superconducting quantum circuits,''
\href{https://doi.org/10.1088/2633-4356/aded66}
{\emph{Mater. Quantum Technol.} \textbf{5}, 035201 (2025)}.


\bibitem{yan2018tunable}
F.~Yan, P.~Krantz, Y.~Sung \emph{et al.}, ``Tunable coupling scheme for implementing high-fidelity two-qubit gates,'' 
\href{https://doi.org/10.1103/PhysRevApplied.10.054062}{\emph{Phys. Rev. Appl.} \textbf{10}, 054062 (2018)}.

\bibitem{ketterer2023characterizing}
A.~Ketterer and T.~Wellens, ``Characterizing crosstalk of superconducting transmon processors,'' 
\href{https://doi.org/10.1103/PhysRevApplied.20.034065}{\emph{Phys. Rev. Appl.} \textbf{20}, 034065 (2023)}.

\bibitem{tripathi2022suppression}
V.~Tripathi, H.~Chen, M.~Khezri \emph{et al.}, ``Suppression of crosstalk in superconducting qubits using dynamical decoupling,'' 
\href{https://doi.org/10.1103/PhysRevApplied.18.024068}{\emph{Phys. Rev. Appl.} \textbf{18}, 024068 (2022)}.

\bibitem{murali2020software}
P.~Murali, D.~C.~McKay, M.~Martonosi, and A.~Javadi-Abhari, ``Software mitigation of crosstalk on noisy intermediate-scale quantum computers,'' 
in \emph{Proc. 25th Int. Conf. on Architectural Support for Programming Languages and Operating Systems (ASPLOS)}, 1001--1016 (2020).

\bibitem{zhao2022quantum}
P.~Zhao, K.~Linghu, Z.~Li \emph{et al.}, ``Quantum crosstalk analysis for simultaneous gate operations on superconducting qubits,'' 
\href{https://doi.org/10.1103/PRXQuantum.3.020301}{\emph{PRX Quantum} \textbf{3}, 020301 (2022)}.

\bibitem{fors2024comprehensive}
S.~Pettersson~Fors, J.~Fern{\'a}ndez-Pend{\'a}s, and A.~F.~Kockum,
``Comprehensive explanation of ZZ coupling in superconducting qubits,''
\href{https://arxiv.org/abs/2408.15402}
{arXiv:2408.15402 [quant-ph] (2024)}.


\bibitem{chow2011simple}
J.~M.~Chow, A.~D.~C{\'o}rcoles, J.~M.~Gambetta \emph{et al.}, ``Simple all-microwave entangling gate for fixed-frequency superconducting qubits,'' 
\href{https://doi.org/10.1103/PhysRevLett.107.080502}{\emph{Phys. Rev. Lett.} \textbf{107}, 080502 (2011)}.

\bibitem{collodo2020implementation}
M.~C.~Collodo, J.~Herrmann, N.~Lacroix \emph{et al.}, ``Implementation of conditional phase gates based on tunable ZZ interactions,'' 
\href{https://doi.org/10.1103/PhysRevLett.125.240502}{\emph{Phys. Rev. Lett.} \textbf{125}, 240502 (2020)}.

\bibitem{sung2021realization}
Y.~Sung, L.~Ding, J.~Braum{\"u}ller \emph{et al.}, ``Realization of high-fidelity CZ and ZZ-free iSWAP gates with a tunable coupler,'' 
\href{https://doi.org/10.1103/PhysRevX.11.021058}{\emph{Phys. Rev. X} \textbf{11}, 021058 (2021)}.

\bibitem{stehlik2021tunable}
J.~Stehlik, D.~M.~Zajac, D.~L.~Underwood \emph{et al.}, ``Tunable coupling architecture for fixed-frequency transmon superconducting qubits,'' 
\href{https://doi.org/10.1103/PhysRevLett.127.080505}{\emph{Phys. Rev. Lett.} \textbf{127}, 080505 (2021)}.

\bibitem{long2021universal}
J.~Long, T.~Zhao, M.~Bal \emph{et al.}, ``A universal quantum gate set for transmon qubits with strong ZZ interactions,'' 
\href{https://arxiv.org/abs/2103.12305}{arXiv:2103.12305 [quant-ph]} (2021).

\bibitem{chu2021coupler}
J.~Chu and F.~Yan, ``Coupler-assisted controlled-phase gate with enhanced adiabaticity,'' 
\href{https://doi.org/10.1103/PhysRevApplied.16.054020}{\emph{Phys. Rev. Appl.} \textbf{16}, 054020 (2021)}.

\bibitem{chen2023voltage}
Y.~Chen, K.~N.~Nesterov, H.~Churchill, J.~Shabani, V.~E.~Manucharyan, and M.~G.~Vavilov, ``Voltage-activated parametric entangling gates based on gatemon qubits,'' 
\href{https://doi.org/10.1103/PhysRevApplied.20.044012}{\emph{Phys. Rev. Appl.} \textbf{20}, 044012 (2023)}.

\bibitem{ganzhorn2020benchmarking}
M.~Ganzhorn, G.~Salis, D.~J.~Egger \emph{et al.}, ``Benchmarking the noise sensitivity of different parametric two-qubit gates in a single superconducting quantum computing platform,'' 
\href{https://doi.org/10.1103/PhysRevResearch.2.033447}{\emph{Phys. Rev. Res.} \textbf{2}, 033447 (2020)}.

\bibitem{mckay2019three}
D.~C.~McKay, S.~Sheldon, J.~A.~Smolin, J.~M.~Chow, and J.~M.~Gambetta, ``Three-qubit randomized benchmarking,'' 
\href{https://doi.org/10.1103/PhysRevLett.122.200502}{\emph{Phys. Rev. Lett.} \textbf{122}, 200502 (2019)}.

\bibitem{bharti2022noisy}
K.~Bharti, A.~Cervera-Lierta, T.~H.~Kyaw \emph{et al.}, ``Noisy intermediate-scale quantum algorithms,'' 
\href{https://doi.org/10.1103/RevModPhys.94.015004}{\emph{Rev. Mod. Phys.} \textbf{94}, 015004 (2022)}.

\bibitem{Bakr2025JJMetasurfaces}
M.~Bakr,
``Dynamic Josephson-junction metasurfaces for multiplexed control of superconducting qubits,''
\href{https://doi.org/10.1103/n6gs-zlhc}
{\emph{Phys. Rev. Appl.} \textbf{24}, 054069 (2025)}.

\bibitem{Bakr2025LongRangeJJM}
M.~Bakr,
``Long-range entangling operations via Josephson junction metasurfaces,''
\href{https://arxiv.org/abs/2506.14958}
{arXiv:2506.14958 [quant-ph] (2025)}.

\bibitem{Bakr2025ReentrantReadout}
M.~Bakr, S.~D.~Fasciati, S.~Cao \emph{et al.},
``Multiplexed readout of superconducting qubits using a three-dimensional reentrant-cavity filter,''
\href{https://doi.org/10.1103/PhysRevApplied.23.054089}
{\emph{Phys. Rev. Appl.} \textbf{23}, 054089 (2025)}.

\bibitem{Fasciati2024GeometricShunt}
S.~D.~Fasciati, B.~Shteynas, G.~Campanaro \emph{et al.},
``Complementing the transmon by integrating a geometric shunt inductor,''
\href{https://arxiv.org/abs/2410.10416}
{arXiv:2410.10416 [quant-ph] (2024)}.

\bibitem{kosen2024signal}
S.~Kosen, H.-X.~Li, M.~Rommel \emph{et al.}, ``Signal crosstalk in a flip-chip quantum processor,'' 
\href{https://doi.org/10.1103/PRXQuantum.5.030350}{\emph{PRX Quantum} \textbf{5}, 030350 (2024)}.

\bibitem{das2024reworkable}
R.~N.~Das, J.~Cummings, T.~Hazard \emph{et al.}, ``Reworkable superconducting qubit package for quantum computing,'' 
in \emph{Proc. IEEE 74th Electronic Components and Technology Conf. (ECTC)}, 427--432 (2024).

\bibitem{huang2021microwave}
S.~Huang, B.~Lienhard, G.~Calusine \emph{et al.}, ``Microwave package design for superconducting quantum processors,'' 
\href{https://doi.org/10.1103/PRXQuantum.2.020306}{\emph{PRX Quantum} \textbf{2}, 020306 (2021)}.

\bibitem{spring2020modeling}
P.~A.~Spring, T.~Tsunoda, B.~Vlastakis, and P.~J.~Leek, ``Modeling enclosures for large-scale superconducting quantum circuits,'' 
\href{https://doi.org/10.1103/PhysRevApplied.14.024061}{\emph{Phys. Rev. Appl.} \textbf{14}, 024061 (2020)}.

\bibitem{spring2022high}
P.~A.~Spring, S.~Cao, T.~Tsunoda \emph{et al.}, ``High coherence and low cross-talk in a tileable 3D integrated superconducting circuit architecture,'' 
\href{https://doi.org/10.1126/sciadv.abl6698}{\emph{Sci. Adv.} \textbf{8}, eabl6698 (2022)}.

\bibitem{Mathews2025}
M.~Mathews, L.~Pahl, D.~Pahl \emph{et al.},
``Placing and routing quantum LDPC codes in multilayer superconducting hardware,''
\href{https://arxiv.org/abs/2507.23011}
{arXiv:2507.23011 [quant-ph] (2025)}.


\bibitem{blaiscqed}
A.~Blais, A.~L.~Grimsmo, S.~M.~Girvin, and A.~Wallraff, 
``Circuit quantum electrodynamics,'' 
\href{https://doi.org/10.1103/RevModPhys.93.025005}{\emph{Rev. Mod. Phys.} \textbf{93}, 025005 (2021)}.

\bibitem{YaoBakrHunter2019Modal}
Y.~Yao, M.~S.~Bakr, and I.~C.~Hunter,
``Modal analysis and same-bandedge response optimization of 3-D lumped networks,''
in \emph{Proc. 49th European Microwave Conference (EuMC)} (2019), pp.~9--12.
\href{https://doi.org/10.23919/EuMC.2019.8910863}{}

\bibitem{MusondaBakr2023Singlet}
E.~Musonda and M.~S.~Bakr,
``The singlet: Direct synthesis of pseudo-elliptic inline filters with frequency-variant couplings,''
\href{https://doi.org/10.1109/TMTT.2023.3269516}
{\emph{IEEE Trans. Microw. Theory Techn.} \textbf{71}, 4969--4981 (2023)}.

\bibitem{swtransform}
S.~Bravyi, D.~P.~DiVincenzo, and D.~Loss, 
``Schrieffer–Wolff transformation for quantum many-body systems,'' 
\href{https://doi.org/10.1016/j.aop.2011.06.004}{\emph{Ann. Phys.} \textbf{326}, 2793–2826 (2011)}.

\bibitem{rw1}
C.~W.~Gardiner and P.~Zoller,
\textit{Quantum Noise: A Handbook of Markovian and Non-Markovian Quantum Stochastic Methods with Applications to Quantum Optics}, 3rd ed.
(Springer, Berlin, 2004).

\bibitem{rw2}
M.~Boissonneault, J.~M.~Gambetta, and A.~Blais,
``Dispersive regime of circuit QED: Photon-dependent qubit dephasing and relaxation rates,''
\href{https://doi.org/10.1103/PhysRevA.79.013819}{\emph{Phys. Rev. A} \textbf{79}, 013819 (2009)}.

\bibitem{AE}
D.~Finkelstein\mbox{-}Shapiro, D.~Viennot, I.~Saideh, T.~Hansen, T\~{o}nu~Pullerits, and A.~Keller,
``Adiabatic elimination and subspace evolution of open quantum systems,''
\href{https://doi.org/10.1103/PhysRevA.101.042102}{\emph{Phys. Rev. A} \textbf{101}, 042102 (2020)}.

\bibitem{Bakr2025IntrinsicMMI}
M.~Bakr, M.~Alghadeer, S.~Pettersson~Fors \emph{et al.},
``Intrinsic multi-mode interference for passive suppression of Purcell decay in superconducting circuits,''
\href{https://arxiv.org/abs/2507.09715}
{arXiv:2507.09715 [quant-ph] (2025)}.

\bibitem{solims}
F.~Solgun, 
``Microwave Engineer’s Guide to the Design of Superconducting Qubit Circuits,'' 
\href{https://doi.org/10.1109/MWSYM.2019.8700862}{in \emph{Proc. IEEE MTT-S Int. Microw. Symp. (IMS)}, 263–266 (2019)}.

\bibitem{solgun2019simple}
F.~Solgun, D.~P.~DiVincenzo, and J.~M.~Gambetta, 
``Simple impedance response formulas for the dispersive interaction rates in the effective Hamiltonians of low anharmonicity superconducting qubits,'' 
\href{https://doi.org/10.1109/TMTT.2018.2875594}{\emph{IEEE Trans. Microw. Theory Techn.} \textbf{67}, 928--948 (2019)}.

\bibitem{solgun2022direct}
F.~Solgun and S.~Srinivasan, 
``Direct calculation of ZZ interaction rates in multimode circuit quantum electrodynamics,'' 
\href{https://doi.org/10.1103/PhysRevApplied.18.044025}{\emph{Phys. Rev. Appl.} \textbf{18}, 044025 (2022)}.

\bibitem{alghadeer2025low}
M.~Alghadeer, S.~Cao, S.~D.~Fasciati \emph{et al.},
``Low crosstalk in a scalable superconducting quantum lattice,''
\href{https://arxiv.org/abs/2505.22276}
{arXiv:2505.22276 [quant-ph] (2025)}.


\bibitem{Mundada2019ZZ}
P.~Mundada, G.~Zhang, T.~Hazard, and A.~A.~Houck,
``Suppression of qubit crosstalk in a tunable coupling superconducting circuit,''
\href{https://doi.org/10.1103/PhysRevApplied.12.054023}
{\emph{Phys. Rev. Appl.} \textbf{12}, 054023 (2019)}.

\bibitem{Cao2025AutoLabAI}
S.~Cao, Z.~Zhang, M.~Alghadeer \emph{et al.},
``Automating quantum computing laboratory experiments with an agent-based AI framework,''
\href{https://doi.org/10.1016/j.patter.2025.101372}
{\emph{Patterns} \textbf{6}, 101372 (2025)}.

\bibitem{petrescuAccurateMethodsAnalysis2023}
A.~Petrescu, C.~L.~Calonnec, C.~Leroux \emph{et al.},
``Accurate methods for the analysis of strong-drive effects in parametric gates,''
\href{https://doi.org/10.1103/PhysRevApplied.19.044003}
{\emph{Phys. Rev. Appl.} \textbf{19}, 044003 (2023)}.


\bibitem{leibNetworksNonlinearSuperconducting2012}
M.~Leib, F.~Deppe, A.~Marx, R.~Gross, and M.~J.~Hartmann,
``Networks of Nonlinear Superconducting Transmission Line Resonators,''
\href{https://doi.org/10.1088/1367-2630/14/7/075024}{\emph{New J. Phys.} \textbf{14}, 075024 (2012)}.

\bibitem{mikhailovOrderingBosonOperator1983a}
V.~V.~Mikhailov,
``Ordering of Some Boson Operator Functions,''
\href{https://dx.doi.org/10.1088/0305-4470/16/16/019}{\emph{J. Phys. A: Math. Gen.} \textbf{16}, 3817 (1983)}.

\end{thebibliography}

\end{document}